\let\ifarxiv\iftrue
\let\ifappendix\iftrue

\ifcsname ifdraft\endcsname\else
  \expandafter\let\csname ifdraft\expandafter\endcsname
                  \csname iffalse\endcsname
\fi

\ifcsname ifappendix\endcsname\else
  \expandafter\let\csname ifappendix\expandafter\endcsname
                  \csname iffalse\endcsname
\fi

\ifcsname ifsub\endcsname\else
  \expandafter\let\csname ifsub\expandafter\endcsname
                  \csname iffalse\endcsname
\fi

\ifcsname ifarxiv\endcsname\else
  \expandafter\let\csname ifarxiv\expandafter\endcsname
                  \csname iffalse\endcsname
\fi

\ifarxiv
\documentclass[acmsmall,screen,nonacm]{acmart}
\pdfoutput=1 
\else
\documentclass[acmsmall,screen]{acmart}
\fi

\setcopyright{rightsretained}
\acmPrice{}
\acmDOI{10.1145/3360587}
\acmYear{2019}
\copyrightyear{2019}
\acmJournal{PACMPL}
\acmVolume{3}
\acmNumber{OOPSLA}
\ifarxiv\acmArticle{}\else\acmArticle{161}\fi
\acmMonth{10}

\usepackage{amssymb,amsmath,amsthm}
\usepackage{latexsym}
\usepackage{graphicx}
\usepackage{textcomp}
\usepackage{verbatim} 
\usepackage{xspace} 
\usepackage{array, booktabs, longtable, makecell} 
\usepackage{afterpage}
\usepackage{mathpartir} 
\usepackage{cancel}
\usepackage{tikz}
\usetikzlibrary{arrows,automata,calc,decorations.text,shapes,cd,arrows,arrows.meta}
\usetikzlibrary{backgrounds,positioning,patterns}
\pgfdeclarelayer{background}
\pgfdeclarelayer{foreground}
\pgfsetlayers{background,main,foreground}
\usepackage{xxcolor}
\usepackage{multirow}
\usepackage{stmaryrd}
\usepackage{dsfont}
\usepackage{enumitem}

\usepackage{booktabs}   
\usepackage{subcaption} 
\usepackage{natbib}                        
\usepackage{dashbox}

\usepackage{skull,ifoddpage,marginnote}

\newcounter{ToDos}
\newcounter{WarnCounts}

\newenvironment{DIFnomarkup}{}{}
           
\newcommand{\act}[1]{\mathsf{#1}}
\newcommand{\hmapsto}{\Mapsto}
\newcommand{\theheap}{\sigma}

\newcommand{\themutex}{\mu}

\newcommand{\thehist}{\tau}
\newcommand{\ldot}{\mathord{.}\,}
\newcommand{\eqdef}{\mathrel{\:\widehat{=}\:}}

\newcommand{\vrf}{\mathsf{vrf}}
\newcommand{\wpeq}[2]{\vrf\ #1\ #2}
\newcommand{\wpeqs}[3]{\wpeq{#1}{#2}\ #3}

\newcommand{\specK}[1]{\ensuremath{\textcolor{blue}{#1}}}

\newcommand{\spec}[1]{\specK{\left\{{#1}\right\}}}

\newcommand{\sspec}[1]{\specK{\{{#1}\}}}
\newcommand{\sspecopen}[1]{\specK{\{{#1}\hphantom{\}}}}
\newcommand{\opensspec}[1]{\specK{\hphantom{\{}{#1}\}}}

\newcommand*{\tyUnit}{\ensuremath{\mathsf{unit}}}

\newcommand{\pcmA}{M}
\newcommand{\pcmF}{\bullet}
\newcommand{\join}{\pcmF}
\newcommand{\pcmU}{\mathds{1}}
\newcommand{\own}{\mathsf{own}}
\newcommand{\nown}{\overline{\own}}

\newcommand{\hflat}[1]{\ulcorner{#1}\urcorner}
\newcommand{\zag}{\lhd}
\newcommand{\zig}{\rhd}
\newcommand{\heaptp}{\mathsf{heap}}

\newcommand{\transpo}[1]{#1^{\scriptscriptstyle\top}}
\newcommand{\sfst}[1]{#1\backslash 1}
\newcommand{\ssnd}[1]{#1\backslash 2}
\newcommand{\sproj}[2]{#2\backslash{#1}}
\newcommand{\spair}[2]{[#1, #2]}
\usepackage{amsmath}
\makeatletter
\newcommand\ostep[2][]{\ext@arrow 0099{\longrightarrowfill@}{#1}{#2}}
\def\longrightarrowfill@{\arrowfill@{\ \ \ }\relbar\longrightarrow}
\makeatother

\makeatletter
\newcommand\osteps[2][]{\ext@arrow 0099{\longrightarrowfillstar@}{#1}{#2}}
\def\longrightarrowfillstar@{\arrowfill@{\ \ \ }\relbar{\longrightarrow^*}}
\makeatother

\makeatletter
\newcommand\mstep[2][]{\ext@arrow 0099{\longrightarrowfill@}{#1}{#2}}
\def\longrightarrowfill@{\arrowfill@{\ \ \ }\relbar\longrightarrow}
\makeatother

\makeatletter
\newcommand\msteps[2][]{\ext@arrow 0099{\longrightarrowfillstar@}{#1}{#2}}
\def\longrightarrowfillstar@{\arrowfill@{\ \ \ }\relbar{\longrightarrow^*}}
\makeatother

\newcommand{\stab}[1]{#1^\bullet}

\newcommand{\smorph}[2]{{#1}\hat{~}{#2}}
\newcommand{\xsmorph}[3]{\smorph{(\fapp{#1}{#2})}{#3}}

\newcommand{\bwedge}{\,{\boldsymbol{\wedge}}\,}
\newcommand{\barrow}{\,{\boldsymbol{\rightarrow}}\,}
\newcommand{\bstar}{\,{\boldsymbol{*}}\,}

\newcommand{\dotof}[2]{{#1}{.}{#2}}
\newcommand{\Sof}[1]{\dotof{#1}{S}}

\newcommand{\TransSet}{\Delta}
\newcommand{\IntSet}{\TransSet_i}

\newcommand{\transof}[1]{\papp{\TransSet}{#1}}
\newcommand{\intof}[1]{\papp{\IntSet}{#1}}

\newcommand{\sigmaof}[1]{\papp{\Sigma}{#1}}
\newcommand{\deltaof}[1]{\papp{\Delta}{#1}}
\newcommand{\morpheq}[3]{\papp{#3}{#2} = {#1}}

\newcommand{\xmorpheq}[4]{\fapp{#3\ {#4}}{#2} = {#1}}

\newcommand{\Spin}{\ensuremath{\mathsf{Spin}}\xspace}
\newcommand{\Counter}{\ensuremath{\mathsf{Counter}}\xspace}
\newcommand{\Xfer}{\ensuremath{\mathsf{CSLX}}\xspace}
\newcommand{\Cnt}{\ensuremath{\mathsf{Counter}}\xspace}

\newcommand{\Priv}{\ensuremath{\mathsf{Priv}}\xspace}

\newcommand{\CSL}{\ensuremath{\mathsf{CSL}}\xspace}

\newcommand{\RWLock}{\ensuremath{\mathsf{RW}}\xspace}

\newcommand{\Stack}{\ensuremath{\mathsf{Stack}}\xspace}

\newcommand{\lockprog}{\ensuremath{\mathsf{lock}}\xspace}
\newcommand{\unlockprog}{\ensuremath{\mathsf{unlock}}\xspace}

\newcommand{\csllockprog}{\ensuremath{\mathsf{exlock}}\xspace}
\newcommand{\cslunlockprog}{\ensuremath{\mathsf{exunlock}}\xspace}

\newcommand{\OCSL}{\textsc{CSL}\xspace}

\newcommand{\FCSL}{\textsc{FCSL}\xspace}

\newcommand{\TaDA}{\textsc{TaDA}\xspace}
\newcommand{\Iris}{\textsc{Iris}\xspace}
\newcommand{\Disel}{\textsc{Disel}\xspace}

\newcommand{\CiC}{\textsc{CiC}\xspace}

\newcommand{\cf}{cf.\xspace}
\newcommand{\dt}[1]{\textbf{\emph{#1}}} 

\newcommand{\para}[1]{\paragraph*{#1}}
\newcommand{\scs}[1]{\ensuremath{{\scriptstyle{#1}}}}

\newcommand{\var}[1]{\ensuremath{\mathit{#1}}}

\newcommand{\thehp}{\chi}
\newcommand{\selfhp}{\self{\thehp}}
\newcommand{\otherhp}{\other{\thehp}}

\newcommand{\ghostcolor}[1]{\colorbox{gray!35}{\raisebox{0pt}[5pt][0pt]{\ensuremath{#1}}}}

\newcommand{\acite}[3]{ \ifappendix
  {Appendix~\ref{#1}}\else{\cite[Appendix~{#2}]{#3}}\fi}

\def\apndxTensor{A}

\def\apndxPCM{B}

\def\apndxMorph{C}

\def\apndxModel{D}

\begin{document}
	    
\title{Specifying Concurrent Programs in Separation Logic: Morphisms and Simulations}         

\def\afIMDEA{
        \affiliation{\institution{IMDEA Software Institute}\country{Spain}}}
\def\afIRIF{
        \affiliation{\institution{IRIF--Universit\'{e} de Paris}\country{France}}}

\author{Aleksandar Nanevski}
\afIMDEA
\email{aleks.nanevski@imdea.org}

\author{Anindya Banerjee}
\afIMDEA
\email{anindya.banerjee@imdea.org}

\author{Germ\'an Andr\'es Delbianco}
\orcid{0000-0002-2249-1168}
\afIRIF
\email{gad@irif.fr}

\author{Ignacio F\'{a}bregas}
\orcid{0000-0002-3045-4180}
\afIMDEA
\email{ignacio.fabregas@imdea.org}

\begin{abstract}
In addition to pre- and postconditions, program specifications in
recent separation logics for concurrency have employed an algebraic
structure of \emph{resources}---a form of state transition
systems---to describe the state-based program invariants that must be
preserved, and to record the permissible atomic changes to program
state.
In this paper we introduce a novel notion of \emph{resource morphism},
i.e.~structure-preserving function on resources, and show how to
effectively integrate it into separation logic, using an associated notion
of morphism-specific \emph{simulation}.
We apply morphisms and simulations to programs verified under one
resource, to compositionally adapt them to operate under another
resource, thus facilitating proof reuse.


\end{abstract}



\maketitle

\newtheorem{notation}[theorem]{Notation}

\ifarxiv
\def\ArtN{161}
\def\TotP{30}

\makeatletter
{\small
\vspace{-2mm}      
\bgroup
\par\medskip\noindent{\bfseries This extended technical report is a companion to:}\par
\noindent\bgroup
\authors\egroup. \@acmYear. \@title.
\textit{\@journalNameShort}
\@acmVolume, \@acmNumber, Article~\ArtN~(\@acmPubDate),
\TotP~pages.
\@formatdoi{\@acmDOI}
\par}
\egroup
\makeatother
\fi
\newcommand{\internal}{\emph{internal}\xspace}
\newcommand{\external}{\emph{external}\xspace}

\newcommand{\tensor}{\otimes}
\newcommand{\cmorph}{\ensuremath{\mathsf{morph}}}
\newcommand{\morph}[2]{\ensuremath{\cmorph\ #1\ #2}}
\newcommand{\lift}{\textsc{Morph}\xspace}

\newcommand{\fapp}[2]{#1\,#2}
\newcommand{\papp}[2]{#1\,(#2)}
\newcommand{\morphtp}[2]{{#1}\rightarrow{#2}}
\newcommand{\rightarrowX}[1]{\overset{#1}{\rightarrow}}
\newcommand{\morphtpX}[3]{{#1}\overset{#3}{\rightarrow}{#2}}

\section{Introduction}\label{sec:intro}
The main problem when formally reasoning about concurrent data
structures is achieving \emph{compositionality} of proofs: how to
ensure that methods of a data structure, once verified, can be used in
a larger context without re-verification. There exist many solutions
to the problem, roughly divided into two kinds:
\emph{linearizability}~\cite{Herlihy-Wing:TOPLAS90}, or more generally
\emph{contextual
  refinement}~\cite{Filipovic-al:TCS10,lia+fen+sha:lics14,lia+fen:popl18},
and \emph{Concurrent Separation Logic
  (\OCSL)}~\cite{OHearn:TCS07,Brookes:TCS07}, and its many recent
extensions to fine-grained (i.e., lock-free)
concurrency~\cite{DinsdaleYoung-al:ECOOP10,Svendsen-al:ESOP13,
  Svendsen-Birkedal:ESOP14,Nanevski-al:ESOP14,
  ArrozPincho-al:ECOOP14,Jung-al:POPL15}. More recently, some
approaches~\cite{Turon-al:ICFP13,fru+kre+bir:lics18} employed variants
of separation logic to establish linearizability and contextual
refinement themselves, suggesting separation logic as a
general-purpose method for reasoning about concurrent programs.

On the other hand, composition is also the cornerstone of \emph{type theory}, where types
serve as the interface that abstracts the internal properties of
programs and proofs. Because both are focused on composition, type
theory and separation logic are closely related.
For example, from the inception of (sequential) separation logic, it
has been understood~\cite{OHearnRY01,Reynolds:LICS02} that its
reasoning power arises from the key property of \emph{fault
  avoidance}, which implies the features usually associated with
separation logic, such as framing and small-footprint semantics. Fault
avoidance states that a program verified against some pre- and
postcondition, doesn't crash (say, by reading from a deallocated
pointer), if started in a state satisfying the precondition. This has
inspired a \emph{stateful} type
theory~\cite{Nanevski-al:ICFP06,nan:oplss16}, where the type
ascription
\[
e : \spec{P} \spec{Q}
\]
signifies that the program $e$ has a precondition $P$ and a
postcondition $Q$ (both predicates over program states), in the sense
of partial correctness. 
The Hoare type $\spec{P} \spec{Q}$ is a form of dependently-typed
state monad, indexed by $P$ and $Q$, which encapsulates the effects of state and divergence,
similarly to monads in Haskell. In the typed setting, fault avoidance
is \emph{forced} onto the formalism by the requirement that
``well-typed programs cannot go wrong''~\cite{Milner78atheory}.
Thus, Hoare types give rise to not just a Hoare logic, but separation
logic specifically. In other words, separation logic is a type theory
of state. Hoare types also enable a formulation and low-overhead
implementation~\cite{Nanevski-al:ICFP08,sve+bir+nan:tlca11} of
separation logic as an extension of, or a shallow embedding into, a
standard type theory (e.g. Coq).

Using the above connection as a guiding principle, this paper derives
a type-based formulation of separation logic for \emph{fine-grained
  concurrency}. Immediately motivated by the form of typeful
specification for fine-grained programs, our contributions are two
novel and foundational abstractions for compositional
verification---the \emph{morphisms} and the \emph{simulations} from
the paper's title---and a way to incorporate them into Hoare-style
reasoning by means of a \emph{single inference rule}.
The upshot is conceptually simple foundations for separation logic for
fine-grained concurrency.

\subsection{Resources}
As proposed by \citet{DinsdaleYoung-al:ECOOP10}, and utilized in
different ways in many recent formalisms~\cite{Svendsen-al:ESOP13,
  Svendsen-Birkedal:ESOP14,Nanevski-al:ESOP14,
  ArrozPincho-al:ECOOP14,Jung-al:POPL15}, the key technical
requirement that fine-grained concurrency imposes on a Hoare-style
logic is enriching the Hoare specifications with \emph{state
  transition systems (STS)} of a specific form---termed
\emph{resources}~\cite{Hoare:ost72,Owicki-Gries:CACM76,OHearn:TCS07}
in this paper.
For example, in our type-based setting, we extend the Hoare type 
with a resource $V$, as in
\[
e : \spec{P} \spec{Q}\,@\,V
\]
to signify that $e$ has a precondition $P$ and a postcondition $Q$,
but also that the atomic state changes that $e$ may carry out are
circumscribed by the transitions of $V$. We also say that $e$ \emph{is
  typed by} $V$, that $e$ \emph{inhabits} $V$, or that $e$ \emph{is
  in} $V$.

Two programs can be composed in parallel (or sequentially), only if
they are typed by the same resource. Thus, the resource in the type
annotation bounds the interference that concurrent threads can perform
on each other's execution, which is essential for reasoning about the
composition.\footnote{The idea of bounding the interference is the
  foundation behind the classic \emph{rely-guarantee}
  method~\cite{Jones:TOPLAS83} as well. In fact, resources may be seen
  as structuring and compactly representing---in the form of
  transitions---the rely and guarantee relations of the rely-guarantee
  method.
}

\newcommand{\spinintro}{%
\begin{tikzpicture}[scale=0.55, transform shape, shorten >= 1pt, node distance=4cm,>=latex']
  \tikzstyle{state}=[circle,draw=black,fill=gray!20,thick, 
                     inner sep=0pt, minimum size=18mm]
  \tikzstyle{dummy}=[circle,inner sep=0pt,minimum size=10mm]
  \tikzset{every loop/.style={min distance=25mm,looseness=12}}
  \node[state] (spin) at (0, 0) {\fontsize{14}{14}$\Spin$};
  \path[->] (spin) edge [loop right, thick] node {\huge$\mathsf{id\_tr}$} ()
                   edge [loop left, thick] node {\huge$\mathsf{unlock\_tr}$} ()
                   edge [loop above, thick] node {\huge$\mathsf{lock\_tr}$} ();
\end{tikzpicture}}

\begin{figure}
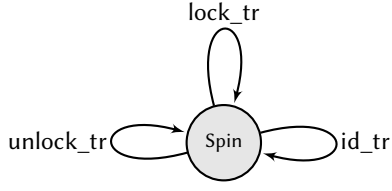

\[ \spinintro\vspace{-3mm}\]
\caption{Resource for spin locks. By convention, the \emph{idle}
  transition $\mathsf{id\_tr}$ will be elided in the future
  diagrams.\vspace{-2mm}}\label{fig:intro}
\end{figure}

To quickly illustrate resources in our particular setting, consider a
spin lock $r$ (a shared Boolean pointer) and a program that locks $r$ 
by setting it to $\mathsf{true}$, and loops if $r$ is already
set\footnote{The \emph{Compare-and-Set} variant of $\mathsf{CAS}(r, a,
  b)$~\cite{Herlihy-Shavit:08} atomically sets the pointer $r$ to $b$
  if $r$ contains $a$, otherwise leaves $r$ unchanged. It moreover
  returns a Boolean value denoting the success or failure of the 
  operation.}:
\[
\mathsf{lock} \eqdef
\mathsf{do}\ x \leftarrow \mathsf{CAS}(r, \mathsf{false}, \mathsf{true})\
\mathsf{while}\ \neg x
\]

Figure~\ref{fig:intro} shows (an abstracted form of)
the resource $\Spin$, suitable to type $\mathsf{lock}$.\footnote{We'll
  define the state space and the transitions of $\Spin$ in
  Section~\ref{sec:overview}, and eventually tie them to the
  implementations of $\mathsf{lock}$ and $\mathsf{unlock}$. At this
  point, it suffices to know that a program in $\Spin$ may transition
  by $\mathsf{lock\_tr}$ only if $r$ is free (thereby locking it), and
  by $\mathsf{unlock\_tr}$ only if $r$ is locked (thereby freeing it).
} Every execution of $\mathsf{lock}$ describes a path
through $\Spin$ consisting of several idle transitions
$\mathsf{id\_tr}$, corresponding to unsuccessful $\mathsf{CAS}$'s,
followed by a locking transition $\mathsf{lock\_tr}$ corresponding to a
successful $\mathsf{CAS}$. Similarly, the $\Spin$ resource also types
the $\mathsf{unlock}$ program
\[
\mathsf{unlock} \eqdef r := \mathsf{false}
\]
which may be seen as taking the $\mathsf{unlock\_tr}$ transition if
$r$ stores $\mathsf{true}$, or staying idle if $r$ stores
$\mathsf{false}$ to begin with. Because $\mathsf{lock}$ and
$\mathsf{unlock}$ are typed by the same resource, they can be
composed, sequentially or in parallel. The typing guarantees that the
concurrent environment of $\mathsf{lock}$ and $\mathsf{unlock}$ is
bound to only ever execute the transitions of $\Spin$, and can't cause
``surprises'', such as deallocating the pointer $r$ while
$\mathsf{lock}$ or $\mathsf{unlock}$ are executing. 

\subsection{Morphisms}

This brings us to the first technical contribution of the paper.  As
soon as resources are introduced into types, it becomes necessary to
coerce a program from one type (i.e., resource) to another.

As one example of coercion, consider two procedures, inhabiting different
resources, each specifying its own concurrent data structure (say, a
stack and a queue). If we want to use the stack and the queue together
in a program, we must coerce the procedures into a common resource
that includes the functionality of both structures, and describes how
the two interact.

As another example, consider refining the behavior of already implemented
resources. Suppose we want to use the bare-bones spin locks
described by $\Spin$ to develop more sophisticated locking protocols: 
\OCSL-style mutually exclusive locks~\cite{OHearn:TCS07}, or
non-mutually-exclusive locks such as readers-writers
locks~\cite{courtois:ACM71,Bornat-al:POPL05} where a reader acquires
a lock to allow access to multiple readers, but not writers.
Both developments can be seen as extending $\Spin$ with additional
ghost state to represent the invariants of the refined locking
protocol, and then coercing the $\lockprog$ and $\unlockprog$
procedures to modify this additional ghost state at the precise moment
when $\lockprog$ transitions by $\mathsf{lock\_tr}$, and $\unlockprog$
transitions by $\mathsf{unlock\_tr}$. The developments thus
compositionally \emph{reuse} the definitions and proofs of $\lockprog$
and $\unlockprog$, each in its own way. We carry out the first
development in Section~\ref{sec:exclusive}, and the second in the Coq
code~\cite{artifact}.

To achieve coercion we introduce morphisms between resources.  A
\emph{resource morphism} $f : \morphtp{V}{W}$ is a
structure-preserving mapping from resource $V$ to resource $W$, which
\emph{acts} on a program $e$ typed by $V$, to derive a program typed
by $W$, essentially by re-interpreting the $V$-transitions that $e$
takes, as $W$-transitions.

Morphisms arise naturally, because a structure in mathematics
typically is associated with an appropriate notion of a
structure-preserving function.
Examples abound: vector spaces and linear maps, groups and their
homomorphisms, complete partial orders and continuous functions, functors and natural
transformations, etc.
Morphisms endow the structure with dynamics and allow studying it
under change. This will be the case for us as well.

More specifically, and akin to how automata
homomorphisms~\cite{ginzburg1968} are defined componentwise, a
resource morphism $f : \morphtp{V}{W}$ consists of two partial
functions $f_\Sigma$ and $f_\Delta$, acting respectively on
\emph{states} ($\Sigma$) and \emph{transitions} ($\Delta$). $f_\Sigma$
takes a state in $W$ and produces a state, if defined, in $V$ (note
the contravariance); and $f_\Delta$ takes a state in $W$ and
transition in $V$ and produces a transition, if defined, in $W$.
Combined, $f_\Sigma$ and $f_\Delta$ \emph{act} on a program $e$
inhabiting $V$ to produce a program inhabiting $W$, using the
following process.

\begin{DIFnomarkup}
\begin{figure}
	\[\hspace{-20mm}
          {\fontsize{13pt}{6pt}\selectfont
	   \begin{tikzpicture}[every label/.append style = {font = \small}]
	   \tikzcdset{arrow style=tikz, diagrams={>=stealth}}
	   \begin{tikzcd}
	   s'_v & \fapp{I}{s'_w} \arrow[l, "f_\Sigma", swap] \\
	   s_v \arrow[u, "t_v"]  & \fapp{I}{s_w} \arrow[l, "f_\Sigma"] \arrow[u,"t_w = f_\Delta\,s_w\,t_v", 
	   swap]
	   \end{tikzcd}
	   \end{tikzpicture}}
	\]\vspace{-9mm}
	\caption{Reinterpreting the transitions of a resource $V$ into
          those of $W$, by a morphism $f : \morphtp V W$ and an
          $f$-simulation $I$. In this diagram, and in the sequel,
          $\fapp{I}{s}$ denotes ``state $s$ such that the
          predicate $I$ holds''.\vspace{-3mm}}\label{fig:diagram}
\end{figure}
\end{DIFnomarkup}

Referring to Figure~\ref{fig:diagram}, the morphed program inhabits
$W$, so we describe it starting with a $W$-state $s_w$ in the
lower-right corner of the diagram (the predicate $I$ that is applied
to $s_w$ in the diagram will be explained promptly). To compute the
next state of the morphed program, we first take $s_v = f_\Sigma\ s_w$
which is a state in $V$ (utilizing contravariance of $f_\Sigma$). If
$e$ takes a transition $t_v$ to step from $s_v$ to $s'_v$ in $V$, the
corresponding transition of the morphed program is $t_w = f_\Delta\ s_w\ t_v$.  If $t_w$ steps from 
$s_w$ to $s'_w$, the process is repeated for $s'_w$ and the next transition of $e$.

The \emph{morphing} process determines a program in $W$ that we denote
$\morph{f}{e}$. Here, $\cmorph$ is a \emph{program constructor}, and a
form of \emph{function application} of $f$ to $e$. In the sequel, we introduce the infrastructure to 
program (and prove!) with morphisms and morphed programs. 

\subsection{Simulations and Inference}

To reason about $\morph{f}{e}$, we introduce our second contribution:
morphism-specific simulations. 
An \emph{$f$-simulation} (or simply, a \emph{simulation}, when $f$ is
clear from the context) is a predicate $I$ over $W$-states that acts
like a loop invariant for the iterative process of morphing by
$f$. Specifically, an $f$-simulation satisfies, among other conditions
presented in Section~\ref{sec:formal}, the key technical property that
the diagram in Figure~\ref{fig:diagram} \emph{commutes}.
Given $s_w$, $s_v$, $t_v$ and $s'_v$ that partially describe the
diagram, if $I\ s_w$, then there exists a state $s'_w$ that completes
the diagram: $t_w = f_\Delta\ s_w\ t_v$ exists, $t_w\ s_w\ s'_w$,
$I\ s'_w$, and $\fapp{f_\Sigma}{s'_w} = s'_v$. Because $I\ s'_w$
holds, simulation $I$ is \emph{preserved} by
$f$. 

Simulations provide a way to reason about $\morph{f}{e}$
compositionally, i.e., out of $e$'s \emph{type}, achieving our
ultimate goal of program and proof re-usability.
The specifics are prescribed by the following \emph{single} inference
rule, which is our third contribution:
\begin{mathpar}
  \inferrule*[Right = Morph] {e : \spec{P} \spec{Q}@V} {\morph{f}{e} :
    \spec{\lambda s_w\ldot \fapp{\smorph f P}{s_w} \wedge
      \fapp{I}{s_w}} \spec{\lambda s_w\ldot \fapp{\smorph f Q}{s_w}
      \wedge \fapp{I}{s_w}}@W}
  \and
      {\textrm{where}\,
    \begin{array}{l}
      {\smorph f R}~{s_w}~{\eqdef}~\exists\ s_v\ldot\ s_v~{=}~
      \fapp{f_\Sigma}{s_w} \wedge \fapp{R}{s_v}
      \end{array}}
\end{mathpar} 

The \TirName{Morph} rule translates the requirements for a morphism
$f : \morphtp{V}{W}$, and a simulation $I$, expressed diagrammatically
in Figure~\ref{fig:diagram}, from properties of states and transitions
of the resources $V$ and $W$, into the specification of a program
$\morph{f}{e}$ in $W$.
The latter spec is structured, both in its pre- and the
postconditions, as a conjunction of (i) the transformation of $e$'s
spec from $V$ to $W$, and (ii) a statement that the simulation $I$,
being in fact an invariant of the resources' transitions, holds at the
boundaries of the \emph{morphed} execution of $e$.
For the first part, we use the predicate ${\smorph f \_}$, to lift the
pre- and postconditions of $e$ from states in $V$ to states of $W$,
which are related via ${f_\Sigma}$.
This motivates the contravariance of $f_\Sigma$: we have verified $e$
in the context of the resource $V$, but we intend to \emph{morph} $e$
and execute it in a new resource context, where states are inhabitants
of $W$.

More concretely, the precondition of $\morph{f}{e}$ in \TirName{Morph}
assumes $\fapp{I}{s_w}$ and the existence of
$s_v = \fapp{f_\Sigma}{s_w}$ such that $\fapp{P}{s_v}$. By fault
avoidance (i.e., type safety), a program (here $e$ in the premise)
that is ascribed a Hoare type isn't stuck. Hence, there exists a transition
$t_v$ by which $e$ steps from $s_v$ into $s'_v$.
Because $I$ is an $f$-simulation, the commuting diagram
implies the existence of $s'_w$ such that $s'_v =
\fapp{f_\Sigma}{s'_w}$ and $\fapp{I}{s'_w}$.
The morphing process is then iterated for $s'_w$ and the subsequent
states. This iteration relies on $f$ preserving $I$, much like a loop relies
on the loop body preserving the loop invariant. Once $e$ terminates in a final state $s''_v$, the 
postcondition in \TirName{Morph} must hold. First, $\fapp{Q}{s''_v}$ must hold as $Q$
is $e$'s postcondition.
Second, $I$ being an $f$-simulation yields the commuting diagram which
implies the existence of $s''_w$ such that $s''_v =
\fapp{f_\Sigma}{s''_w}$ (hence $\fapp{\smorph f Q}{s''_w}$) and
$\fapp{I}{s''_w}$.

The paper can thus be seen as introducing simulations into separation
logic in a simple\footnote{Our formalization exports nine
  rules for Hoare-style reasoning, each addressing an
  orthogonal linguistic feature.}, but also constructive
manner.
Customarily, an STS $W$ simulates another STS $V$ if whenever $V$
takes a transition, there \emph{exists} a transition for $W$ to
take~\cite{LynchV+IC95}. For us, a morphism $f$ \emph{computes the
  witness} of this existential (via $f_\Delta$), in the style of
constructive logic and type theory.
Moreover, a simulation is usually defined as a relation between the
states of $V$ and $W$. For us, an $f$-simulation is a predicate on
$W$-states alone, as $f_\Sigma$ \emph{deterministically} computes the
unique $V$-state that corresponds to a $W$-state in the simulation.
Finally, simulations are also customarily required to relate a
distinguished set of initial states of the source and target STSs.
Our resources and $f$-simulations, on the contrary, don't need to
consider specific initial states of $V$ and $W$, because the initial
states of any program are described by its precondition, and the
\TirName{Morph} rule checks that the simulation holds on the pre-state
of every invocation of $\cmorph$.

The paper can also be seen as introducing a form of \emph{refinement
  mappings}~\cite{aba+lam:91} into separation logic, since refinement
mappings, like morphisms, are functions between STSs. The two,
however, have very different technical details, largely imposed by our
connection to separation logic. This includes the introduction of the
morphism action on programs and the \TirName{Morph} rule, but also the
treatment of state ownership, ownership transfer, and framing
(\cf~Section~\ref{sec:formal}), none of which have been considered
with refinement mappings.

\newcommand{\ftntLift}{As apparent from the \TirName{Morph} rule, we
  explicitly bind the state $s_w$, in contrast to the classic
  presentation of separation logics where this state is implicit. This
  is merely a syntactic distinction.}

Moreover, resource morphisms go beyond mere program specification and
proof, as they also support generic constructions over resources, such
as ``tensoring'' two resources, adjoining an invariant to a
resource, or forgetting a ghost field from a resource (the last one
with a mild generalization to \emph{indexed morphism
  families}).
Morphisms relate a construction to its components, much as arrows in
category theory relate objects of universal constructions, and are
thus essential for the constructions to compose. This is why we see
resource morphisms as a step towards a general type-theoretic calculus
of concurrent constructions.

We formalize the development in Coq, using Coq's predicates over
states as assertions\footnote{\ftntLift}, building on the code base of
Fine-grained Concurrent Separation Logic
(\FCSL)~\cite{Nanevski-al:ESOP14}.
\ifappendix
{The sources are available online as an
  Artifact~\cite{artifact}.\footnotemark{}}
\else
{The sources are available online as an Artifact~\cite{artifact},
  along with the extended version of the
  paper~\cite{extended}.\footnotemark{}}
\fi

\footnotetext{In addition to the examples from the paper, the sources
  include further benchmarks such as Treiber stack~\cite{Treiber:TR},
  flat combiner~\cite{Hendler-al:SPAA10}, a concurrent allocator, a
  concurrent graph spanning tree algorithm~\cite{Sergey-al:PLDI15},
  ticketed~\cite{mcs91} and readers-writers~\cite{courtois:ACM71}
  locks.}

\para{Readmap} The rest of the paper is organized as
follows. Section~\ref{sec:overview} introduces resources and
associated notions of ghost state and transitions, via the spin lock
example. Section~\ref{sec:formal} develops the theory formally,
including our specific notion of framing. Section~\ref{sec:exclusive}
illustrates how to morph spin locks into exclusive
locks. Section~\ref{sec:param} introduces indexed morphism families
and applies them to ``forgetting'' the ghost state of a resource. This
models what is often referred to as
\emph{quiescence}~\cite{Aspnes-al:JACM94,Derrick-al:TOPLAS11,
  Jagadeesan-Riely:ICALP14,Nanevski-al:ESOP14, sergey:oopsla16}, most
commonly used when installing one concurrent structure into the
private state of another.
Section~\ref{sec:related} discusses related work and
Section~\ref{sec:conclusions} concludes.

\newcommand{\SC}{\ensuremath{\mathsf{SC}}\xspace}

\newcommand{\unlock}{\pi}
\newcommand{\lock}{r}
\newcommand{\lockval}{\mathsf{L}}
\newcommand{\unlockval}{\mathsf{U}}
\newcommand{\self}[1]{#1_s}
\newcommand{\other}[1]{#1_o}
\newcommand{\joint}[1]{#1_j}
\newcommand{\total}[1]{\hat{#1}}
\newcommand{\selfhist}{\self{\thehist}}
\newcommand{\otherhist}{\other{\thehist}}
\newcommand{\totalhist}{\total{\thehist}}
\newcommand{\thepu}{\pi}
\newcommand{\selfpu}{\self{\thepu}}
\newcommand{\otherpu}{\other{\thepu}}
\newcommand{\totalpu}{\total{\thepu}}
\newcommand{\orth}{\mathbin{\bot}}
\newcommand{\hempty}{\emptyset}
\newcommand{\lastkey}{\mathsf{last\_stamp}}
\newcommand{\lastval}{\mathsf{last\_op}}
\newcommand{\dom}{\mathsf{dom}}
\newcommand{\selfa}{\self{a}}
\newcommand{\jointa}{\joint{a}}
\newcommand{\othera}{\other{a}}
\newcommand{\totala}{\total{a}}
\newcommand{\selfb}{\self{b}}
\newcommand{\jointb}{\joint{b}}
\newcommand{\otherb}{\other{b}}
\newcommand{\totalb}{\total{b}}

\newcommand{\thecnt}{\kappa}
\newcommand{\selfcnt}{\self{\thecnt}}
\newcommand{\othercnt}{\other{\thecnt}}
\newcommand{\jointcnt}{\joint{\thecnt}}

\newcommand{\locktr}{\mathsf{lock\_tr}}
\newcommand{\unlocktr}{\mathsf{unlock\_tr}}
\newcommand{\idtr}{\mathsf{id\_tr}}
\newcommand{\acqpu}{\mathsf{acqperm\_tr}}
\newcommand{\relpu}{\mathsf{relperm\_tr}}
\newcommand{\fresh}{\mathsf{fresh}}
\newcommand{\islocked}{\omega}
\newcommand{\rstate}{\mathsf{state}}
\newcommand{\pcmof}[1]{U(#1)}
\newcommand{\tpof}[1]{T(#1)}

\newcommand{\auth}{\mathsf{auth}}
\newcommand{\theauth}{\alpha}
\newcommand{\selfmu}{\self{\theauth}}
\newcommand{\othermu}{\other{\theauth}}
\newcommand{\totalmu}{\total{\theauth}}
\newcommand{\selfheap}{\self{\theheap}}
\newcommand{\otherheap}{\other{\theheap}}
\newcommand{\totalheap}{\total{\theheap}}
\newcommand{\jointheap}{\joint{\theheap}}
\newcommand{\opentr}{\mathsf{open\_tr}}
\newcommand{\incrtr}{\mathsf{incr\_tr}}
\newcommand{\closetr}{\mathsf{close\_tr}}
\newcommand{\acqmu}{\mathsf{acqauth\_tr}}
\newcommand{\relmu}{\mathsf{relauth\_tr}}
\newcommand{\openprog}{\mathsf{open}}
\newcommand{\closeprog}{\mathsf{close}}
\newcommand{\lkv}{\themutex}
\newcommand{\selflkv}{\self{\themutex}}
\newcommand{\otherlkv}{\other{\themutex}}
\newcommand{\totallkv}{\total{\themutex}}
\newcommand{\morphspincsl}{\Phi}
\newcommand{\morphsc}{f}
\newcommand{\morphxfercsl}{\Psi}
\newcommand{\atomic}[1]{\langle{#1}\rangle}
\newcommand{\couple}{\bowtie}
\newcommand{\mxown}{\mathsf{own\_of}}
\newcommand{\mxundef}{\mathsf{no\_auth}}
\newcommand{\kernel}[1]{\mathsf{ker}\ #1}
\newcommand{\eqlz}[2]{\mathsf{eql}\ #1\ #2}
\newcommand{\submorph}{\Pi}
\newcommand{\restrict}[2]{{#1}/{#2}}

\newcommand{\alternate}{\mathsf{alternate}}

\newcommand{\separate}{compatible\xspace}
\newcommand{\separateness}{compatibility\xspace}
\newcommand{\Separate}{Compatible\xspace}
\newcommand{\Separateness}{Compatibility\xspace}

\newcommand{\authorization}{authorization\xspace}
\newcommand{\Authorization}{Authorization\xspace}
\newcommand{\authorizations}{authorizations\xspace}
\newcommand{\Authorizations}{Authorizations\xspace}

\newcommand{\histset}{\mathsf{Hist}}
\newcommand{\heapset}{\mathsf{Heap}}

\section{Background and overview}\label{sec:overview}
We illustrate our specification idiom, resources, and resource
morphisms, by fleshing out the example of spin locks. 

\subsection{Histories}

To specify the locking and unlocking methods over spin lock, we build
on the idea of linearizability~\cite{Herlihy-Wing:TOPLAS90}, and
record the operations on $r$ in the linear sequence in which they
occurred. We do so in Hoare triples, but in a \emph{thread-local} way,
i.e.~from the point of view of the specified thread, which we refer to
as ``us''~\cite{LeyWild-Nanevski:POPL13}.

Specifically, a program state $s$ contains a \emph{ghost} component
that we project as $\fapp{\selfhist}{s}$, and which keeps ``our''
\emph{history} of lock operations. Dually, the projection
$\fapp{\otherhist}{s}$ keeps the collective history of all ``other''
(i.e.~environment) threads. Each thread has these two components in
scope, but they may have different values in different threads. We
refer to $\selfhist$ and $\otherhist$ as \emph{self} and \emph{other}
histories, respectively~\cite{Sergey-al:ESOP15,Nanevski-al:ESOP14}.

A history is a timestamped log of the locking and unlocking
operations.
Mathematically, it's a \emph{finite map} from timestamps (strictly
positive nats) to the set $\{\lockval, \unlockval\}$. For example, the
self history $\fapp{\selfhist}{s}$ defined as 
$
2 \hmapsto \unlockval \join 7 \hmapsto \lockval \join 9\hmapsto\lockval,
$
signifies that ``we'' have unlocked at time 2, and locked at
times 7 and 9. The timestamp gaps indicate the activity of the
interfering threads, e.g., another thread must have locked at time 1,
otherwise we couldn't have unlocked at time 2. Similarly, another
thread must have unlocked at time 8. The entries such as $2 \hmapsto
\unlockval$ are \emph{singleton} maps, and $\join$ is \emph{disjoint}
union, undefined if operand histories share a timestamp. 
We abbreviate by $\fapp{\totalhist}{s}$ the history
$\fapp{\selfhist}{s} \join \fapp{\otherhist}{s}$, which is the
combined history of all threads, and use $\histset$ for the collection
of all histories.

\newcommand{\histvar}{h}

\newcommand{\spinstatedef}{%
\!\!\!\begin{array}[t]{l}
 \textrm{$s$ contains fields $\fapp{\selfhist}{s}, \fapp{\otherhist}{s} \in \histset$, and}\\
 \fapp{\selfhist}{s} \orth \fapp{\otherhist}{s} \wedge 
                                         \papp{\alternate}{\fapp{\totalhist}{s}} \wedge r \neq \mathsf{null}
\end{array}}

\newcommand{\spinerasure}{%
\lock \hmapsto \papp{\islocked}{\fapp{\totalhist}{s}}}

\newcommand{\locktrdef}{%
\!\!\!\begin{array}[t]{l}
\locktr\ s\ s' \eqdef \neg \papp{\islocked}{\fapp{\totalhist}{s}} \wedge 
\fapp{\selfhist}{s'}=\fapp{\selfhist}{s}\join \papp{\fresh}{\fapp{\totalhist}{s}} \hmapsto \lockval
\end{array}}

\newcommand{\unlocktrdef}{%
\!\!\!\begin{array}[t]{l}
\unlocktr\ s\ s' \eqdef \papp{\islocked}{\fapp{\totalhist}{s}} \wedge 
\fapp{\selfhist}{s'}=\fapp{\selfhist}{s}\join \papp{\fresh}{\fapp{\totalhist}{s}} \hmapsto \unlockval\\
\end{array}}

\newcommand{\lastvaldef}{%
\left\{\!\!\!\begin{array}{l@{\, }l}
 \papp{\histvar}{\fapp{\lastkey}{\histvar}}, & \textrm{if $\fapp{\lastkey}{\histvar} \neq 0$}\\
  \unlockval, & \textrm{otherwise}
\end{array}\right.}

\newcommand{\continuousdef}{%
\!\!\!\begin{array}[t]{l}
  (\forall t>0\ldot \papp{\histvar}{t+1} = \lockval \rightarrow \fapp{\mathsf{even}}{t} \wedge \fapp{\histvar}{t}= \unlockval) \wedge \hbox{}\\
  (\forall t>0\ldot \papp{\histvar}{t+1} = \unlockval \rightarrow \fapp{\mathsf{odd}}{t} \wedge \fapp{\histvar}{t}= \lockval) 
\end{array}
}

\newcommand{\continuousdefshort}{%
\! (h = 1 \hmapsto \lockval \join 2 \hmapsto \unlockval \join 3
\hmapsto \lockval \join \cdots \join \fapp{\lastkey}{h} \hmapsto \fapp{\lastval}{h})
}

\newcommand{\spinblurb}{%
\!\!\!\begin{array}[t]{l}
\textrm{\textbf{State space}\ $\papp{\Sigma}{\Spin}$: $s \in
  \papp{\Sigma}{\Spin}$ iff}\\ \quad \spinstatedef\\
\textrm{\textbf{Erasure:}} \quad \hflat{s} \eqdef \spinerasure\\
\textrm{\textbf{Transitions}\ $\TransSet(\Spin)$:}\\
\locktrdef\\
\unlocktrdef
\end{array}}

\newcommand{\abbreviations}{%
\!\!\!\begin{array}[t]{l}
\begin{array}[t]{c@{}c}
\!\!\!\begin{array}[t]{r@{\,}c@{\,}l}
\fapp{\dom}{\histvar} & \eqdef & \{t \mid \papp{\histvar}{t}\ \mathsf{defined}\}\\
\fapp{\lastkey}{\histvar} & \eqdef & \papp{\max}{\{0\} \cup \fapp{\dom}{\histvar}}\\
\fapp{\fresh}{\histvar} & \eqdef & 1+\fapp{\lastkey}{\histvar}
\end{array} & 
\!\!\!\begin{array}[t]{r@{\,}c@{\,}l}
\fapp{\lastval}{\histvar} & \eqdef & \lastvaldef\\
\fapp{\islocked}{\histvar} & \eqdef & (\fapp{\lastval}{\histvar} = \lockval)
\end{array}
\end{array}\\
\hspace{4.2mm}\fapp{\alternate}{\histvar} \eqdef \continuousdefshort
\end{array}}

\newcommand{\xferstatedef}{%
\!\!\!\begin{array}{l}
\textrm{$s$ contains fields $\fapp{\selfcnt}{s}, \fapp{\othercnt}{s} \in \mathbb{N}$}\\
\textrm{with no additional constraints}
\end{array}}

\newcommand{\xfererasure}{%
\textrm{empty heap}}

\newcommand{\incrtrdef}{%
\!\!\!\begin{array}[t]{l}
\incrtr\ n\ s\ s' \eqdef \fapp{\selfcnt}{s'} = \fapp{\selfcnt}{s} + n
\end{array}}

\newcommand{\xferblurb}{%
\!\!\!\begin{array}[t]{l}
\textrm{\textbf{State space} $\papp{\Sigma}{\Cnt}$: $s \in \papp{\Sigma}{\Cnt}$ iff}\\
\quad \xferstatedef\\
\textrm{\textbf{Erasure}: \quad $\hflat{s} \eqdef \xfererasure$}\\
\textrm{\textbf{Transitions} $\papp{\TransSet}{\Cnt}$:}\\
\incrtrdef
\end{array}}

\newcommand{\spinfig}{%
\begin{tikzpicture}[scale=0.55, transform shape, shorten >= 1pt, node distance=4cm,>=latex']
  \tikzstyle{state}=[circle,draw=black,fill=gray!20,thick, 
                     inner sep=0pt, minimum size=18mm]
  \tikzstyle{dummy}=[circle,inner sep=0pt,minimum size=10mm]
  \tikzset{every loop/.style={min distance=25mm,looseness=12}}
  \node[state] (spin) at (0, 0) {\fontsize{14}{14}$\Spin$};
  \node[state] (xfer) at (6, 0) {\fontsize{14}{14}$\Cnt$};
  \path[->] (spin) edge [loop above, thick] node {\huge$\locktr$} () 
                   edge [loop left, thick] node {\huge$\unlocktr$} () 
            (xfer) edge [loop above, thick] node {\huge$\incrtr\ n$} (); 
  \node[dummy] at (3,  2) {\fontsize{20}{20}$\couple$};
\end{tikzpicture}}

\begin{figure}[t]
\[
\begin{array}{c}
\hspace{-2cm}\spinfig\\[1ex]
\fontsize{8.5}{8.5}
\hspace{-5mm}\begin{array}{c@{}c}
\spinblurb & \xferblurb
\end{array}\\\\
\fontsize{8.5}{8.5}
\begin{array}[t]{l}
\textrm{\textbf{Abbreviations:} (in the abbreviations below, $\histvar$ is a bound variable ranging over histories)}\\
\abbreviations
\end{array}
\end{array}\vspace{-1mm}
\]
\caption{$\Spin$ and $\Cnt$ resources (with $\idtr$
  elided).}\label{fig:spin}
\vspace{-2mm}
\end{figure}

\subsection{Resources}

We next define the resource $\Spin$ from Section~\ref{sec:intro}, that
types spin lock methods. It is pictorially shown in
Figure~\ref{fig:spin} on the left. 

The \emph{state space} of $\Spin$, denoted $\papp{\Sigma}{\Spin}$,
makes explicit the assumptions about the components: that the
histories are disjoint (denoted $\fapp{\selfhist}{s} \orth
\fapp{\otherhist}{s}$), that the entries in $\fapp{\totalhist}{s}$
alternate between $\lockval$ and $\unlockval$, and that $r$ isn't the
$\mathsf{null}$ pointer.

The \emph{erasure} $\hflat{s}$ shows how the state $s$ maps to a heap
once the ghost histories are removed. The expression $\lock \hmapsto
\papp{\islocked}{\fapp{\totalhist}{s}}$ denotes a heap with only the
pointer $\lock$, storing the Boolean value
$\papp{\islocked}{\fapp{\totalhist}{s}}$. The latter computes the lock
status out of the combined history $\fapp{\totalhist}{s}$; it equals
$\mathsf{true}$ if the last log in the combined history is a lock entry $\lockval$, and $\mathsf{false}$ 
otherwise.

The \emph{set of transitions} of $\Spin$, denoted $\transof{\Spin}$,
contains $\locktr$, $\unlocktr$, and the (elided) idle transition. The
transition $\locktr$ adds a fresh $\lockval$ entry to
$\fapp{\selfhist}{s}$ if $\neg
\papp{\islocked}{\fapp{\totalhist}{s}}$, i.e., if the lock is free in
the pre-state. Similarly, $\unlocktr$ adds a fresh $\unlockval$ entry
if $\papp{\islocked}{\fapp{\totalhist}{s}}$, i.e., the lock is
taken. The lock can by taken by ``us'' or by ``others'', as
$\islocked$ is computed from the combined history
$\fapp{\totalhist}{s}$. If the locking protocol insists that the
thread that unlocks is the same thread that last locked, then the
precondition of $\unlocktr$ should be changed to
$\papp{\islocked}{\fapp{\selfhist}{s}}$. 
We don't want to impose such behavior at this stage, but show how to
achieve it \emph{a posteriori}, together with additional
functionality, in Section~\ref{sec:exclusive}.

\subsection{Method Specifications}\label{sec:methspec}

We now give the following pidgin code for $\lockprog$ and
$\unlockprog$, intended to further the intuition about transitions. The
actual implementation of the methods will be shown in
Section~\ref{sec:formal}, once we have formally introduced our system.
\[
\begin{array}[t]{r@{\ }c@{\ }l}
\lockprog & \eqdef & \mathsf{do}\ \langle x \leftarrow \mathsf{CAS}(r, \mathsf{false}, \mathsf{true});\, \ghostcolor{\mathsf{if}\ x\ \mathsf{then}\ \fapp{\selfhist}{s} := \fapp{\selfhist}{s}\join \papp{\fresh}{\fapp{\totalhist}{s}} \hmapsto \lockval} 
    \rangle; \mathsf{while}\ \neg x\\[2mm]
\unlockprog & \eqdef & 
\langle \ghostcolor{x \leftarrow\,!r;}\ r := \mathsf{false};\, \ghostcolor{\mathsf{if}\ x\ \mathsf{then}\ \fapp{\selfhist}{s}
               := \fapp{\selfhist}{s} \join
               \papp{\fresh}{\fapp{\totalhist}{s}} \hmapsto
               \unlockval}
\rangle
\end{array}
\]
The brackets $\langle - \rangle$ denote atomic execution (i.e.,
uninterrupted by other threads) of real and ghost code, the latter
given in gray.
Note how the bracketed code in $\lockprog$ implicitly describes a
choice, depending on the contents of $r$, between executing $\locktr$
or the idle transition in the resource. The former, when considered on
\emph{erased} states, corresponds to $\mathsf{CAS}$ successfully
setting $r$, the latter to $\mathsf{CAS}$ failing.
Similarly, $\unlockprog$ chooses between $\unlocktr$ and the idle
transition.
Thus, we shall abstractly view the atomic executions as a choice
between transitions of the corresponding resource, rather than as
bracketing of ghost with real code.

We can now explain the history-based specs for $\mathsf{lock}$ and
$\mathsf{unlock}$.
\[
\begin{array}[t]{r@{\ }c@{\ }l}
\lockprog & : &
   [h, k]\ldot\!\!\!\begin{array}[t]{l}
     \spec{\lambda s\ldot \fapp{\selfhist}{s} = h \wedge k \leq \papp{\lastkey}{\fapp{\totalhist}{s}}}\\
     \spec{\lambda s\ldot \exists t\ldot \fapp{\selfhist}{s} = h \join t \hmapsto \lockval \wedge k < t} @ \Spin
          \end{array} 
\\
\unlockprog & : & 
 [h, k]\ldot\!\!\!\begin{array}[t]{l}
  \spec{\lambda s\ldot \fapp{\selfhist}{s} = h \wedge k \leq \papp{\lastkey}{\fapp{\totalhist}{s}}}\\
  \spec{\lambda s\ldot \exists t\ldot \fapp{\selfhist}{s} = h \join t \hmapsto \unlockval \wedge k < t \vee 
          \fapp{\selfhist}{s} = h \wedge \fapp{\fapp{\totalhist}{s}}{\,t} = \unlockval \wedge k \leq t} @ \Spin
  \end{array}
\end{array}
\]

The precondition of $\mathsf{lock}$ starts with self history
$\fapp{\selfhist}{s}$ equal to $h$\footnote{As customary in Hoare
  logic, $h$ and $k$ are \emph{logical variables}, used to relate the
  pre and post-state. They are universally quantified, scoping over
  pre and post-condition, and the syntax $[\cdots]$ makes the binding
  explicit.}, which is increased in the postcondition to log a locking
event at time $t$. The conjunct $k < t$ in the postcondition claims
that $t$ is fresh, because it's larger than any $k$ generated prior to
the call (as $k \leq \papp{\lastkey}{\fapp{\totalhist}{s}}$ is a
conjunct in the precondition, and $k$ is universally quantified
outside of the pre- and postcondition). The natural numbers ordering
on timestamps gives the linear sequence in which the events logged in
$\fapp{\totalhist}{s}$ occurred. Notice that the spec is
\emph{stable}, i.e., invariant under interference. Intuitively, other
threads can't modify the $\fapp{\selfhist}{s}$ field, as it's private
to ``us''. They can log new events into $\fapp{\otherhist}{s}$, which
features in the comparison $k \leq
\papp{\lastkey}{\fapp{\totalhist}{s}}$, but this only increases the
right-hand side of the comparison and doesn't invalidate it.

Similarly, $\unlockprog$ starts with history $h$, which is either
increased to log a fresh unlocking event at time $t$, or remains
unchanged if the unlocking \emph{fails} because $\unlockprog$
encounters $\lock$ already freed at time $t$ (conjunct
$\fapp{\fapp{\totalhist}{s}}{\,t} = \unlockval$). The conjunct $k\leq
t$ captures that another thread may have freed $\lock$ after the
invocation of $\unlockprog$ ($k < t$), or that we invoked
$\unlockprog$ with $\lock$ already freed ($k = t$). 

Observe that the spec for $\unlockprog$ doesn't require that the
unlocking thread is the one that last locked, or even that the lock is
taken when unlocking is attempted. This is so because we intend the
specs to capture only the basic mechanics of spin locks, and leave it
to the clients to supply application-specific policies, via morphing,
as we illustrate on exclusive locks in Section~\ref{sec:exclusive}
(and on readers-writers locks in~\cite{artifact}).

\subsection{Morphisms} 
Consider next how to express a client of $\lockprog$ that,
simultaneously with a successful lock, adds $n$ to the ghost component
$\fapp{\selfcnt}{s}$ of resource $\Cnt$ (right half of
Figure~\ref{fig:spin}). Intuitively, we desire something like $\langle
\lockprog; \ghostcolor{\fapp{\selfcnt}{s} := \fapp{\selfcnt}{s} +
  n}\rangle$
but this isn't quite right. Indeed, bracketing would prevent other
programs from running during the iterations of $\lockprog$'s loop,
thus changing the granularity of the program. We want to model that
addition occurs only upon the successful $\mathsf{CAS}$ of the last
iteration in $\lockprog$. To do so, we use morphisms as follows.

\newcommand*{\ftntTensor}{\ifappendix
We elide the definition of tensoring, as it isn't required to follow
the presentation. It can be found
in~Appendix~\ref{appendix:tensor}.%
\else
We elide the definition of tensoring, as it isn't required to follow
the presentation. It can be found in the online 
appendices~\cite[Appendix~\apndxTensor]{extended}.%
\fi}

First, we ``tensor'' the resources $\Spin$ and $\Cnt$, as graphically
indicated\footnote{\ftntTensor}  in
Figure~\ref{fig:spin}; that is, we create a new resource $\SC$ whose
state is a pair of $\Spin$ and $\Cnt$ states, and transitions are
$\locktr \couple \incrtr\ n$ and $\unlocktr \couple \idtr$. Operator
$\couple$ (pronounced ``couple'') indicates that the operand
transitions are executed simultaneously on their respective state
halves. It's defined as follows, where $\sfst{s}$ and $\ssnd{s}$
project state $s$ to its $\Spin$ and $\Counter$ components,
respectively:
\[
(t_1 \couple t_2)\ s\ s' \eqdef t_1\ (\sfst{s})\ (\sfst{s'}) \wedge
t_2\ (\ssnd{s})\ (\ssnd{s'})
\]
Second, for each $n \in \mathbb{N}$, we define the morphism
$\morphsc_n : \morphtp{\Spin}{\SC}$ as follows:
\[
\begin{array}{l@{\ }l}
\begin{array}[t]{l@{\ }c@{\ }l}
(\morphsc_n)_\Sigma\ s & \eqdef & \sfst{s} \\[1ex]
(\morphsc_n)_\Delta\ s\ \locktr & \eqdef & \locktr \couple \incrtr\ n\\
(\morphsc_n)_\Delta\ s\ \unlocktr & \eqdef & \unlocktr \couple \idtr\\
\end{array}
\end{array}
\]
This definition captures: (1) starting from an $\SC$ state
$s$,
we can obtain a $\Spin$ state by taking the first projection; (2) a
$\Spin$ program can be lifted to $\SC$ by changing the transition
$\locktr$ by $(\morphsc_n)_\Delta$ on the fly, to increment
$\fapp{\selfcnt}{s}$ \emph{simultaneously} with the lock acquisition; and (3)
$\unlocktr$ is coupled with the idle transition in $\Counter$, thus
$\fapp{\selfcnt}{s}$ is unchanged by unlocking.

Now, our desired program is
%
\[
\morph {\morphsc_n} {\lockprog}
\]
which is typed by $\SC$, and executes $\locktr \couple \incrtr\ n$
whenever $\lockprog$ executes $\locktr$, thus incrementing
$\fapp{\selfcnt}{s}$ precisely, and only, upon a successful
$\mathsf{CAS}$.

\subsection{Inference} 

The \TirName{Morph} rule provides a way to reason about morphed
programs. To illustrate the proofs, we consider the following simple
program
\[
\!\!\!\begin{array}{l}
\morph{\morphsc_1} \lockprog;\\
\morph{\morphsc_{42}} \unlockprog;\\
\morph{\morphsc_2} \lockprog
\end{array}
\]
which, in addition to locking and unlocking, increments
$\fapp{\selfcnt}{s}$ by $1$ in the first line, and by $2$ in the third
line.\footnote{Strictly speaking, we should write
  $\papp{\selfcnt}{\ssnd{s}}$ (resp. $\papp{\selfhist}{\sfst{s}}$) to
  extract the \emph{self} component of the $\Cnt$ (resp.~$\Spin$)
  ''sub-resource'' of $\SC$. However, the components have different
  names, so there's no confusion which projection of $s$ they come
  from. We thus abbreviate $\papp{\selfcnt}{\ssnd{s}}$ with
  $\fapp{\selfcnt}{s}$, $\papp{\selfhist}{\sfst{s}}$ with
  $\fapp{\selfhist}{s}$, and similarly for $\othercnt$ and
  $\otherhist$.}
The second line morphs $\unlockprog$ \emph{vacuously}, as unlocking
leaves $\fapp{\selfcnt}{s}$ unchanged. Nevertheless, \emph{some}
morphing of $\unlockprog$ is necessary, to bring the commands under
the same resource type.

\begin{figure}
\[
\!\!\!\begin{array}{r@{\ \ }l}
\scs 1. & \spec{\fapp{\selfcnt}{s} = n}\\
\scs 2. & \spec{\fapp{\selfcnt}{s} = n \wedge \fapp{\selfhist}{s} = h}\\
\scs 3. & \morph {\morphsc_1} \lockprog;\ /\!/\ \fapp {I_1} s \eqdef \fapp{\selfcnt}{s} = n + \papp{\sharp_\lockval}{\fapp{\selfhist}{s}} - \fapp{\sharp_\lockval}{h}\\
\scs 4. & \spec{\fapp{\selfcnt}{s} = n + \papp{\sharp_\lockval}{\fapp{\selfhist}{s}} - \fapp{\sharp_\lockval}{h} \wedge \fapp{\selfhist}{s} = h \join t \hmapsto \lockval}\\
\scs 5. & \spec{\fapp{\selfcnt}{s} = n + 1 \wedge \fapp{\selfhist}{s} = h'}\\
\scs 6. & \morph {\morphsc_{42}} \unlockprog;\ /\!/\ \fapp {I_{42}} s \eqdef \fapp{\selfcnt}{s} = n + 1\\
\scs 7. & \spec{\fapp{\selfcnt}{s} = n + 1 \wedge (\fapp{\selfhist}{s} = h' \vee 
                                                \fapp{\selfhist}{s} = h' \join t' \hmapsto \unlockval)}\\
\scs 8. & \spec{\fapp{\selfcnt}{s} = n + 1 \wedge \fapp{\selfhist}{s} = h''}\\
\scs 9. & \morph {\morphsc_2} \lockprog\ /\!/ \ \fapp {I_2} s \eqdef \fapp{\selfcnt}{s} = n + 1 + 2(\papp{\sharp_\lockval}{\fapp{\selfhist}{s}} - \fapp{\sharp_\lockval}{h''})\\
\scs{10}. & \spec{\fapp{\selfcnt}{s} = n + 1 + 2(\papp{\sharp_\lockval}{\fapp{\selfhist}{s}} - \fapp{\sharp_\lockval}{h''}) \wedge 
                  \fapp{\selfhist}{s} = h'' \join t'' \hmapsto \lockval}\\
\scs{11}. & \spec{\fapp{\selfcnt}{s} = n+3}\\
\end{array}
\]\vspace{-3mm}
\caption{Using the \TirName{Morph} rule to show that
  $\fapp{\selfcnt}{s}$ increments by $3$. $\papp{\sharp_\lockval}{-}$
  is the number of $\lockval$-entries in a
  history.}\label{fig:morphoutline}
\end{figure}

The proof outline in Figure~\ref{fig:morphoutline} shows that
$\fapp{\selfcnt}{s}$ increments by $3$, and we discuss its main points
next. In the outline, $\fapp{\sharp_\lockval}$ is a function on
history that computes the number of $\lockval$ entries in the history.
The outline starts with the precondition $\fapp{\selfcnt}{s} = n$,
where $n$ snapshots ``our'' current count. Line 2 uses $h$ to snapshot
``our'' history.  Line 3 applies $\mathsf{morph}$ to lock, and
correspondingly, the \TirName{Morph} rule in the proof.
At this point, we choose the simulation $I_1$ as indicated in line 3,
to state that the counter $\selfcnt$ increments $n$ by the number of
\emph{fresh} $\lockval$-entries in the history. Intuitively, $I_1$ is
an $\morphsc_1$-simulation because it is preserved under incrementing
$\fapp{\selfcnt}{s}$ by $1$ while simultaneously adding an
$\lockval$-entry to $\fapp{\selfhist}{s}$ (Figure~\ref{fig:f1}).
It's easy to see that $I_1$ holds in line 2, thus by \TirName{Morph},
it holds in line 4 as well. But, in line 4, by postcondition of
$\mathsf{lock}$, the history $\fapp{\selfhist}{s}$ has one more
locking entry. Thus, $\fapp{\selfcnt}{s}$ is increased by 1 (line
5). The remainder of the outline proceeds similarly.

\begin{figure}[t]
\[\tikzcdset{arrow style=tikz, diagrams={>=stealth}}
\begin{tikzcd}[every label/.append style = {font = \scriptsize},sep=3em]
	\sfst{s'}  & \fapp{I_1}{s'} \arrow[l, "(\morphsc_1)_\Sigma", swap] & 
     \fapp{\selfcnt}{s'} = n + \papp{\sharp_\lockval}{\fapp{\selfhist}{s'}} - \fapp{\sharp_\lockval}{h} \\
     \sfst{s} \arrow[u, "\mathsf{lock\_tr}"]  & \fapp{I_1}{s} \arrow[l, "(\morphsc_1)_\Sigma"] 
	\arrow[u, "\begin{array}{c}\locktr \hbox{}\\ \couple \hbox{}\\ \incrtr\ 1\end{array}", swap] & 
     \fapp{\selfcnt}{s} = n + \papp{\sharp_\lockval}{\fapp{\selfhist}{s}} - \fapp{\sharp_\lockval}{h}
     \arrow[u, phantom, "{\scriptsize\begin{array}{c}
        \fapp{\selfhist}{s'} = \fapp{\selfhist}{s}\join \papp{\fresh}{\fapp{\totalhist}{s}} \hmapsto \lockval\\
          \fapp{\selfcnt}{s'} = \fapp{\selfcnt}{s} + 1
        \end{array}}"]\\[-.5cm]
\end{tikzcd}
\]
\vspace{-7mm}
\caption{Diagram showing that $\fapp{I_1}{s} \eqdef \fapp{\selfcnt}{s}
  = n + \papp{\sharp_\lockval}{\fapp{\selfhist}{s}} -
  \fapp{\sharp_\lockval}{h}$ is an $\morphsc_1$-simulation (case of $\locktr$
  transition). 
The diagram specializes Figure~\ref{fig:diagram} to $\morphsc_1$, $I_1$ and
$\locktr$.}\label{fig:f1}
\end{figure}

We close the discussion with the observation that the property of
being a simulation (i.e., making diagrams in Figures~\ref{fig:diagram}
and~\ref{fig:f1} commute) relies only on the resource in the program's
type, and the morphism in question, not on the program's code, as
required for compositional reasoning. In this respect, the simulations
are different from loop invariants, which are properties of
programs. The \TirName{Morph} rule ties the simulations to the morphed
program by conjoining them with the program's pre- and the
postcondition. Specifically above, $I_1$ enables computing the
end-value of $\selfcnt$ from the end-value of $\selfhist$, and
$\selfhist$ is given by the spec of $\lockprog$.

\newcommand{\aof}[1]{\papp{\pcmA}{#1}}
\newcommand{\thread}{\theta}

\section{Definitions of the Formal Structures}\label{sec:formal}

To develop the notions of morphisms and simulations, we first require
a number of auxiliary definitions, such as states, transitions, and
resources on which morphisms act. This section defines all the
concepts formally, culminating with the inference rules of our system.

\subsection{States}\label{sec:statedefs}

\subsubsection{Subjective Components}
Different resources may contain different state components, e.g.,
$\thehist$ of $\Spin$ and $\thecnt$ of $\Cnt$. In general, a state is
parametrized by two types: $\pcmA$ classifies the \emph{self} and
\emph{other} components, and $T$ classifies the \emph{joint} (aka.,
\emph{shared}) state. Thus, $s = (\selfa, \jointa, \othera)$ is a
\dt{state} if $\selfa, \othera\!  \in\!  \pcmA$, and $\jointa\!\in\!
T$. If we want to be explicit about the types, we say that $s$ is an
\dt{$(\pcmA, T)$-state}.
We use $\fapp{\selfa}{s}$, $\fapp{\jointa}{s}$ and $\fapp{\othera}{s}$
as generic projections out of $s$, but rename them in specific cases,
for readability.
For example, in the case of $\Spin$: $\pcmA$ is $\histset$, $T$ is
$\tyUnit$ type, and $\selfhist/\otherhist$ renames
$\selfa/\othera$. In the case of $\Cnt$: $\pcmA$ is $\mathbb{N}$, $T$
is $\tyUnit$ type, and $\selfcnt/\othercnt$ renames
$\selfa/\othera$.

Because $\selfa$ and $\othera$ represent thread-specific views of the
state, we refer to them as \emph{subjective components}, and to $s$ as
\emph{subjective state}~\cite{LeyWild-Nanevski:POPL13}.

\subsubsection{Algebra of Subjectivity}

The specs must often combine the subjective components, \cf how
histories were unioned by $\join$ to express timestamp freshness in
the spec of $\lockprog$. To make the combination uniform, $\pcmA$ is
endowed with the structure of a \dt{partial commutative monoid (PCM)}.
A PCM is triple $(\pcmA, \join, \pcmU)$ where $\join$ (\emph{join}) is
a partial, commutative, associative, binary operation on $\pcmA$, with
$\pcmU$ as the unit. 
As a generic notation, we write $x \orth y$ to denote that $x \join y$
is defined.

Example PCMs are $\histset$ with disjoint union and the empty history
$\hempty$, and $\mathbb{N}$ with $+$ and $0$. Another common PCM is
the set of heaps (denoted $\heapset$). Heaps map pointers to values,
and are thus similar to histories, which map timestamps to
operations. We can therefore reuse the history notation, and write,
e.g.:
\[
x \hmapsto 3 \join y \hmapsto \mathsf{false}
\]
to describe the heap containing pointers $x$ and $y$, storing $3$ and
$\mathsf{false}$, respectively.\footnote{We silently already used
  this notation to define the erasure function for $\Spin$ in
  Figure~\ref{fig:spin}.} $\heapset$ is a PCM with disjoint union and
the empty heap $\hempty$, similar to $\histset$.  
Cartesian product of PCMs is a PCM, so PCMs can be combined, \cf the
PCM of $\SC$ is constructed out of those of $\Spin$ and $\Cnt$ in
Section~\ref{sec:overview}.

\begin{figure}
\[
\begin{array}{c@{}c@{}c}
\begin{array}{c}
\begin{tikzpicture}[scale=0.94, transform shape]
  \filldraw[fill=gray!70, draw=black] (0, 0) -- (135:1cm) arc (135:-135:1cm) -- cycle; 
  \filldraw[fill=gray!20, draw=black] (0, 0) -- (135:1cm) arc (135:225:1cm) -- cycle; 
  \draw[thin] (-45:1cm) -- (0, 0);
  \filldraw[fill=white, draw=black] (0,0) circle (0.53cm);
  \node[scale=0.9] at (-0.75cm, 0) {$a_1$};
  \node[scale=0.9] at (0, -0.75cm) {$a_2$};
  \node[scale=0.9] at (0.545cm, 0.545cm) {$a_3$};
  \node[scale=0.9] at (0, 0) {$\jointa$};
  \newlength\stxtwdl%
  \settowidth{\stxtwdl}{$\selfa = a_1 \hspace{3mm}$}%
  \node[scale=0.8, anchor=east] at (-225:1cm) {\makebox[\stxtwdl][l]{$s_1$:}};
  \node[scale=0.8, anchor=east] at (225:1cm) {$\selfa = a_1\quad$};
  \node[scale=0.8, anchor=west] at (45:1cm) {$\quad \othera = a_2 \join a_3$};
\end{tikzpicture}\\
{\fontsize{9}{9}\textrm{(1) left thread $\thread_1$}}
\end{array}
&
\begin{array}{c}
\begin{tikzpicture}[scale=0.94, transform shape]
  \filldraw[fill=gray!70, draw=black] (0, 0) -- (-45:1cm) arc (-45:225:1cm) -- cycle; 
  \filldraw[fill=gray!20, draw=black] (0, 0) -- (225:1cm) arc (225:315:1cm) -- cycle; 
  \draw[thin] (135:1cm) -- (0, 0);
  \filldraw[fill=white, draw=black] (0,0) circle (0.53cm);
  \node[scale=0.9] at (-0.75cm, 0) {$a_1$};
  \node[scale=0.9] at (0, -0.75cm) {$a_2$};
  \node[scale=0.9] at (0.545cm, 0.545cm) {$a_3$};
  \node[scale=0.9] at (0, 0) {$\jointa$};
  \newlength\stxtwdr%
  \settowidth{\stxtwdr}{$\selfa = a_1 \hspace{3mm}$}%
  \node[scale=0.8, anchor=east] at (-225:1cm) {\makebox[\stxtwdr][l]{$s_2$:}};
  \node[scale=0.8, anchor=east] at (225:1cm) {$\selfa = a_2\quad$};
  \node[scale=0.8, anchor=west] at (45:1cm) {$\quad \othera = a_3 \join a_1$};
\end{tikzpicture}\\
{\fontsize{9}{9}\textrm{(2) right thread $\thread_2$}}
\end{array}
&
\begin{array}{c}
\begin{tikzpicture}[scale=0.94, transform shape]
  \filldraw[fill=gray!20, draw=black] (135:1cm) arc (135:315:1cm);
  \filldraw[fill=gray!70, draw=black] (135:1cm) -- (-45:1cm) arc (-45:135:1cm);
  \draw[thin] (225:1cm) -- (0, 0);
  \filldraw[fill=white, draw=black] (0,0) circle (0.53cm);
  \node[scale=0.9] at (-0.75cm, 0) {$a_1$};
  \node[scale=0.9] at (0, -0.75cm) {$a_2$};
  \node[scale=0.9] at (0.545cm, 0.545cm) {$a_3$};
  \node[scale=0.9] at (0, 0) {$\jointa$};
  \newlength\stxtwd%
  \settowidth{\stxtwd}{$\selfa = a_1 \join a_2 \hspace{3mm}$}%
  \node[scale=0.8, anchor=east] at (-225:1cm) {\makebox[\stxtwd][l]{$s = s_1 * s_2$:}};
  \node[scale=0.8, anchor=east] at (225:1cm) {$\selfa = a_1 \join a_2 \quad$};
  \node[scale=0.8, anchor=west] at (45:1cm) {$\quad \othera = a_3$}; 
\end{tikzpicture}\\
{\fontsize{9}{9}\textrm{(3) parent thread $\thread = \thread_1 \parallel \thread_2$}}
\end{array}
\end{array}\vspace{-1em}
\]
\caption{Values of \emph{self} component $\selfa$ (light shade) and
  \emph{other} component $\othera$ (dark shade) in the states of
  parallel threads and their parent. The inner white circle represents
  the \emph{joint} component, and is equal for all
  threads.}\label{fig:subjectivity}
\end{figure}
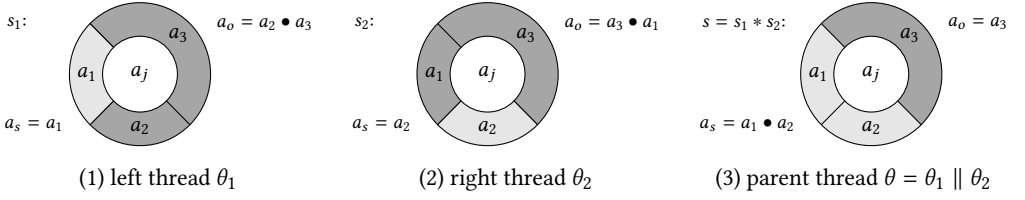

\subsubsection{Subjectivity and Parallel Composition}\label{sec:subjectivity}

The subjective components are local, in the sense that they have
different values in different threads. However, despite the locality,
the components of different threads aren't independent, but are
inter-related as shown in Figure~\ref{fig:subjectivity}.

Imagine three threads $\thread_1$, $\thread_2$ and $\thread_3$ running
concurrently. Their respective states must have the forms
$s_1 = (a_1, \jointa, a_2 \join a_3)$,
$s_2 = (a_2, \jointa, a_3 \join a_1)$ and
$s_3 = (a_3, \jointa, a_1 \join a_2)$. Indeed, any two of the threads
combined are the environment for the third thread. Thus, the PCM join
of the \emph{self}'s of any two threads must equal the \emph{other} of
the third thread.
Figures~\ref{fig:subjectivity}(1) and~\ref{fig:subjectivity}(2)
illustrate this property for threads $\thread_1$ and $\thread_2$,
with $\thread_3$ being their implicit environment.

If $\thread$ is the parent thread of $\thread_1$ and $\thread_2$, then
its state is $s = (a_1 \join a_2, \jointa, a_3)$, since $\thread$ is
the combination of $\thread_1$ and $\thread_2$, and has $\thread_3$ as
the environment. We abbreviate as $s = s_1 * s_2$ the relationship
between the parent state $s$, and the children states $s_1$ and $s_2$,
and illustrate it in Figure~\ref{fig:subjectivity}(3).

\subsubsection{Globality}

A property or a function is \emph{global} if it remains invariant
under moving PCM values between subjective components. In light of
Figure~\ref{fig:subjectivity}, such properties and functions obtain
equal valuations across \emph{all} concurrent threads, thus justifying
the name.
We introduce several operations for surgery on subjective states, and
then use them to define globality and conditional globality, where the
invariance holds only under a (global) condition.

\begin{definition}\label{def:framing}
  Let $p \in \pcmA$ and $s=(\selfa, \jointa, \othera)$ be an
  $(\pcmA, T)$-state. The \dt{\emph{self}-framing} of $s$ with the
  \dt{frame} $p$ is the state
  $s \zag p = (\selfa \join p, \jointa, \othera)$. Dually, the
  \dt{\emph{other}-framing} of $s$ with $p$ is the state
  $s \zig p = (\selfa, \jointa, p \join \othera)$. 
\end{definition}

\begin{definition}\label{def:global}
A predicate $P$ is \dt{global}, if $\papp{P}{s \zig p} \leftrightarrow
\papp{P}{s \zag p}$ for every $p$ and state $s$ such that
$\fapp{\selfa}{s}\orth p \orth \fapp{\othera}{s}$.
A (partial) function $f$ on states is global, if $\papp{f}{s \zig p} =
\papp{f}{s \zag p}$ under the same conditions.
\end{definition}

Examples of global predicates from Section~\ref{sec:overview} are
$\fapp{P}{s} = \fapp{\selfhist}{s} \orth \fapp{\otherhist}{s}$, and
$\fapp{Q}{s} = \papp{\alternate}{\fapp{\totalhist}{s}}$ used in
Figure~\ref{fig:spin} to characterize $\Spin$ histories. $P$ and $Q$
are both defined in terms of $\fapp{\totalhist}{s} =
\fapp{\selfhist}{s} \join \fapp{\otherhist}{s}$; $Q$ directly so, and
$P$ because $\fapp{\selfhist}{s} \orth \fapp{\otherhist}{s}$ iff
$\fapp{\selfhist}{s} \join \fapp{\otherhist}{s}$ is itself defined.
In other words, both $P$ and $Q$ express a property of the collective
history of all threads operating over $\Spin$, taken
together. Clearly, the value of this history is invariant across all
the threads, and therefore, so are $P$ and $Q$. Specifically, they are
invariant under shuffling timestamps between $\selfhist$ and
$\otherhist$, as this doesn't alter the total.
In fact, $\totalhist$ itself is a global function, so we proceed to
refer to $\totalhist$ as the global history.

\begin{definition}\label{def:globalunder}
  Let $X$ be a global predicate. A predicate $P$ is \dt{global under 
    $X$}, if $\papp{P}{s \zig p} \leftrightarrow \papp{P}{s \zag p}$
  for every $p$ and state $s$ such that
  $\fapp{\selfa}{s}\orth p \orth \fapp{\othera}{s}$ \emph{and
    $\papp{X}{s \zag p}$}. Similarly for functions.
\end{definition}

\subsubsection{Subjectivity and Framing}\label{sec:subjective-framing}

Subjective state makes framing work somewhat differently than in the
customary, non-subjective, separation logics. The latter may be viewed
as having the \emph{self} component, but lacking \emph{other}.
To illustrate the difference, we give $\lockprog$
(Section~\ref{sec:overview}) the following spec, which is
small~\cite{OHearnRY01} wrt.~the history $\fapp{\selfhist}{s}$,
\[
\begin{array}[t]{r@{\,}c@{\,}l}
\lockprog & : &
   [k]\ldot\!\!\!\begin{array}[t]{l}
          \spec{\lambda s\ldot \fapp{\selfhist}{s} = \hempty \wedge k \leq \papp{\lastkey}{\fapp{\otherhist}{s}}}\\
          \spec{\lambda s\ldot \exists t\ldot \fapp{\selfhist}{s} = t \hmapsto \lockval \wedge k < t} @ \Spin
          \end{array} 
\end{array}
\]
and then we frame the history $h$ onto $\fapp{\selfhist}{s}$ to obtain
the equivalent large spec we actually presented:
\[
\begin{array}[t]{r@{\,}c@{\,}l}
\lockprog & : &
   [h, k]\ldot\!\!\!\begin{array}[t]{l}
          \spec{\lambda s\ldot \fapp{\selfhist}{s} = h \wedge k \leq \papp{\lastkey}{\fapp{\totalhist}{s}}}\\ 
          \spec{\lambda s\ldot \exists t\ldot \fapp{\selfhist}{s} = h \join t \hmapsto \lockval \wedge k < t} @ \Spin
          \end{array} 
\end{array}
\]
As expected in separation logic, framing increased the starting
$\fapp{\selfhist}{s}$ from $\hempty$ to $h$, which is the key
distinction between small and large specs.
But this isn't all it did; it also deducted $h$ from
$\fapp{\otherhist}{s}$. Indeed, had $\fapp{\otherhist}{s}$ been
unchanged (as might also be expected in separation logic), then both
specs would contain the same conjunct $k \leq
\papp{\lastkey}{\fapp{\otherhist}{s}}$. But the large spec contains $k
\leq \papp{\lastkey}{\fapp{\totalhist}{s}} = \papp{\lastkey}{h \join
  \fapp{\otherhist}{s}}$, where $h$ is joined to
$\fapp{\otherhist}{s}$ to compensate for the deduction.

To explain the deduction, notice that in \emph{any} separation logic,
framing is a special case of parallel composition. To add a frame $h$
to the state of a program $e$, it suffices to compose $e$ in parallel
with the idle thread having $h$ as its \emph{self}. The composition
executes like $e$, but with \emph{self} enlarged by $h$, and $h$
remains unchanged.
In the subjective setting, parallel composition joins the
\emph{self}'s of two threads, \emph{but also decreases the
  \emph{other} of the parent}, as illustrated in
Figure~\ref{fig:subjectivity}. It is this decrease that is evidenced
in the large spec.

Therefore, framing enlarges \emph{self} by $h$, and simultaneously
removes $h$ from \emph{other}, which \emph{must already contain} $h$.
Framing shuffles existing state between components, but doesn't
introduce new state, in contrast to the usual separation logic
formulations. This preserves the values of global functions, and
facilitates their use in specs (e.g., the global history
$\fapp{\totalhist}$ in $\lockprog$).

\subsection{Resources}\label{sec:resdefs}

Resources consist of state spaces and transitions. The state spaces
describe the properties that hold for all threads of the resource, so
we use global predicates and functions to represent them.

\begin{definition}\label{def:coh}
A \dt{state space} is a pair $\Sigma = (P, \hflat{-})$, where $P$ is a
global predicate and $\hflat{-}$ is a partial function into heaps,
global under $P$, called \emph{erasure}, such that for every state
$s$, $\fapp{P}{s}$ implies $\selfa(s) \orth \othera(s)$ and
$\hflat{s}$ is defined. We write $s \in \Sigma$ to mean $\fapp{P}{s}$.
\end{definition}

Transitions describe the allowed atomic modifications on state.  We
require the following properties of them, to facilitate
separation-style reasoning.

\begin{definition}\label{def:trans}
A \dt{transition} $t$ over state space $\Sigma$ is a binary relation
on $\Sigma$ states, such that:
\begin{enumerate}
\item[(1)]\label{def:func} \emph{(partial function)} if $t\ s\ s'_1$ and $t\ s\ s'_2$ then $s'_1 = s'_2$.
\item[(2)] \emph{(other-fixity)} if $t\ s\ s'$, then $\fapp{\othera}{s} = \fapp{\othera}{s'}$
\item[(3)] \emph{(transition locality)} if $t\ (s \zig p)\ x$, 
then there exists $s'$ such that $x = s' \zig p$ and
  $t\ (s\zag p)\ (s'\zag p)$
\end{enumerate}
A state $s$ is \dt{safe} for $t$, if there exists $s'$ such that
$t\ s\ s'$.
\end{definition}

%
Property (2) captures that a transition can't
change the \emph{other}-component, as it's private to other
threads. However, a transition can read this component, cf.~how
$\locktr$ in Figure~\ref{fig:spin} uses $\fapp{\otherhist}{s}$ as part
of $\fapp{\totalhist}{s}$ to compute a fresh timestamp.

Transition locality (3) essentially says that transitions can be
framed. To see how, let $\thread_1$ be a thread in the state
$s \zig p$, whose sibling $\thread_2$ has \emph{self}-component
$p$. Their parent $\thread$ is thus in the state $s \zag p$, by
Figure~\ref{fig:subjectivity}. If $\thread_1$ performs a transition
$t\ (s \zig p)\ x$, then by (3), the move can be seen as a transition
of $\thread$ in the state $s \zag p$. In other words, the transition
of a child can be seen as a transition of the parent, but with
\emph{self} enlarged by $p$, and \emph{other} suitably reduced by $p$.
This is precisely the view of framing described in
Section~\ref{sec:subjective-framing}.
Hence, transition locality is the base case of, and gives rise to,
framing on programs, as a program's execution is a sequence of
transitions.

\begin{definition}\label{def:internal}
A $\Sigma$-transition $t$ is \dt{footprint preserving} if $t\ s\ s'$
implies that $\hflat{s}$ and $\hflat{s'}$ contain the same pointers.
\end{definition}

Transitions that preserve footprints are important because they can be
coupled with other such transitions without imposing side conditions
on the combination.
For example, consider the $\incrtr$ transition of $\Cnt$ in
Figure~\ref{fig:spin}, which is footprint preserving, as it doesn't
allocate or deallocate any pointers. Were it also to allocate, we will
have a problem when combining $\Spin$ and $\Cnt$, as we must impose
that $\incrtr$ won't allocate the pointer $r$, already taken by
$\Spin$.
For simplicity, we here present the theory with only
footprint-preserving transitions, but have added non-preserving
(aka.~external) transitions as well~\cite{artifact}. External
transitions encode transfer of data in and out of a
resource~\cite{intauto:2001}, of which allocation and deallocation are
an instance.
When a resource requires allocation or deallocation, it can be
tensored with an allocator resource to exchange pointers through
ownership transfer~\cite{FilipovicOTY10,Nanevski-al:ESOP14} via
external transitions. We elide further discussion, but refer to the
Coq files for the implementation of an allocator resource and example
programs that use it.

\begin{definition}\label{def:resource}
  A \dt{resource} is a tuple $V = (\pcmA, T, \Sigma, \TransSet)$,
  where $\Sigma$ is a space of $(M, T)$-states, and $\TransSet$ a set
  of footprint preserving $\Sigma$ transitions. We refer to $V$'s
  components as projections, e.g.~$\sigmaof{V}$ for the state space,
  $\deltaof{V}$ for the transitions, $\aof{V}$ for the PCM, etc. A
  state $s$ is $V$-state iff $s \in \sigmaof{V}$.
\end{definition}

We close the discussion on resources by defining actions---atomic
operations on (combined real and ghost) state, which are the basic
building blocks of programs.

\begin{definition}
An \dt{action} of type $A$ in a resource $V$ is a partial function $a
: \sigmaof{V} \rightharpoonup \transof{V} \times A$, mapping input
state to output transition and value, which is \dt{local}, in the
sense that it is invariant under framing. Formally, if $a\ (s \zig p)
= (t, v)$ then $a\ (s \zag p) = (t, v)$; that is, if $a$ is performed
by a child thread, it behaves the same when viewed by the parent.

The \dt{effect} of $a$ is the partial function $[a] :
\sigmaof{V}\rightharpoonup \sigmaof{V} \times A$ mapping input state
to output \emph{state} and value, defined as $[a]\ s = (s', v)$ iff
$\exists t\ldot a\ s = (t, v) \wedge t\ s\ s'$. Note that $[a]$ is a
(partial) function because $a$ and $t$ are.
\end{definition}

For example, we model the bracketed code used in the $\lockprog$ loop
in Section~\ref{sec:overview}, as the following action of type
$\mathsf{bool}$:
\vspace{-0.5mm}
\begin{equation}
\label{act:trylock}  \mathsf{trylock\_act}\ s \eqdef
\begin{cases}
    (\locktr, \mathsf{true}) & \textrm{if $\neg \papp{\islocked}{\fapp{\totalhist}{s}}$}\\
  (\idtr, \mathsf{false}) & \textrm{otherwise}
\end{cases}
\end{equation}
The action is local, as it depends only on $\fapp{\totalhist}{s}$,
which is invariant under framing.  

We say that $a$ \dt{erases} to an atomic read-modify-write (RMW)
command $c$~\cite{Herlihy-Shavit:08}, if $[a]$ behaves like $c$ when
the states are erased to heaps. In other words, if $[a]\ s = (s', v)$,
then $c\ \hflat{s} = (\hflat{s'}, v)$. One may check that
$\mathsf{trylock\_act}$ erases to $\mathsf{CAS}(r, \mathsf{false},
\mathsf{true})$, as expected.\footnote{All the actions we use in this
  paper and in the Coq code erase to some RMW command. However, we
  proved this only by hand, as our formalism and the Coq
  implementation don't currently issue proof obligations to check
  this. In general, we currently treat code and ghost code equally,
  and, as customary in type theory, equally to proofs. Differentiating
  between these \emph{formally} is an orthogonal issue that we plan to
  address in the future by making a type distinction between them,
  such as in the work on proof irrelevance in type
  theory~\cite{pfe:lics01,gil+coc+soz+tab:popl19,bar+ber:fossacs08}.}
Similarly,
\vspace{-0.5mm}
\begin{equation}
 \label{act:unlock} \mathsf{unlock\_act}\ s \eqdef
 \begin{cases}
  (\unlocktr, ()) & \textrm{if $\papp{\islocked}{\fapp{\totalhist}{s}}$}\\
   (\idtr, ())     & \textrm{otherwise}
 \end{cases}
\end{equation}
is an action of $\tyUnit$ type, which erases to $r := \mathsf{false}$.

\subsection{Morphisms}\label{sec:morphdefs}

\begin{definition}\label{def:morph}
A \dt{resource morphism} $f : \morphtp{V}{W}$ consists of two partial
functions $f_\Sigma : \sigmaof{W} \rightharpoonup \sigmaof{V}$ (note
the contravariance), and $f_\Delta : \sigmaof{W} \rightharpoonup
\transof{V} \rightharpoonup \transof{W}$, such that:
\begin{enumerate}
\item[(1)]\label{morf:frame} \emph{(locality of $f_\Sigma$)} there
  exists a function $\phi : \aof{W} \rightarrow \aof{V}$ such that if
  $\morpheq{s_v}{s_w \zig p}{f_\Sigma}$,
%
%
  then there exists $s'_v$ such that $s_v = s'_v \zig \papp{\phi}{p}$,
  and $\morpheq{s'_v \zag \papp{\phi}{p}}{s_w \zag p}{f_\Sigma}$.
\item[(2)]\label{morph:loc} \emph{(locality of $f_\Delta$)} if
  $\papp{f_\Delta}{s_w \zig p}{(t_v)} = t_w$, 
%
%
  then $\papp{f_\Delta}{s_w \zag p}{(t_v)} = t_w$.
\item[(3)]\label{morf:other} \emph{(other-fixity)} if
  $\papp{\othera}{s_w} = \papp{\othera}{s'_w}$ and
  $\papp{f_\Sigma}{s_w}$, $\papp{f_\Sigma}{s'_w}$ exist, then
  $\papp{\othera}{\papp{f_\Sigma}{s_w}} =
  \papp{\othera}{\papp{f_\Sigma}{s'_w}}$.
\end{enumerate}
\end{definition}
A morphism $f$ transforms a $V$-program $e$ into a $W$-program, as
follows. When $\morph{f}{e}$ is in a $W$-state $s_w$, it has to
determine a $W$-transition to take. It does so by obtaining a
$V$-state $s_v = \papp{f_\Sigma}{s_w}$.  Next, out of $s_v$, $e$ can
determine the transition $t_v$ to take. The morphed $W$-program then
takes the $W$-transition $\papp{f_\Delta}{s_w}{(t_v)}$.

The properties (1) and (2) of Definition~\ref{def:morph} provide basic
technical conditions for this process to be invariant under
framing. Property (1) is a form of ``simulation of framing'', i.e., a
frame $p$ in $W$ can be matched with a frame $\papp{\phi}{p}$ in
$V$. Thus, framing a morphed program can be viewed as framing the
original program. Property (2) says that framing doesn't change the
transition that $f_\Delta$ produces; thus it doesn't influence the
behavior of morphed programs.
The property (3) restricts the choice of $s'_v$ in (1) so that 
$\papp{\othera}{s'_v}$ is uniquely determined by
$\papp{\othera}{s_w}$, much as how $\papp{\phi}{p}$ in (1) is uniquely
determined by $p$. This is a technical condition which we required to
prove the soundness of the frame rule.

\textit{Example.} Properties (1)-(3) are all satisfied by the
morphisms $\morphsc_n : \Spin \rightarrow \SC$ from
Section~\ref{sec:overview}. Indeed, $\aof{\SC} = \aof{\Spin} \times
\aof{\Cnt} = \histset \times \mathbb{N}$. Thus, a frame in $\SC$ is a
pair of a history and a nat; it is transformed into a frame in $\Spin$
just by taking the history component. We thus instantiate $\phi$ in
(1) with the first projection function, and it is easy to see that it
satisfies the rest of (1). Property (2) holds because
$(\morphsc_n)_\Delta$ doesn't depend on the state argument, hence
framing this state doesn't change the output. Finally, in (3), the
values $\papp{\othera}{s_w}$ and $\papp{\othera}{s'_w}$ are also pairs
of a history and a nat. If the pairs are equal, then their history
components are equal too, deriving (3).

Finally, resources and their morphisms support a basic categorical
structure, under the following notions of morphism \emph{identity} and
\emph{composition}. We have proved in the Coq files that morphism
composition is associative, with the identity morphism as the unit,
where two morphisms are equal if their $\Sigma$ and $\Delta$
components are equal as partial functions.
\begin{definition}
The \dt{identity} morphism $\mathsf{id} : V \rightarrow V$ is defined 
by $\mathsf{id}_\Sigma\ s = s$ and $\mathsf{id}_\Delta\ s\ t = t$.  
The \dt{composition} of morphisms $f : U \rightarrow V$ and
$g : V \rightarrow W$ is the morphism $g \circ f : U \rightarrow W$
defined by:
\[
\begin{array}{l@{\ }c@{\ }l}
(g \circ f)_\Sigma\ s_w  & \eqdef & f_\Sigma\ (g_\Sigma\ s_w)\\
(g \circ f)_\Delta\ s_w\ t_u & \eqdef & g_\Delta\ s_w\ (f_\Delta\ (g_\Sigma\ s_w)\ t_u) 
\end{array}
\]
\end{definition}

\subsection{Simulations}\label{sec:simdefs}

Because $f_\Sigma$ and $f_\Delta$ are partial, a program lifted by a
morphism isn't immediately guaranteed to be safe (i.e., doesn't get
stuck). For example, the state $s_v = f_\Sigma\ {s_w}$, whose
computation is the first step of morphing, needn't exist. Even if
$s_v$ does exist, and the original program takes the transition $t_v$
in $s_v$, then $t_w = f_\Delta\ s_w\ t_v$ needn't exist. Even if
$t_w$ does exist, there is no guarantee that $s_w$ is safe for $t_w$.
An $f$-simulation is a condition that guarantees the existence of
these entities, and their mutual agreement (e.g., that $s_w$ is safe
for $t_w$), so that a morphed program that typechecks against the
$\TirName{Morph}$ rule doesn't get stuck.

\begin{definition}\label{def:sim}
Given a morphism $f : \morphtp{V}{W}$, an \dt{$f$-simulation} is a
predicate $I$ on $W$-states such that:
\begin{enumerate}
\item [(1)]\label{morf:simWV} 
  if $\fapp{I}{s_w}$, and $s_v = \papp{f_\Sigma}{s_w}$ exists, and 
  $t_v\ s_v\ s'_v$, then there exist $t_w = {f_\Delta}\ {s_w}\ {t_v}$
  and $s'_w$ such that $\fapp{I}{s'_w}$ and
  $s'_v = \papp{f_\Sigma}{s'_w}$, and $t_w\ s_w\ s'_w$. 
\item [(2)]\label{morf:simVW} 
if $\fapp{I}{s_w}$, and $s_v = \papp{f_\Sigma}{s_w}$ exists, and $s_w
\osteps[W]{} s'_w$, then $\fapp{I}{s'_w}$, and $s'_v = \papp{f_\Sigma}{s'_w}$
exists, and $s_v \osteps[V]{} s'_v$. 
%
Here, the relation $s \ostep[W]{} s'$ denotes that $s$
\dt{other-steps} by $W$ to $s'$, i.e., that there exists a transition
$t \in \transof{W}$ such that $t\ {\transpo{s}}\ {\transpo{s'}}$.  The
\dt{transposition} $s^\top = (\fapp{\othera}{s}, \fapp{\jointa}{s},
\fapp{\selfa}{s})$ swaps the subjective components of $s$, to
obtain the view of \emph{other} threads. The relation
$\osteps[W]{}$ is the reflexive-transitive closure of $\ostep[W]{}$,
allowing for an arbitrary number of steps.
\end{enumerate}
\end{definition}

\begin{figure}[t]
\vspace{-4mm}
\[
\begin{array}{cc}
\tikzcdset{arrow style=tikz, diagrams={>=stealth}}
\begin{tikzcd}[every label/.append style = {font = \normalsize},sep=3em]
s'_v  & \fapp{I}{s'_w} \arrow[l, "f_\Sigma", swap] 
\\ 
s_v \arrow[u, "t_v"]  & \fapp{I}{s_w} \arrow[l, "f_\Sigma"] \arrow[u, "f_\Delta\,s_w\,t_v", swap] 
\end{tikzcd} &
\tikzcdset{
 arrow style=tikz, 
 diagrams={>=stealth}, 
 to*/.style={
   shorten >=.25em,#1-to, 
   to path={-- node[inner sep=0pt,at end,sloped] {${}^*$} (\tikztotarget) \tikztonodes}}, 
 to*/.default={arrow style=tikz, diagrams={>=stealth}}
}
\begin{tikzcd}[every label/.append style = {font = \normalsize},sep=3em]
s'_v 
  & \fapp{I}{s'_w} \arrow[l, "f_\Sigma", swap] \\ 
s_v \arrow[u, "V", swap, to*]  & \fapp{I}{s_w} \arrow[l, "f_\Sigma"]
\arrow[u, "W", swap, to*] 
\end{tikzcd}
\end{array}\vspace{-2mm}
\]
\caption{Commutative diagrams for the properties (1) and (2) of 
  Definition~\ref{def:sim} for $I$ to be an 
  $f$-simulation.\label{fig:sim}}
\vspace{-4mm}
\end{figure}
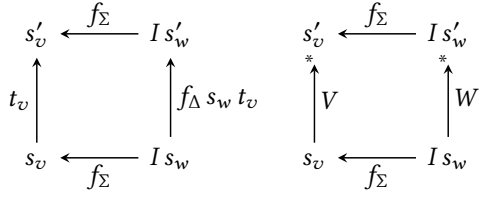

Property~(1) says that $W$ simulates $V$ on states satisfying $I$.
Property~(2) states the simulation in the opposite direction, i.e., of
$W$ by $V$, but allowing many \emph{other}-steps to match many
\emph{other}-steps. Notice that \emph{other}-stepping transitions over
\emph{transposed} states; that is, it changes the \emph{other}, but,
by Definition~\ref{def:trans}(2), preserves the \emph{self} of the
states. Intuitively, (2) ensures that interference in $W$ may be
viewed as interference in $V$, so that stable Hoare triples in $V$ can
be transformed into stable Hoare triples in $W$, which is required for
the soundness of the $\TirName{Morph}$ rule. 
Property (1) has already been shown in Figure~\ref{fig:diagram}; we
repeat it in Figure~\ref{fig:sim}, together with a diagram for
property (2). 

\subsection{Inference Rules}\label{sec:infdefs}

\newcommand*{\ftntDem}{\ifappendix
Appendix~\ref{sec:model} defines the denotational
semantics, in \CiC, for these notions,
and states a theorem, proved in Coq, that the inference rules are
sound wrt.~the denotational semantics.%
\else
The extended version of the
paper~\cite[Appendix~\apndxModel]{extended} defines the denotational
semantics, in \CiC, for these notions,
and states a theorem, proved in Coq, that the inference rules are
sound wrt.~the denotational semantics.%
\fi%
}

We present the system using the Calculus of Inductive Constructions
(\CiC) as an environment logic, hence as a shallow embedding in
Coq. We inherit from \CiC the useful concepts of higher-order
functions and substitution principles, and only present the notions
specific to Hoare logic\footnote{\ftntDem}.

We differentiate between two different notions of program types:
$\mathsf{ST}\,V\,A$ and $[\Gamma]\ldot \spec{P}\,A\,\spec{Q}@V$. The first
type circumscribes programs that respect the transitions of the
resource $V$, and return a value of type $A$ if they terminate.  The
second, Hoare type, is a subset of $\mathsf{ST}\,V\,A$, selecting only
those programs that satisfy the precondition $P$ and postcondition
$Q$, under the context $\Gamma$ of logical variables. To accommodate
for the return values, the postcondition $Q$ is now a predicate over
values of type $A$ \emph{and} states
(if $A=\tyUnit$, we elide it from the Hoare type, as we did in
Section~\ref{sec:overview}).

The key concept in the inference rules is the predicate transformer
$\vrf~e~Q$, which takes a program $e : \mathsf{ST}\,V\,A$, and a
postcondition $Q$,
and returns the set of $V$-states from which $e$ is safe to
run\footnote{Thus ensuring fault avoidance.} and produces a result $v$
and ending state $s'$ such that $Q\ v\ s'$. 
Hoare types are then defined in terms of $\vrf$, as follows.
\begin{align}
  [\Gamma]\ldot \spec{P}\,A\,\spec{Q}@V = \{ e : \mathsf{ST}\,V\,A \mid
     \forall \Gamma\ldot\forall s \in \papp{\Sigma}{V}\ldot \fapp{P}{s} \rightarrow \wpeqs e Q s\}\label{eq:hoare}
\end{align}
We formulate the system using both $\mathsf{vrf}$ and the Hoare
types. The former is useful, as it leads to compact presentation,
avoiding a number of structural rules of Hoare logic. The latter is
useful because it lets us easily combine Hoare reasoning with
higher-order concepts. For example, having inherited higher-order
functions from \CiC, we can immediately give the following type to the
fixed-point combinator, where $T$ is the dependent type $T =
\Pi_{x:A}\ldot[\Gamma]\ldot\spec{P}\,B\,\spec{Q}@ V$:
\begin{align}
\mathsf{fix} : (T \rightarrow T) \rightarrow T \tag*{\TirName{Fix}}
\end{align}
Here, $T$ serves as a loop invariant; in $\mathsf{fix}~(\lambda f\ldot
e)$ we assume that $T$ holds of $f$, but then have to prove that it
holds of $e$ as well, i.e., it is preserved upon the end of the
iteration.

In reasoning about programs, we keep the transformer $\vrf$ abstract,
and only rely on the following minimal set of rules. These, together
with the above definition of Hoare types and typing for
$\mathsf{fix}$, are the only Hoare-related rules of the system. In the
rules we assume that $e : \mathsf{ST}\,V\,A$, $e_i :
\mathsf{ST}\,V\,A_i$, $a$ is a $V$-action, $f : \morphtp{V}{W}$ is a
morphism, $I$ is an $f$-simulation,
%
%
$s \in \sigmaof{V}$, and $s_w \in \sigmaof{W}$.
\[
\!\!\!\begin{array}{l@{\ }c@{\ }l}
\mathsf{\vrf\_post} & : & 
(\forall v~s\ldot \fapp{J}{s} \rightarrow Q_1~v~s 
        \rightarrow Q_2~v~s) \rightarrow 
 \fapp{J}{s} \rightarrow \wpeqs e {Q_1} s \rightarrow \wpeqs e {Q_2} s\\
\mathsf{\vrf\_ret} & : & \fapp{\stab{(Q\ v)}}{s} \rightarrow \wpeqs {(\mathsf{ret}\ v)} Q s\\
\mathsf{\vrf\_bnd} & : & \wpeqs {e_1} {(\lambda x\ldot \wpeq {(e_2\ x)} Q)} s \rightarrow 
   \wpeqs {(x\leftarrow e_1; (e_2\ x))} Q s\\
\mathsf{\vrf\_par} & : & ((\wpeq {e_1} {Q_1})\,{\bstar}\,(\wpeq {e_2} {Q_2}))\ s \rightarrow 
   \wpeqs {(e_1 \parallel e_2)} {(\lambda v{:}A_1{\times}A_2\ldot (Q_1\,v.1){\bstar}(Q_2\,v.2))} s\\
& & \textrm{where $\fapp{(P\,{\bstar}\,Q)}{s} \eqdef \exists s_1\, s_2\ldot s = s_1 * s_2 \wedge \fapp{P}{s_1} \wedge \fapp{Q}{s_2}$}\\
\mathsf{\vrf\_frame} & : & 
 ((\wpeq {e} {Q_1}) \bstar {\stab Q_2})\ s \rightarrow \wpeqs {e} {(\lambda v\ldot (Q_1\ v) \bstar Q_2)} s\\
\mathsf{\vrf\_act} & : & \fapp{\stab{(\lambda s'\ldot \exists s''\ v\ldot [a]\ s' = (s'', v) \wedge \stab{(Q\ v)}\ s'')}}{s} \rightarrow 
 \wpeqs {\atomic{a}} Q s\\
\mathsf{\vrf\_morph} & : & \smorph f {(\wpeq e Q)}\ s_w \rightarrow \fapp{I}{s_w} \rightarrow 
 \wpeqs {(\mathsf{morph}\ f\ e)} {(\lambda v~s'_w\ldot \smorph f {(Q\ v)}\ s'_w \wedge \fapp{I}{s'_w})} s_w\\
& & \textrm{where $\fapp{\smorph{f}{R}}{s_w} \eqdef \exists s_v\ldot s_v = \fapp{f_\Sigma}{s_w} \wedge \fapp{R}{s_v}$}
\end{array}
\vspace{-1mm}
\]
In English:
\begin{itemize}
\item The $\mathsf{\vrf\_post}$ rule weakens the postcondition,
  similar to the well-known rule of Consequence in Hoare logic. The
  rule allows assuming a property $J$ when establishing a
  postcondition $Q_2$ out of $Q_1$. Here $J$ is an \dt{invariant},
  i.e., a property preserved by the transitions of $V$; an
  $\mathsf{id}$-simulation. Thus, invariants can be elided from program specs,
  and invoked by $\mathsf{\vrf\_post}$ when needed. 

\item The $\mathsf{\vrf\_ret}$ rule applies to an idle program
  returning $v$. When we want an idle program that returns no value,
  we simply take $v$ to be of $\tyUnit$ type. The rule
  explicitly \dt{stabilizes} the postcondition $Q$ to allow for the
  state $s$ to be changed by interference of other threads in between
  the invocation of the idle program and its termination. Here,
  stabilization of a predicate $Q$ is $\papp{\stab{Q}}{s} \eqdef
  \forall s'\ldot s \osteps[V]{} s' \rightarrow \papp{Q}{s'}$. The
  predicate $Q$ is \dt{stable} if $Q = \stab{Q}$, and it is easy to
  see that $\stab{Q}$ is stable for every $Q$.

\item The $\mathsf{\vrf\_bnd}$ rule is a Dijkstra-style rule for
  sequential composition. In order to show that the sequential
  composition $x \leftarrow e_1; (e_2\ x)$ has a postcondition $Q$, it
  suffices to show that $e_1$ has a postcondition
  $\lambda x\ldot \wpeq {(e_2\ x)} Q$. In other words, $e_1$
  terminates with a value $x$ and in a state satisfying
  $\wpeq {(e_2\ x)} Q$, so that running $e_2\ x$ in that state yields
  $Q$. 

\item The $\mathsf{\vrf\_par}$ and $\mathsf{\vrf\_frame}$ rules are
  predicate transformer variants of the rules for parallel composition
  and framing from separation logic. The separating conjunction $P
  \bstar Q$ is defined as customary in separation logic, except that
  we use the subjective splitting of state, as explained in
  Section~\ref{sec:subjectivity} and Figure~\ref{fig:subjectivity}.
  The $\mathsf{\vrf\_frame}$ rule can be seen as an instance of
  $\mathsf{\vrf\_par}$, where $e_2$ is taken to be the idle programs
  returning no value. Thus, $Q_2$ is explicitly stabilized in
  $\mathsf{\vrf\_frame}$, to match the precondition of the
  $\mathsf{vrf\_ret}$ rule for idle programs. 

\item The $\mathsf{\vrf\_act}$ rule says that $Q$ holds after
  executing action $a$ in state $s$, if $s$ steps to $s'$ by
  interfering threads, and then $[a]\ s'$ returns the pair $(s'', v)$
  of output state and value. The latter satisfy the stabilization of
  $Q$, to allow for interference on $s''$ after the termination of
  $a$.

\item The $\mathsf{vrf\_morph}$ rule is a straightforward
  casting of the $\TirName{Morph}$ rule from Section~\ref{sec:intro}
  into a predicate transformer style.

\end{itemize}

Finally, we also inherit all the CiC logical and programming
constructs as well, which has important consequences for Hoare-style
reasoning. For example, in CiC one can form conditionals over any type,
including propositions and $\mathsf{ST}\,V\,A$ types.  Thus, given a
Boolean $b$ and $e_1, e_2 : \mathsf{ST}\,V\,A$, the following rule,
\emph{derivable} by case analysis on $b$, allows us to write programs
that use conditionals, and verify them in the usual Hoare-logic style.
\[
  \mathsf{\vrf\_cond} : (\mathsf{if}\ b\ \mathsf{then}\ \wpeqs 
  {e_1}{Q}{s}\ \mathsf{else}\ \wpeqs{e_2}{Q}{s})\rightarrow 
  \wpeqs{(\mathsf{if}\ b\ \mathsf{then}\ e_1\ \mathsf{else}\ e_2)}{Q}{s}
\]
All the other customary rules of Hoare logic also become
derivable. For example, if $e : \spec{P}\,\spec{Q}$ and
$\forall s\in \sigmaof{V}\ldot \fapp{P'}{s} \rightarrow \fapp{P}{s}$,
then also $e : \spec{P'}\,\spec{Q}$. Similarly, if $e$ depends on a
logical variable $x: A$ (i.e.,
$e : [x : A]\ldot\spec{\fapp{P}{x}}\,\spec{\fapp{Q}{x}}$), then $x$
can be specialized by $v : A$, to derive
$e : \spec{\fapp{P}{v}}\,\spec{\fapp{Q}{v}}$. The latter follows
because the logical variables are universally quantified in the
definition of Hoare types (context $\Gamma$ in~\eqref{eq:hoare}), and
can thus be specialized just like any other universally quantified
variable.\footnote{We perform the described type changes silently in
  the paper. In Coq, they aren't silent, but must be marked by a
  constructor. Our implementation minimizes the number of such
  constructors, and makes them unobtrusive, but describing how is
  beyond the scope of the paper.} 

\subsection{Revisiting Spinlocks}

\newcommand{\lcA}{1}
\newcommand{\lcB}{2}
\newcommand{\lcC}{3}
\newcommand{\lcD}{4}
\newcommand{\lcE}{5}
\newcommand{\lcF}{6}
\newcommand{\lcG}{7}
\newcommand{\lcH}{8}
\newcommand{\lcI}{9}
\newcommand{\lcJ}{10}
\newcommand{\lcK}{11}
\newcommand{\lcL}{12}
\newcommand{\lcM}{13}
\newcommand{\lcN}{14}

\newcommand{\unA}{12}
\newcommand{\unB}{13}
\newcommand{\unC}{14}
\newcommand{\unD}{15}

\newcommand*{\tyLock}{\ensuremath{\var{tyLck}}}

\newcommand*{\bndlp}{\var{loop}{\,:\,}\mathsf{unit}{\rightarrow}\tyLock}

\newcommand*{\ifbspc}{\quad \mathsf{if}~b~\mathsf{then}~~}
\newcommand*{\elsespc}{\quad \mathsf{else}~~}

\begin{figure}
\[
\begin{array}[t]{l}
\begin{array}[t]{r@{~~}l}
\multicolumn{2}{c}{\tyLock{}{\eqdef}{}[h, k]
  {\ldot}\spec{\lambda s\ldot \fapp{\selfhist}{s} = h
    \wedge  k \leq \papp{\lastkey}{\fapp{\totalhist}{s}}}
  \spec{\lambda s\ldot \exists t\ldot \fapp{\selfhist}{s} = h \join t \hmapsto \lockval
    \wedge k < t} @ \Spin}\\[1em]
\scs{\lcA}. &
       ~~ \mathsf{fix}~(\lambda \bndlp\ldot~\lambda\_:\mathsf{unit}\ldot~\\
\scs{\lcB}. & \quad
\spec{\fapp{\selfhist}{s} = h \wedge k \leq \papp{\lastkey}{\fapp{\totalhist}{s}}}\\       
\scs{\lcC}. & \quad
   \sspecopen{\exists s'\, b\ldot
   [\mathsf{trylock\_act}]\ s = \left(s', b\right)\wedge \hbox{}}\\
   & \opensspec{
      \hphantom{\exists s'\, b\ldot~~}\quad \mathsf{if}\ b\ \mathsf{then}\ \exists
     t\ldot\fapp{\selfhist}{s'} = h \join t \hmapsto \lockval \wedge k < t\
     \mathsf{else}\
     \fapp{\selfhist}{s'} = h \wedge k \leq \papp{\lastkey}{\fapp{\totalhist}{s'}}}\\
\scs{\lcD}. & \quad b \leftarrow \atomic{\mathsf{trylock\_act}}; \\
\scs{\lcE}. & \quad
     \spec{\mathsf{if}~b~\mathsf{then}~
     \exists t\ldot \fapp{\selfhist}{s} = h \join t \hmapsto \lockval 
     \wedge k < t\
     \mathsf{else}~ \fapp{\selfhist}{s} = h
     \wedge k \leq\papp{\lastkey}{\fapp{\totalhist}{s}}}\\  
\scs{\lcF}. &
\quad \mathsf{if}~b~\mathsf{then}~
      \spec{\exists t\ldot \fapp{\selfhist}{s} = h \join t \hmapsto \lockval
        \wedge k < t}~~\mathsf{ret}~()~~
      \spec{\exists t\ldot \fapp{\selfhist}{s} = h \join t \hmapsto \lockval
        \wedge k < t}\\
\scs{\lcG}. &
      \quad \mathsf{else}~
      \spec{\fapp{\selfhist}{s} = h \wedge k \leq \papp{\lastkey}{\fapp{\totalhist}{s}}}
      ~~\var{loop}\,()~~\spec{\exists\, t\ldot
        \fapp{\selfhist}{s} = h \join t \hmapsto \lockval \wedge k < t}\\
\scs{\lcH}. & \quad
       \spec{\exists\, t\ldot
         \fapp{\selfhist}{s} = h \join t \hmapsto \lockval \wedge k < t})~()     
\end{array}
\end{array}\vspace{-1em}
\]
\caption{Proof outline (and implementation) for $\lockprog$. Here,
  $\tyLock$ binds the spec given to $\lockprog$ in
  Section~\ref{sec:methspec}.}\label{fig:outlines}
\vspace{-2mm}
\end{figure}

To illustrate the inference rules, the proof outline in
Figure~\ref{fig:outlines} shows the proper implementation of
$\lockprog$ and the proof that $\lockprog$ has the type from
Section~\ref{sec:methspec} (the type is named $\tyLock$ in the
figure).
The program is a loop executing $\mathsf{CAS}$ until it succeeds to
lock. This is as in Section~\ref{sec:methspec}, except there we
informally bracketed $\mathsf{CAS}$ with the ghost code for
manipulating histories, whereas here we explicitly invoke the
$\mathsf{trylock\_act}$ action, which erases to $\mathsf{CAS}$.
The outline uses stable assertions only: for example, the precondition
in $\tyLock$ is stable, as argued in Section~\ref{sec:methspec}. Thus,
we dispense with explicit stabilization of assertions, i.e., applying
$\stab{(-)}$.

Given the \TirName{Fix} rule, in order to show that $\lockprog$ has
the type $\tyLock$, we must first prove that $\tyLock$ is a loop
invariant for $\mathsf{fix}$, i.e., that it holds of the body of
$\lockprog$. Thus, the outline starts with the precondition of
$\tyLock$ in line~{\lcB}, and derives the postcondition of $\tyLock$
in line~{\lcH}.
Line~{\lcC} derives immediately from {\lcB} and the definition of
$\mathsf{trylock\_act}$ in Section~\ref{sec:resdefs},
equation~\eqref{act:trylock}, to expose that $\mathsf{trylock\_act}$
either succeeds to lock adding an $\lockval$ to the \emph{self}
history, of fails to lock keeping the history unchanged.\footnote{In
  the case of failure, we could also derive that the lock was taken at
  the moment $\mathsf{trylock\_act}$ was attempted, i.e.  $\exists
  t\ldot \fapp{\selfhist}{s'} = h \wedge
  \fapp{\fapp{\totalhist}{s'}}{t} = \lockval \wedge k \leq t \leq
  \papp{\lastkey}{\fapp{\totalhist}{s'}}$. However, the rest of the
  proof doesn't require the additional detail.}
Notice that Line~{\lcC} has exactly the form required of a premise for
the $\mathsf{vrf\_act}$ rule, with stabilization elided. Thus, the
$\mathsf{if{-}then{-}else}$ conjunct in Line~{\lcC} is also a
postcondition of $\mathsf{trylock\_act}$, and therefore holds in
line~{\lcE}.
Next we branch on $b$, which corresponds to applying the rule
$\mathsf{vrf\_cond}$. Line~{\lcF} considers the case $b =
\mathsf{true}$, and the postcondition immediately follows by the rule
$\mathsf{vrf\_ret}$ (again, eliding stabilization).
Line~{\lcG} considers the case $b = \mathsf{false}$, and the
postcondition immediately follows as the recursive call to
$\var{loop}$, by assumption, already has the desired type $\tyLock$.
As both branches of the conditional have the same postcondition, the
postcondition propagates to line~{\lcH} to complete the proof.

\newcommand{\heap}{\mathsf{heap}}
\newcommand{\hist}{\mathsf{hist}}
\newcommand{\cons}[2]{#1\,{::}\,#2}
\newcommand{\thesent}{\pi}
\newcommand{\thestack}{\alpha}
\newcommand{\islist}[2]{\mathsf{layout}\,#1\,#2}
\newcommand{\liftx}{\textsc{MorphX}\xspace}
\newcommand{\indexhist}{x}
\newcommand{\jointalpha}{\total{\alpha}}
\newcommand{\morphspinx}{f}

\newcommand{\thepriv}{\chi}
\newcommand{\selfpriv}{\self{\thepriv}}
\newcommand{\otherpriv}{\other{\thepriv}}
\newcommand{\totalpriv}{\total{\thepriv}}

\section{Exclusive locking via morphing and the need for permissions}\label{sec:exclusive}

We next illustrate a more involved application of morphisms and
simulation: how to derive a resource and methods for \emph{exclusive}
locking, \`a la \OCSL~\cite{OHearn:TCS07}, from the resource for spin
locks from Section~\ref{sec:overview}. An exclusive lock protects a
shared heap, satisfying a user-supplied predicate $R$
(\emph{aka.}~resource invariant). Upon successful locking, the shared
heap is transferred to the private ownership of the locking thread,
where it can be modified at will, potentially violating $R$. Before
unlocking, the owning thread must re-establish $R$ in its private
heap, after which, the part of the heap satisfying $R$ is moved back
to the shared status. The idea is captured by the following methods
and specs, which we name $\csllockprog$ and $\cslunlockprog$ to
differentiate from $\lockprog$ and $\unlockprog$ in
Section~\ref{sec:overview}.
\[
\begin{array}[t]{r@{\,}c@{\,}l}
\csllockprog & : & 
\!\!\!\begin{array}[t]{l}
    \spec{\lambda s\ldot \fapp{\selflkv}{s}=\nown \wedge \fapp{\selfhp}{s} = \hempty}\ 
    \spec{\lambda s\ldot \fapp{\selflkv}{s}=\own \wedge \papp{R}{\fapp{\selfhp}{s}}}@\CSL\\
  \end{array}\\
\cslunlockprog & : & \!\!\!\begin{array}[t]{l} \spec{\lambda s\ldot
  \fapp{\selflkv}{s}=\own \wedge \papp{R}{\fapp{\selfhp}{s}}}\ 
  \spec{\lambda s\ldot \fapp{\selflkv}{s}=\nown \wedge \fapp{\selfhp}{s} = \hempty}@\CSL
  \end{array}
\end{array}
\]
Here $\fapp{\selflkv}{s}$ is a ghost of type $O = \{\own,
\nown\}$, signifying whether ``we'' own the lock or not, and
$\fapp{\selfhp}{s}$ is ``our'' private heap. $O$ has PCM structure
with the join defined by $x \join \nown = \nown \join x = x$, so that
$\nown$ is the unit of the operation. We leave $\own \join \own$
undefined, to capture that the locking is exclusive, i.e., the lock
can't be owned by a thread and its environment simultaneously.
Our goal in this section is to \emph{derive} $\csllockprog$ and
$\cslunlockprog$ using morphing and simulations to ``attach'' to
$\lockprog$ and $\unlockprog$ the functionality of transferring the
protected heap between shared and private state.

\newcommand{\spinstatedeff}{%
\!\!\!\!\!\begin{array}[t]{l}
 \textrm{$\fapp{\selfhist}{s}, \fapp{\otherhist}{s} \in \histset$, and $\fapp{\selfpu}{s}, \fapp{\otherpu}{s} \in \mathbb{N}$, and \hbox{}}\\
 \fapp{\selfhist}{s} \orth \fapp{\otherhist}{s} \wedge 
 \papp{\alternate}{\fapp{\totalhist}{s}} \wedge 
  r \neq \mathsf{null}\\
  \hbox{}
\end{array}}

\newcommand{\spinerasuree}{%
\lock \hmapsto \papp{\islocked}{\fapp{\totalhist}{s}}}

\newcommand{\locktrdeff}{%
\!\!\!\begin{array}[t]{l}
\locktr\ s\ s' \eqdef \neg \papp{\islocked}{\fapp{\totalhist}{s}} \wedge \hbox{}\\
\quad \fapp{\selfhist}{s'}=\fapp{\selfhist}{s}\join
        \papp{\fresh}{\fapp{\totalhist}{s}} \hmapsto \lockval \wedge 
\fapp{\selfpu}{s'} = \fapp{\selfpu}{s}+1
\end{array}}

\newcommand{\unlocktrdeff}{%
\!\!\!\begin{array}[t]{l}
\unlocktr\ s\ s' \eqdef \papp{\islocked}{\fapp{\totalhist}{s}} \wedge \fapp{\selfpu}{s}>0 \wedge \hbox{}\\
\quad \fapp{\selfhist}{s'}=\fapp{\selfhist}{s} \join
        \papp{\fresh}{\fapp{\totalhist}{s}}\hmapsto \unlockval \wedge 
\fapp{\selfpu}{s'} = \fapp{\selfpu}{s}-1

\end{array}}

\newcommand{\acqpudef}{%
\!\!\!\begin{array}[t]{l}
\papp{\islocked}{\fapp{\totalhist}{s}}=\own \wedge \fapp{\totalpu}{s}=\nown \wedge 
\fapp{\selfpu}{s'}=\own 
\end{array}}

\newcommand{\relpudef}{%
\!\!\!\begin{array}[t]{l}
\fapp{\selfpu}{s}=\own \wedge \fapp{\selfpu}{s'}=\nown 
\end{array}}

\newcommand{\lastvaldeff}{%
\left\{\!\!\!\begin{array}{l@{\, }l}
 \papp{\thehist}{\fapp{\lastkey}{\thehist}}, & \textrm{if $\fapp{\lastkey}{\thehist} \neq 0$}\\
  \unlockval, & \textrm{otherwise}
\end{array}\right.}

\newcommand{\spinblurbb}{%
\!\!\!\begin{array}[t]{l}
\textrm{\textbf{State space $\papp{\Sigma}{\Spin}$: $s \in \papp{\Sigma}{\Spin}$}\ \textrm{iff}\ }\\
\quad \spinstatedeff\\
\textrm{\textbf{Erasure:}} \quad \hflat{s} \eqdef \spinerasuree\\
\textrm{\textbf{Transitions $\TransSet(\Spin)$:}}\\
\locktrdeff\\
\unlocktrdeff
\end{array}}

\newcommand{\xfercomps}{%
\!\!\!\begin{array}[t]{l}
\papp{A}{\Xfer} = \{\nown,\auth,\own\}\times\heaptp, \papp{T}{\Xfer} = \heaptp\\
s = ((\fapp{\selfmu}{s},\fapp{\selfheap}{s}), \fapp{\jointheap}{s}, (\fapp{\othermu}{s}, 
\fapp{\otherheap}{s}))
\end{array}}

\newcommand{\xferstatedeff}{%
\!\!\!\begin{array}[t]{l}
\fapp{\selfmu}{s}, \fapp{\othermu}{s} \in O, \textrm{and}\ 
\fapp{\selfheap}{s}, \fapp{\otherheap}{s}, \fapp{\jointheap}{s} \in \heapset, \textrm{and}\hbox{}\\
\fapp{\selfmu}{s} \orth \fapp{\othermu}{s} \wedge \fapp{\selfheap}{s}\orth\fapp{\jointheap}{s}\orth\fapp{\otherheap}{s} \wedge \hbox{}\\
\mathsf{if}\ \fapp{\totalmu}{s} = \own\ \mathsf{then}\ \fapp{\jointheap}{s}=\hempty\ 
\mathsf{else}\ \papp{R}{\fapp{\jointheap}{s}}
\end{array}}

\newcommand{\xfererasuree}{%
\fapp{\totalheap}{s} \join \fapp{\jointheap}{s}}

\newcommand{\opentrdef}{%
\!\!\!\begin{array}[t]{l}
\opentr\ s\ s' \eqdef \fapp{\selfmu}{s} = \nown \wedge \fapp{\selfmu}{s'}=\own \wedge \hbox{}\\
\quad \fapp{\selfheap}{s'}=\fapp{\selfheap}{s}\join\fapp{\jointheap}{s} \wedge \fapp{\jointheap}{s'}=\hempty
\end{array}}

\newcommand{\closetrdef}{%
\!\!\!\begin{array}[t]{l}
\closetr\ s\ s' \eqdef \fapp{\selfmu}{s}=\own \wedge \fapp{\selfmu}{s'} = \nown \wedge \hbox{}\\
\quad \fapp{\selfheap}{s}=\fapp{\selfheap}{s'}\join\fapp{\jointheap}{s'} \wedge \papp{R}{\fapp{\jointheap}{s'}} 
\end{array}}

\newcommand{\xferblurbb}{%
\!\!\!\begin{array}[t]{l}
\textrm{\textbf{State space $\papp{\Sigma}{\Xfer}$: $s \in \papp{\Sigma}{\Xfer}$}\ \textrm{iff}\ }\\
\quad \xferstatedeff\\
\textrm{\textbf{Erasure:}} \quad\hflat{s} \eqdef \xfererasuree\\
\textrm{\textbf{Transitions $\papp{\TransSet}{\Xfer}$:}}\\
\opentrdef\\
\closetrdef
\end{array}}

\newcommand{\spinfigg}{%
\begin{tikzpicture}[scale=0.55, transform shape, shorten >= 1pt, node distance=4cm,>=latex']
  \tikzstyle{state}=[circle,draw=black,fill=gray!20,thick, 
                     inner sep=0pt, minimum size=18mm]
  \tikzstyle{dummy}=[circle,inner sep=0pt,minimum size=10mm]
  \tikzset{every loop/.style={min distance=25mm,looseness=12}}
  \node[state] (spin) at (0, 0) {\huge$\Spin$};
  \node[state] (xfer) at (10, 0) {\huge$\Xfer$};
  \path[->] (spin) edge [above, in=30, out=60, loop, thick] node {\huge$\locktr$} ()
                   edge [below, in=-60, out=-30, loop, thick] node {\huge$\unlocktr$} ()
            (xfer) edge [above, in=120, out=150, loop, thick] node {\huge$\opentr$} ()
                   edge [below, in=-150, out=-120, loop, thick] node {\huge$\closetr$} ();
  \node[dummy] (b) at (5, 1.8) {\huge$\couple$};
  \node[dummy] (c) at (5, -1.8) {\huge$\couple$};
\end{tikzpicture}}

\begin{figure}[t]
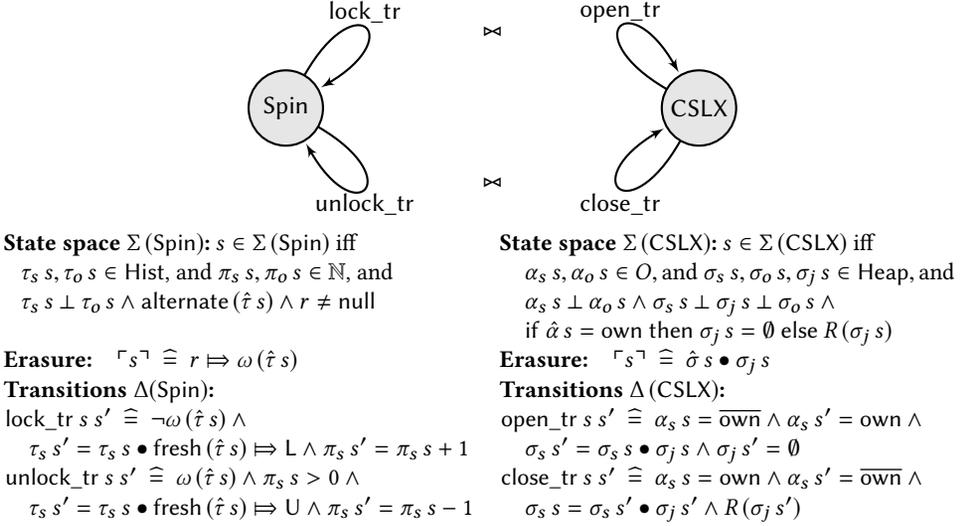

\[
\!\!\!\begin{array}{c}
\spinfigg\vspace{-2mm}\\
\begin{array}{c@{}c}
\small{\spinblurbb}
& \small{\xferblurbb}
\end{array}
\end{array}\vspace{-3.1mm}
\]
\caption{Redefinition of $\Spin$, and $\Xfer$ resource for heap
  transfer in exclusive locking (in our implementation, $\Spin$
  contains external transitions for receiving and giving away
  permissions to unlock, and $\Xfer$ contains transitions for reading
  and writing pointers in $\selfheap$; we elide both for
  simplicity).}\label{fig:csl}
\vspace{-1mm}
\end{figure}

The idea for doing so is pictorially shown in Figure~\ref{fig:csl},
where the $\Xfer$ resource contains the state components and
transitions describing the functionality needed for heap transfers. In
particular, $\fapp{\selfmu}{s} \in O$ keeps track of whether we own
the lock or not, $\fapp{\selfheap}$ and $\fapp{\jointheap}$ are the
private and shared heap respectively, and the transitions $\opentr$
and $\closetr$ move the heap from shared to private and back,
respectively.
We want to combine the $\Spin$ and $\Xfer$ resources as shown in the
figure, by combining their state spaces, and coupling $\opentr$ with
$\locktr$, and $\closetr$ with $\unlocktr$, so that the transitions
execute simultaneously. This will give us an intermediate resource
$\CSL'$, and a morphism $\morphspinx : \Spin \rightarrow \CSL'$
defined similarly to the morphisms in Section~\ref{sec:overview}:
\[
\begin{array}{l@{\ }c@{\ }l}
\morphspinx_\Sigma\ s & \eqdef & \sfst{s}\\[1ex]
\morphspinx_\Delta\ s\ \locktr & \eqdef & \locktr \couple \opentr\\
\morphspinx_\Delta\ s\ \unlocktr & \eqdef & \unlocktr \couple \closetr
\end{array} 
\]
We will then restrict $\CSL'$ into the $\CSL$ resource that we used in
the specs for $\csllockprog$ and $\cslunlockprog$, as we shall
describe. The components $\selflkv$ and $\selfhp$ used in these specs
will be functions out of the state components of $\CSL$.

However, if we try to carry out the above construction using the
$\Spin$ resource from Section~\ref{sec:overview}, we run into the
following problem. 
Recall that $\Spin$ can execute $\unlocktr$ whenever the lock is
taken, irrespective of which thread took it. On the other hand,
$\closetr$ can execute only if ``we'' hold the lock. But,
$\morphspinx_\Delta\ s\ \unlocktr = \unlocktr \couple \closetr$, and
therefore, in states where others hold the lock, $\Spin$ may
transition by $\unlocktr$, with $\CSL'$ unable to follow by
$\morphspinx_\Delta$. Moreover, it's impossible to avoid such
situations by choosing a specific $\morphspinx$-simulation $I$ that
will allow $\unlocktr$ to execute only if we hold the lock. Simply,
there is no way to define such $I$ because we can't differentiate in
$\Spin$ between the notions of $\unlocktr$ being ``enabled for us'',
vs. ``enabled for others, but not for us'', as $\unlocktr$ is enabled
whenever the lock is taken.

The analysis implies that we should have defined $\Spin$ in a more
general way, as shown in Figure~\ref{fig:csl}. In particular, $\Spin$
should contain the integer components $\fapp{\selfpu}/\fapp{\otherpu}$
which indicate if $\unlocktr$ is ``enabled for us''
($\fapp{\selfpu}{s} > 0$), or not ($\fapp{\selfpu}{s} = 0$), and
dually for others. These will give us the distinction we seek, as we
shall see.  In line with related work, we call $\thepu$
\emph{permission to unlock}.\footnote{%
  In general, the design of resource's permissions obviously and
  essentially influences how that resource composes with others. Some
  systems, such as CAP~\cite{DinsdaleYoung-al:ECOOP10} and
  iCAP~\cite{Svendsen-Birkedal:ESOP14}, although they don't consider
  morphisms and simulations, by default provide a permission for each
  transition of a resource. In our example, that would correspond to
  also having a permission for $\locktr$. Full generality also
  requires external transitions that move permissions to and from an
  outside resource. In our Coq code, these are used in the
  readers-writers example, to support non-exclusive locking. For
  simplicity, we elide such generality here, and consider only the
  permission to unlock, which suffices to illustrate morphisms and
  simulations.}\textsuperscript{,}\footnote{Similar concepts arise in
  other concurrency models as well. For example, a transition in a
  Petri net fires only if there are sufficient tokens---akin to
  permissions---in its input places. The tokens are consumed upon
  firing.}

A thread may have more than one permission to unlock, which it can
distribute among its children upon forking, who can then race to
unlock.
The addition of the new fields leads to the following minimal
modification of the specs from Section~\ref{sec:overview}, to indicate
that $\lockprog$ enables $\unlocktr$, and a successful $\unlockprog$
consumes one permission. Note that the specs don't assume that having
a permission to unlock implies that it was ``us'' who last locked, or
even that the lock is taken. We will impose such locking-protocol
specific properties on $\CSL$, but there is no need for them in
$\Spin$.\footnote{Following Section~\ref{sec:subjective-framing}, we
  make the specs small wrt.~$\fapp{\selfpu}{s}$, for simplicity. By
  framing, $\lockprog'$ can be invoked when $\fapp{\selfpu}{s} \geq
  0$, in which case it increments $\fapp{\selfpu}{s}$ by $1$.}
\[
\!\!\!\begin{array}[t]{r@{\,}c@{\,}l}
\lockprog' & : &
   [h,k]\ldot\!\!\!\begin{array}[t]{l}
     \spec{\lambda s\ldot \fapp{\selfhist}{s} = h \wedge k \leq \papp{\lastkey}{\fapp{\totalhist}{s}} \wedge \fapp{\selfpu}{s}=0}\\ 
     \spec{\lambda s\ldot \exists t\ldot \fapp{\selfhist}{s} = h \join t \hmapsto \lockval \wedge k < t \wedge \fapp{\selfpu}{s}=1} @ \Spin
          \end{array} 
\\
\unlockprog' & : & 
 [h,k]\ldot\!\!\!\begin{array}[t]{l}
  \spec{\lambda s\ldot \fapp{\selfhist}{s} = h \wedge k \leq \papp{\lastkey}{\fapp{\totalhist}{s}} \wedge \fapp{\selfpu}{s}=1}\\
  \sspecopen{\lambda s\ldot \exists t\ldot \fapp{\selfhist}{s} = h \join t \hmapsto \unlockval \wedge k < t \wedge \fapp{\selfpu}{s}=0 \vee \hbox{}}\\
  \opensspec{\hphantom{\lambda s\ldot\exists t\ldot}\hspace{-6mm}
             \fapp{\selfhist}{s} = h \wedge \fapp{\fapp{\totalhist}{s}}{\,t} = \unlockval \wedge k \leq t \wedge \fapp{\selfpu}{s}=1} @ \Spin
  \end{array}
\end{array}
\]
Let us now consider the combination $\CSL'$ of $\Spin$ and $\Xfer$ as
defined in Figure~\ref{fig:csl}. The combination has a
number of state components with overlapping roles. For example,
$\theauth$ from $\Xfer$ keeps the status of the lock, and is needed in
$\Xfer$ in order to describe the heap-transfer functionality
independently of $\Spin$.
On the other hand, $\Spin$ keeps the locking
histories in $\thehist$. Thus, once $\Spin$ and $\Xfer$ are combined,
the two components must satisfy
\vspace{-1mm}
\begin{align}
\papp{\islocked}{\fapp{\selfhist}{s}} & = (\fapp{\selfmu}{s} = \own) \label{eq:Invcond1} \\[-.1cm] 
\papp{\islocked}{\fapp{\otherhist}{s}} & = (\fapp{\othermu}{s} = \own)\nonumber
\end{align}
\vspace{-1mm}
as a basic coherence property.
Furthermore, we want to encode exclusive locking, so we must require
that only the thread that holds the lock has the permission to unlock:
\begin{align}
\fapp{\selfpu}{s} & = (\mathsf{if}\ \fapp{\selfmu}{s} = \own\ \mathsf{then}\ 1\ \mathsf{else}\ 
0) \label{eq:Invcond2} \\[-.1cm]
\fapp{\otherpu}{s} & = (\mathsf{if}\ \fapp{\othermu}{s} = \own\ \mathsf{then}\ 1\ \mathsf{else}\ 
0)\nonumber
\end{align} 
(thus, $\fapp{\selfpu}{s}, \fapp{\otherpu}{s} \in \{0, 1\}$, and at
most one of them is $1$).  

Most importantly, the events recorded in the histories of $\Spin$
should correspond to exclusive locking, and thus:
\begin{equation}
\fapp{\selfhist}{s} \orth_\islocked \fapp{\otherhist}{s}\label{eq:Invcond3}
\end{equation}
where $h \orth_\islocked k$ is defined as \label{sep_J1} 
\[
\begin{array}{l}
  (\fapp{\islocked}{h} \rightarrow \fapp{\lastkey}{k} < \fapp{\lastkey}{h}) \wedge \hbox{}\\
  (\fapp{\islocked}{k} \rightarrow \fapp{\lastkey}{h} < \fapp{\lastkey}{k}) \wedge h \orth k
  \end{array}
\]
to say that if $h$ (resp.~$k$) indicates that a thread holds the lock,
then another thread couldn't have proceeded to add logs to its own
history $k$ (resp.~$h$), and unlock itself.\footnote{Requirement~(\ref{eq:Invcond3})
  restricts only the \emph{last} timestamp in $h$ and $k$, not all
  timestamps hereditarily. This suffices for our proof.}

\newcommand{\Jone}{\ensuremath{Inv}}

It is now easy to see that $\Jone = \eqref{eq:Invcond1} \wedge
\eqref{eq:Invcond2} \wedge \eqref{eq:Invcond3}$ is an invariant of
$\CSL'$. The critical point is that (\ref{eq:Invcond3}) is preserved
by the transition $t = \unlocktr \couple \closetr$. Indeed, if in
state $s \in \Jone$ we transition by $t$, it must be
$\fapp{\selfpu}{s} > 0$ by $t$'s definition, and thus
$\fapp{\selfpu}{s}=1$, and $\fapp{\otherpu}{s}=0$, by
(\ref{eq:Invcond2}). Also, we add a fresh $\unlockval$ entry to the
ending state $s'$, thus making $\papp{\lastkey}{\fapp{\selfhist}{s'}}
> \papp{\lastkey}{\fapp{\otherhist}{s'}}$. For (\ref{eq:Invcond3}) to
be preserved, it must then be $\papp{\islocked}{\fapp{\otherhist}{s'}}
= \papp{\islocked}{\fapp{\otherhist}{s}}=\mathsf{false}$,
i.e., the lock \emph{wasn't} held by another thread. But this is
guaranteed by (\ref{eq:Invcond1}), (\ref{eq:Invcond2}) and
$\fapp{\otherpu}{s}=0$.
In other words, by using the permissions to unlock, we have
precisely achieved the distinction that our previous definition of
$\Spin$ couldn't make.

Because $\Jone$ is invariant, we can construct a resource $\CSL$ out
of $\CSL'$, where $\Jone$ is imposed as an additional property of the
underlying PCM and state space of $\CSL'$. Indeed, our theory ensures
that the set $\sigmaof{\CSL'} \cap \Jone$ can be made a global
predicate, and thus be used as a state space of a new resource $\CSL$.
By Definition~\ref{def:global}, globality depends on the underlying
PCM, hence the construction involves restricting the PCM of $\CSL'$ by
$\Jone$. The mathematical underpinnings of such restrictions involve
developing the notions of \emph{sub-PCMs}, \emph{PCM morphisms} and
\emph{\separateness relations}, which we carry out
in~\acite{sec:pcm-morph}{\apndxPCM}{extended}. Here, it suffices to
say that the construction leads to the situation summarized by the
following diagram:
\makeatletter
\newcommand\diag[2][]{\ext@arrow 0099{\longrightarrowfill@}{#1}{#2}}
\makeatother
\[
\Spin \diag{\ \morphspinx}{} \CSL' \diag{\ \iota}{} \CSL
\]
where morphism $\iota$\label{iota_morph} is defined by $\iota_\Sigma\ s = s$ and
$\iota_\Delta\ s\ t = t$. Intuitively, $\CSL$ states are a subset of
$\CSL'$ states satisfying $\Jone$, and $\iota_\Sigma$ is the injection
from $\sigmaof{\CSL}$ to $\sigmaof{\CSL'}$.

\newcommand{\Jtwo}{\ensuremath{Sim}}

This gives us the $\CSL$ resource, but we still need to transform
$\lockprog'$/$\unlockprog'$ into $\csllockprog$/$\cslunlockprog$,
respectively. We thus introduce the following property on $\CSL$
states:
\[
\Jtwo\ s \eqdef \mathsf{if}\ \fapp{\selfmu}{s} = \own\ \mathsf{then}\ \papp{R}{\fapp{\selfheap}{s}}\ \mathsf{else}\ \fapp{\selfheap}{s}=\hempty
\]
which says that the \emph{self} heap satisfies the resource invariant
$R$ iff the thread owns the lock.
$\Jtwo$, unlike $\Jone$, is not an invariant, because it is
perfectly possible for a thread to own the lock, but for its heap to
not satisfy $R$, because the thread has modified the acquired heap
after locking it. However, $\Jtwo$ \emph{is} an $(\iota \circ
\morphspinx)$-simulation, as it satisfies the commuting diagrams from
Figure~\ref{fig:sim}. For example, when $\Spin$ executes $\locktr$,
then $\CSL$ sets $\fapp{\selfmu}{s} = \own$ and acquires the shared
heap, thus making the self heap satisfy $R$. When $\Spin$ executes
$\unlocktr$, then $\CSL$ returns the shared heap, making the self heap
empty. In other words, $\Jtwo$ describes the state of $\CSL$
immediately after locking, and immediately before unlocking, which
suffices for the morphing of $\lockprog'$ and $\unlockprog'$. We only
show the derivation for $\cslunlockprog = \morph{(\iota \circ
  \morphspinx)}{\unlockprog'}$, and refer to the Coq
code~\cite{artifact} for the derivation of $\csllockprog$, which is
similar.
\[
\!\!\!\begin{array}{r@{\ \ }l}
\scs 1. & \spec{\fapp{\selfmu}{s} = \own \wedge \papp{R}{\fapp{\selfheap}{s}}}\\
\scs 2. & \spec{\fapp{\selfhist}{s} = h \wedge \fapp{\selfpu}{s} = 1 \wedge \fapp{\selfmu}{s} = \own \wedge \papp{R}{\fapp{\selfheap}{s}}}\\
\scs 3. & \spec{\fapp{\selfhist}{s} = h \wedge k = \papp{\lastkey}{\fapp{\totalhist}{s}} \wedge \fapp{\selfpu}{s} = 1 \wedge \fapp{\Jtwo}{s}}\\
\scs 4. & \morph{(\iota \circ f)}{\unlockprog'} \quad \textrm{// using simulation $\Jtwo$}\\
\scs 5. & \sspecopen{(\fapp{\selfhist}{s} = h \join t \hmapsto \unlockval \wedge k < t \wedge \fapp{\selfpu}{s} = 0 \vee \hbox{}}\\
        & \opensspec{\hphantom{(}\fapp{\selfhist}{s} = h \wedge k \leq t \wedge \fapp{\fapp{\totalhist}{s}}{\,t} = \unlockval \wedge \fapp{\selfpu}{s} = 1) \wedge \fapp{\Jtwo}{s}}\\
\scs 6. & \spec{\fapp{\selfhist}{s} = h \join t \hmapsto \unlockval \wedge k < t \wedge \fapp{\selfpu}{s} = 0 \wedge \fapp{\Jtwo}{s}}\\
\scs 7. & \spec{\fapp{\selfmu}{s} = \nown \wedge \fapp{\selfheap}{s} = \hempty}
\end{array}
\]
The key step is in line 6, where we must derive that the second
disjunct in line 5 is false; that is, no thread could have unlocked
before us in line 4. We infer this by reasoning about the histories
$\fapp{\selfhist}{s}$ and $\fapp{\otherhist}{s}$ in line 5. From
$\fapp{\selfpu}{s} = 1$, it must be $\fapp{\selfmu}{s} = \own$, and
then $\papp{\islocked}{\fapp{\selfhist}{s}} = \mathsf{true}$, by the
invariant $\Jone$ which holds throughout, as $s$ is a $\CSL$ state.
By $\Jone$ again, $\fapp{\selfhist}{s} \orth_\islocked
\fapp{\otherhist}{s}$, so $\papp{\lastkey}{\fapp{\otherhist}{s}} <
\papp{\lastkey}{\fapp{\selfhist}{s}}$. Thus, it must be
$\papp{\lastkey}{\fapp{\selfhist}{s}} =
\papp{\lastkey}{\fapp{\totalhist}{s}} = k$, because
$\papp{\lastkey}{\fapp{\totalhist}{s}}$ is the maximum of
$\papp{\lastkey}{\fapp{\selfhist}{s}}$ and
$\papp{\lastkey}{\fapp{\otherhist}{s}}$. But this contradicts that
$\fapp{\totalhist}{s}$ contains entry $\unlockval$ at $t \geq k$.  We
can now derive line 7: $\fapp{\selfmu}{s} = \nown$ follows from
$\fapp{\selfpu}{s}=0$ and $\Jone$, and $\fapp{\selfheap}{s} = 0$
follows by $\Jtwo$. Finally, we obtain the desired spec of
$\cslunlockprog$ from the beginning of the section, by letting
$\themutex$ be $\theauth$ and $\thehp$ be $\theheap$.


\newcommand{\xmorph}[3]{\ensuremath{\cmorph\ #1\ #2\ #3}}
\newcommand{\morphq}{g}

\section{Quiescence and indexed morphism families}\label{sec:param}

The previous examples were about extending $\Spin$ the functionality
of another resource, $\Cnt$ or $\Xfer$. In this section, we apply
resource morphism not to extend a resource, but to restrict it,
specifically by ``forgetting'' its ghost state. This is a feature
commonly required when installing one resource into a private state of
another.
We need a slight generalization, however, to \emph{indexed morphism
  families} (or just \emph{families}, for short), as follows. 

A family $f : \morphtpX{V}{W}{X}$ introduces a type $X$ of indices for
$f$. The state component $f_\Sigma :
X\,{\rightarrow}\,\papp{\Sigma}{W} \rightharpoonup \papp{\Sigma}{V}$
and the transition component $f_\Delta :
X\,{\rightarrow}\,\sigmaof{V}\,{\rightarrow}\,\transof{V}\,{\rightharpoonup}\,\transof{W}$
now allow input $X$, and satisfy a number of properties, listed
in~\acite{sec:famdef}{\apndxMorph}{extended}, that reduce to
Definition~\ref{def:morph} when $X$ is the $\tyUnit$ type. Similarly,
$f$-simulations must be indexed too, to be predicates over $X$ and
$\sigmaof{W}$, satisfying a number of properties which reduce to
Definition~\ref{def:sim} when $X=\tyUnit$.

The $\mathsf{morph}$ constructor and its rule are generalized
to receive the initial index $x$, and postulate the existence of
an ending index $y$ in the postcondition, as follows:
\begin{mathpar}
\inferrule*[Right = MorphX]{e : \spec{P}\,\spec{Q}@V}
    {\xmorph{f}{x}{e} : \spec{\lambda s_w\ldot \xsmorph {f}{x}{P}\ s_w \wedge I\ x\ s_w}~
     \spec{\lambda s_w\ldot \exists y\ldot \xsmorph{f}{y}{Q}\ s_w \wedge I\ y\ s_w}@W}
\and
{\textrm{where}\,
 \begin{array}{l}
   {\xsmorph f x R}~{s_w}~{\eqdef}~\exists\ s_v\ldot\ s_v~{=}~
    \fapp{\fapp{f_\Sigma}{x}}{s_w} \wedge \fapp{R}{s_v}
     \end{array}}
\end{mathpar}


\newcommand{\stkstate}{	
	\begin{tikzpicture}[scale=1, transform shape]
	\tikzset{block/.style={shape=rectangle, draw, node distance=0pt, minimum width =.4cm, minimum 
	height=.4cm}}
	\filldraw[fill=gray!70, draw=black] (135:1cm) -- (-45:1cm) arc (-45:135:1cm); 
	\filldraw[fill=gray!20, draw=black] (135:1cm) arc (135:315:1cm); 
	\draw[draw=black] (0,0) circle (0.7cm);
	\filldraw[fill=white, draw=black] (0,0) circle (0.35cm);
	\node[scale=0.41] at (180:0.85cm) {\Huge$\selfhist$};
	\node[scale=0.41] at (180:-0.85cm) {\Huge$\otherhist$};
	\node[scale=0.41] at (70:-0.52cm) {\Huge$\selfheap$}; 
	\node[scale=0.41] at (70:0.52cm) {\Huge$\otherheap$};
	\node[scale=0.41] at (0, 0cm) {\Huge$\jointheap$};
	\node[scale=1, anchor=east] at (140:1.2cm) {\fontsize{7}{7}$\Stack$:}; 
	\node[scale=1, anchor=west] at (1.1,0.1) {\fontsize{7}{7}{\begin{array}[t]{l}
                        \quad \papp{\dom}{\fapp{\totalhist}{s}} = \{1,\ldots,\papp{\lastkey}{\fapp{\totalhist}{s}}\}\\
                        \quad \islist{(\fapp{\jointalpha}{s})}{(\fapp{\jointheap}{s})}
                        \end{array}}};
\end{tikzpicture}
}

\begin{figure}
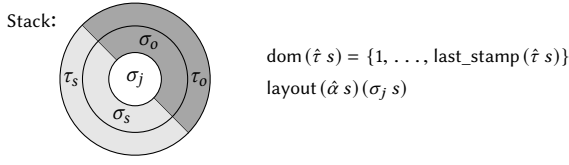

\[\stkstate
\]
\caption{Representation of the state components of the $\Stack$
  resource. The \emph{self} components ($\selfheap$ for the heap, and
  $\selfhist$ for the history) are in light shade, the \emph{other}
  components are in dark; the \emph{joint} component ($\jointheap$ for
  the heap storing the stack's physical layout) is white. The
  abbreviation $\fapp{\jointalpha}{s}$ is the abstract value of the
  stack (computed out of $\fapp{\totalhist}{s}$).}\label{fig:stack}
\end{figure}

To illustrate, consider the resource $\Stack$ (Figure~\ref{fig:stack})
implementing concurrent stacks, and the following spec for the stack's
$\mathsf{push}$ method, similar to that of $\lockprog$ from
Section~\ref{sec:overview}.
\[
\!\!\!\begin{array}{c}
\act{push}(v) : [k]\ldot 
  \!\!\!\begin{array}[t]{l} 
        \sspec{\lambda s\ldot \fapp{\selfheap}{s} = \hempty \wedge \fapp{\selfhist}{s} = \hempty \wedge 
        k \leq \papp{\lastkey}{\fapp{\otherhist}{s}}}\\
        \sspec{\lambda s\ldot \fapp{\selfheap}{s} = \hempty \wedge \exists t\ vs\ldot \fapp{\selfhist}{s} = t \hmapsto (vs, v\,{::}\,vs) \wedge k < t}@\Stack
 \end{array}\\
\end{array}
\]
Here $\selfhist$, and $\otherhist$ are histories of stack's
operations, as in the case of $\Spin$, and $\selfheap$ is the
thread-private heap.  The spec says that $\mathsf{push}$ starts with
$\fapp{\selfhist}{s} = \hempty$ (by framing, any history) and ends
with $\fapp{\selfhist}{s} = t \hmapsto (vs, v::vs)$ to indicate that a
push of $v$ indeed occurred, and \emph{after} all the timestamps from
the pre-state. The joint heap $\jointheap$ stores the stack's physical
layout, and $\fapp{\jointalpha}{s}$ is the abstract contents of the
stack as a mathematical sequence (computed out of
$\fapp{\totalhist}{s}$). Intuitively, $\mathsf{push}$ first allocates
a new node in $\selfheap$, then moves it to $\jointheap$ where it is
enlinked to the top of the laid-out stack, after which $\mathsf{push}$
updates $\selfhist$ and $\jointalpha$ to reflect the addition of the
node. We also elide the full definition of $\Stack$ as it isn't
essential here; it suffices to know that predicate $\islist$ describes
how $\jointalpha$ is laid out in $\jointheap$ (i.e., $\forall s \in
\papp{\Sigma}{\Stack}\ldot
\islist{(\fapp{\jointalpha}{s})}{(\fapp{\jointheap}{s})}$), and that
the global history $\fapp{\totalhist}{s}$ has no timestamp gaps (i.e.,
$\forall s \in
\papp{\Sigma}{\Stack}\ldot\papp{\dom}{\fapp{\totalhist}{s}} =
\{1,\ldots,\papp{\lastkey}{\fapp{\totalhist}{s}}\}$).\footnote{We
  mechanized this development in Coq for the Treiber variant of
  stacks, with some minor Treiber-specific modifications.}

\newcommand{\aval}{a}
\newcommand{\bval}{b}

Consider now the program $e = \mathsf{push}(\aval) \parallel
\mathsf{push}(\bval)$ of the following type (also derived in Coq):
\[
\!\!\!e : \!\!\!\begin{array}[t]{l}
   \spec{\lambda s\ldot \fapp{\selfheap}{s} = \hempty \wedge \fapp{\selfhist}{s} = \hempty}\\ 
   \sspec{\lambda s\ldot \fapp{\selfheap}{s} = \hempty \wedge \exists t_1\ {vs_1}\ t_2\ {vs_2}\ldot 
    \fapp{\selfhist}{s} = t_1 \hmapsto (vs_1, \aval\,{::}\,{vs_1}) \join 
                                     t_2 \hmapsto (vs_2, \bval\,{::}\,{vs_2})}@\Stack
   \end{array}
\]
The specification reflects that $e$ pushes $\aval$ and $\bval$, to
change the stack contents from $vs_1$ to $\aval\,{::}\,{vs_1}$ at time
$t_1$, and from $vs_2$ to $\bval\,{::}\,{vs_2}$ at time $t_2$. The
order of pushes is unspecified, so we don't know if $t_1 < t_2$ or
$t_2 < t_1$ (as $\join$ is commutative, the order of $t_1$ and $t_2$
in the binding to $\fapp{\selfhist}{s}$ in the post doesn't imply an
ordering between $t_1$ and $t_2$).  Moreover, we don't know that
$t_1$ and $t_2$ occurred in immediate succession (i.e., $t_2 = t_1 + 1
\vee t_1 = t_2 + 1$), as threads concurrent with $e$ could have
executed between $t_1$ and $t_2$, changing the stack
arbitrarily. Thus, we also can't infer that the ending state of $t_1$
equals the beginning state of $t_2$, or vice versa.

But what if we knew that $e$ is invoked without interfering threads,
i.e., \emph{quiescently}~\cite{Aspnes-al:JACM94,Derrick-al:TOPLAS11,
  Jagadeesan-Riely:ICALP14,Nanevski-al:ESOP14, sergey:oopsla16}?
For example, imagine a resource $\Priv$ with only heaps $\selfpriv$
and $\otherpriv$, and no other components
(Figure~\ref{fig:privmorph}(1)), and transitions that allow modifying
the self heap by reading, writing, CAS-ing, or executing any other
read-modify-write command~\cite{Herlihy-Shavit:08}.
A program working over $\Priv$ can install an empty stack in
$\selfpriv$ and then invoke $e$ over it. Because the stack is
installed \emph{privately}, no threads other than the two children of
$e$ can race on it. Could we exploit quiescence, and derive \emph{just
  out of the specification} of $e$ that the stack at the end stores
either the list $[\aval, \bval]$, or $[\bval, \aval]$? This fact can
be stated even without histories, using solely heaps, as follows:
\begin{equation}
\spec{\lambda s\ldot \islist {\mathsf{nil}} {(\fapp{\selfheap}{s})}}
\sspec{\lambda s\ldot \islist {[\aval, \bval]} {(\fapp{\selfpriv}{s})} \vee 
\islist {[\bval, \aval]} {(\fapp{\selfpriv}{s})}}@\Priv\label{eq:layout}
\end{equation}
The move from $\Stack$ to $\Priv$ thus essentially \emph{forgets} the
ghost state of histories, and the distinction in $\Stack$ between
shared and private heaps. These components and distinctions are
visible when in the scope of $\Stack$, but hidden when in
$\Priv$.
We would like to obtain the spec (\ref{eq:layout}) by applying the $\TirName{Morph}$
rule to the $\Stack$ spec of $e$, with a morphism $\morphq :
\morphtp{\Stack}{\Priv}$ that forgets the histories. Unfortunately,
such a morphism can't be constructed as-is. Were it to exist, then
$\morphq_\Sigma$, being contravariant, should map a state $s_\Priv$,
containing only heaps, to a state $s_\Stack$, containing heaps
\emph{and histories}; thus $\morphq_\Sigma$ must ``invent'' the
history component out of thin air.

\newcommand{\prvstate}{	
	\begin{tikzpicture}[scale=1, transform shape]
	\filldraw[fill=gray!70] (135:1cm) -- (-45:1cm) arc (-45:135:1cm); 
	\filldraw[pattern=bricks, pattern color=gray, draw=black] (135:1cm) -- (-45:1cm) arc (-45:135:1cm); 
	\filldraw[fill=gray!20] (135:1cm) arc (135:315:1cm); 
	\filldraw[pattern=vertical lines, pattern color=gray, draw=black] (135:1cm) arc (135:315:1cm); 
	\node[scale=0.41] at (50:-0.54cm) {\Huge$\chi_s$}; 
	\node[scale=0.41] at (50:0.54cm) {\Huge$\chi_o$};
	\node[scale=1, anchor=east] at (140:1.2cm) {\fontsize{7}{7}$\Priv$:}; 
	\end{tikzpicture}
}

\newcommand{\combstate}{	
	\begin{tikzpicture}[scale=1, transform shape]
	\filldraw[fill=gray!70, draw=black] (135:1cm) -- (-45:1cm) arc (-45:135:1cm);
	\filldraw[fill=gray!70, draw=black] (135:0.70cm) -- (-45:0.70cm) arc (-45:135:0.70cm);
	\fill[pattern=bricks,pattern color=gray] (135:0.70cm) -- (-45:0.70cm) arc (-45:135:0.70cm);
	\filldraw[fill=gray!20, draw=black] (135:1cm) arc (135:315:1cm); 
	\filldraw[pattern=vertical lines, pattern color=gray!90] (135:0.70cm) arc (135:315:0.70cm); 
	\filldraw[fill=white, draw=black] (0,0) circle (0.35cm);
	\filldraw[pattern=vertical lines, pattern color=gray] (0,0) circle (0.35cm);
	\node[scale=0.41] at (180:0.85cm) {\Huge$\indexhist$};
	\node[scale=0.41] at (180:-0.85cm) {\Huge$\emptyset$};
	\node[scale=0.41] at (70:-0.52cm) {\Huge$\selfheap$}; 
	\node[scale=0.41] at (70:0.52cm) {\Huge$\otherheap$};
	\node[scale=0.41] at (0,0) {\Huge$\jointheap$};
	\node[scale=1, anchor=east] at (140:1.2cm) {\fontsize{7}{7}$\morphq_\Sigma~\indexhist~s_\Priv$:}; 
	\end{tikzpicture}
}

\begin{figure}
	\[
	\begin{array}{c@{\hspace{1cm}}c}
	 \prvstate &\combstate\\
	 {\fontsize{9}{9}\textrm{\hspace{.55cm}(1)}} & {\fontsize{9}{9}\textrm{\hspace{1.2cm}(2)}}
	\end{array}
	\]\vspace{-3mm}
	\caption{The left figure shows the state components of
          $\Priv$. The right figure shows how $\Stack$ from
          Figure~\ref{fig:stack} is related to $\Priv$ by
          $\morphq_\Sigma$. Striped and bricked regions show the
          $\Priv$'s \emph{self} heap $\selfhp = \selfheap \join
          \jointheap$ and \emph{other} heap $\otherhp = \otherheap$,
          respectively. Solid regions show the $\Stack$'s \emph{self}
          history $\selfhist = x$ and \emph{other} history $\otherhist
          = \hempty$.}\label{fig:privmorph}
\end{figure}

This is where families come in. We make $\morphq :
\morphtpX{\Stack}{\Priv}{\histset}$ a family over $X = \histset$,
thereby passing to $\morphq_\Sigma$ the history $\indexhist$ that
should be added to $s_\Priv$ to produce an $s_\Stack$
(Figure~\ref{fig:privmorph}(2)).
\[
\begin{array}{r@{\,}c@{\,}l}
\morphq_\Sigma~\indexhist~s_\Priv = s_\Stack & \eqdef &
\papp{\selfpriv}{s_\Priv} =
  \papp{\selfheap}{s_\Stack} \join \papp{\jointheap}{s_\Stack} \wedge \hbox{}\\
  & & \papp{\otherpriv}{s_\Priv} = \papp{\otherheap}{s_\Stack} \wedge \hbox{}\\
  & & \papp{\selfhist}{s_\Stack} = \indexhist \wedge \papp{\otherhist}{s_\Stack} =
  \hempty
\end{array}
\]
The first conjunct directly states that $\Stack$ is installed in
$\papp{\selfpriv}{s_\Priv}$ by making $\papp{\selfpriv}{s_\Priv}$ be
the join of the heaps $\papp{\jointheap}{s_\Stack}$ and
$\papp{\selfheap}{s_\Stack}$.\footnote{As we want to build $s_\Stack$
  out of $s_\Priv$, we have to identify the part of
  $\papp{\selfpriv}{s_\Priv}$ which we want to assign to
  $\papp{\jointheap}{s_\Stack}$. This part has to be uniquely
  determined, else $\morphq_\Sigma$ won't be a function. We ensure
  uniqueness by insisting that the predicate $\islist$ is precise -- a
  property commonly required in separation logics.}
The second conjunct says that the heap $\papp{\otherpriv}{s_\Priv}$ of
the interfering threads is propagated to
$\papp{\otherheap}{s_\Stack}$. The third conjunct captures that the
history component of $s_\Stack$ is set to the index $\indexhist$, as
discussed immediately above. In the last conjunct, the
$\papp{\otherhist}{s_\Stack}$ history is declared $\hempty$, thus
formalizing quiescence. We elide the definition of $\morphq_\Delta$;
it suffices to know that it maps a $\Stack$ transition (relation over
heap and ghost state of $\Stack$) to its ``erasure'', i.e. a relation
over heaps of $\Priv$ representing a single-pointer operation such as
read, write and $\mathsf{CAS}$, but ignoring the ghost histories.

We can now obtain the quiescent spec (\ref{eq:layout}) by applying the
$\TirName{MorphX}$ rule to $e$, as shown below, with $x = \hempty$,
and $I~x$ being the always-true predicate on $\Priv$ states (the
outline expands the definition of ${\xsmorph \morphq x \_}$, and
elides the always-true $I~x$).
\[
\begin{array}{r@{\ \ }l}
\scs 1. & \spec{\islist {\mathsf{nil}} {(\papp{\selfpriv}{s_\Priv})}}\\
\scs 2. & \spec{\exists s_\Stack\ldot 
   s_\Stack = \fapp{\fapp{\morphq_\Sigma}{\hempty}}{s_\Priv} \wedge \papp{\selfheap}{s_\Stack} = \hempty \wedge \papp{\selfhist}{s_\Stack} = \hempty}\\  
\scs 3. & \xmorph \morphq \hempty e\\
\scs 4. & \sspecopen{\exists y\ s_\Stack\ldot s_\Stack = \fapp{\fapp{\morphq_\Sigma}{y}}{s_\Priv} \wedge \papp{\selfheap}{s_\Stack} = \hempty \wedge \hbox{}} \\
        & \opensspec{\hphantom{\exists y\ s_\Stack\ldot} \exists t_1\ vs_1\ t_2\ vs_2\ldot \papp{\selfhist}{s_\Stack} = t_1 \hmapsto (vs_1, \aval\,{::}\,{vs_1}) \join 
   t_2 \hmapsto (vs_2, \bval\,{::}\,{vs_2})} \\ 
\scs 5. & \spec{\islist {[\aval, \bval]} {(\papp{\selfpriv}{s_\Priv})} \vee \islist {[\bval, \aval]} {(\papp{\selfpriv}{s_\Priv})}}
\end{array}
\]
Line 2 derives trivially from line 1, as the state $s_\Stack$ is
uniquely determined; i.e., take $\papp{\jointheap}{s_\Stack}$ to be
the heap storing the empty stack, and all the other components of
$s_\Stack$ to be empty.
We next derive the postcondition in line 5.
From the properties of $\papp{\Sigma}{\Stack}$, we know
$\islist{(\papp{\jointalpha}{s_\Stack})}{(\papp{\jointheap}{s_\Stack})}$.
From $s_\Stack = \fapp{\fapp{\morphq_\Sigma}{y}}{s_\Priv}$ and other
conjuncts in line 4, we know $\papp{\selfpriv}{s_\Priv} =
\papp{\jointheap}{s_\Stack}$, and
$\papp{\selfhist}{s_\Stack}\,{=}\,y$, and $\papp{\otherhist}{s_\Stack}
= \hempty$.
Thus, it is also $\papp{\totalhist}{s_\Stack} = y\,{=}\,t_1 \hmapsto
(vs_1,\aval\,{::}\,{vs_1}) \join t_2 \hmapsto (vs_2,
\bval\,{::}\,{vs_2})$, and:
\begin{equation}
\islist{(\papp{\jointalpha}{s_\Stack})}{(\papp{\selfpriv}{s_\Priv})}\label{eq:layout2}
\end{equation}
From the properties of $\papp{\Sigma}{\Stack}$, we also know that
$\papp{\totalhist}{s_\Stack}$ has no timestamp gaps; thus $\{t_1,
t_2\} = \papp{\dom}{y} = \{1, 2\}$, i.e., $t_1$ and $t_2$ are the
only, and consecutive, events in $y$. But then
$\papp{\jointalpha}{s_\Stack}$ must be either $[\aval, \bval]$ or
$[\bval, \aval]$, which, with (\ref{eq:layout2}), derives the postcondition.
\section{Related work}\label{sec:related}
\quad\emph{Coalgebra morphisms and refinement mappings.\ }  
Transition systems are mathematically representable as
coalgebras~\cite{RUTTEN20003,jacobs_2016}; thus, resource morphisms
are closely related to coalgebra morphisms, with differences arising
from our application to concurrent separation logic and types. For
example, for us, given $f : V \rightarrow W$, $f_\Sigma$ is
contravariant, but in the coalgebraic case, $f$ is covariant on
states~\cite{RUTTEN20003,hasuo09}.
A coalgebra morphism $f$ doesn't have the $f_\Delta$ component, but
requires that $f$ preserve and reflect the $V$-transitions (i.e., if
$x\rightarrow y$ is a $V$-transition, then $f(x)\rightarrow f(y)$ is a
$W$-transition, and if $f(x)\rightarrow t$ is a $W$-transition, then
there exists $y$ such that $x\rightarrow y$ is a $V$-transition, and
$f(y)=t$). These properties are similar in spirit to the two clauses
of our Definition~\ref{def:sim} of $f$-simulations.

Similarly, refinement mappings~\cite{aba+lam:91,LynchV+IC95}, like
coalgebra morphisms, are covariant functions on STSs, and differ from
resource morphisms in the intended use.
For example, with refinement mapping, the key question is how to
extend the source STS with ghost state. The extension may be
necessary, as a refinement mapping need not exist otherwise.
In contrast, we seek to give a definitive resource type to a program,
which suffices for all reasoning. The program can be morphed, to
change the type, but can't be re-typed by an extended resource, as
that violates compositionality in our setting.
Given a morphism $f : V \rightarrow W$, the resource $W$ will most
commonly generalize and include $V$'s functionality. Thus, $f_\Sigma$
can compute $s_v \in \papp{\Sigma}{V}$ out of $s_w \in
\papp{\Sigma}{W}$, without needing to extend $V$ or $W$. When $s_w$
lacks information to compute $s_v$ (\cf the quiescence example in
Section~\ref{sec:param}), we don't extend the resources, but pass the
missing information by an index in a morphism family. Morphisms also
exhibit a form of (weak) simulation of $W$ by $V$ on the transposed
states (Definition~\ref{def:sim}(2)).

We further establish the action of $f$ on programs, and provide a
Hoare logic rule to reason about it, supporting the usual
compositionality notions from separation logic, such as framing and
ownership transfer, which haven't been considered in the context of
coalgebra morphisms and refinement mappings.

\vspace{1mm}

\emph{Linearizability.\ } In a relational flavor of
separation
logics~\cite{Liang-al:POPL12,Turon-al:ICFP13,fru+kre+bir:lics18}, and
more generally, in the work on proving
linearizability~\cite{HenzingerSV+CONCUR13,Liang-Feng:PLDI13,
  SchellhornWD+CAV12,BouajjaniEEM+cav17,gu:popl15,
  gu:pldi18,khy+dod+got+par:esop17},
the goal is to explicitly relate two \emph{programs}, typically one
concurrent, the other sequential. The sequential program then serves
as a spec for the concurrent one, and can replace it in any larger
context.
Our goal in this paper is somewhat different;
we seek to identify the concurrent program's \emph{type}, which for us
takes the form of a Hoare triple enriched with a resource. The type
serves as the program's interface, and, as standard in type theory,
any two programs with the same type can be interchanged in clients'
code and proofs. As clients can already reason about the program via
this type, the program \emph{shouldn't need} a spec in the form of
another program.

There are several advantages of our approach over
linearizability. First, a spec in the form of a Hoare triple with a
resource is much simpler, and thus easier to establish than a spec in
the form of another program. Indeed, resources \emph{aren't programs};
they are STSs and don't admit programming constructs such as
conditionals, loops, initial or local state, or function calls. A
Hoare spec is also immediately useful in proofs, whereas with
linearizability, one also has to verify the sequential program
itself. 
Second, in linearizability it has traditionally been difficult to
address ownership transfer of heaps between data
structures~\cite{Gotsman-Yang:CONCUR12,Cerone-al:ICALP14}, whereas for
us (and other extensions of
\OCSL~\cite{DinsdaleYoung-al:ECOOP10,Liang-Feng:PLDI13,
  ArrozPincho-al:ECOOP14,Jung-al:POPL15,jung:jfp18}), ownership
transfer is directly inherited from separation logic.
We also inherit from separation logic a way to dynamically nest
parallel compositions of threads, whereas linearizability is typically
considered on programs with a fixed, though arbitrary, number of
threads.

That said, we note that our specs of $\lockprog$ and $\unlockprog$ in
Section~\ref{sec:overview} are actually very close to what one gets
from linearizability, as they essentially establish a linear order
between locking and unlocking events in the history PCM. In fact, our
approach seems sufficiently powerful to \emph{directly specify}, in
the state space of a resource, the general property of linearizability
as a user-level proposition over the resource's subjective histories,
which is the direction we intend to pursue in the future. Such a
development would generalize the current paper, in that the events
tracked by histories wouldn't be instantaneous, as was the case in
Section~\ref{sec:overview}, but would have non-zero duration. The
histories would have to record the events' beginning as well as 
ending times.

Morphisms and simulations will play a key role in such a setting. For
example, two different resources will have two different spaces of
timestamps (i.e., two different clocks). Combining the two into a
larger resource, will require constructing a history for the
combination, of which the first step is reconciling the clocks of the
two components into a common clock. Morphisms and simulations will be
a necessary abstraction to relate the common clock, timestamps, and
history to the clocks, timestamps, and histories of the
components.

In future work, we thus expect to incorporate general linearizable 
programs and resources; note, however, that the PCM and history-based 
approach is further general still, and capable of compositionally 
specifying and verifying \emph{non-linearizable} programs as 
well~\cite{sergey:oopsla16}.

\quad\emph{State transition systems and abstract atomicity.\ }
Abstract atomicity refers to extending the functionality of a
concurrent program so that it operates over other STSs.
One of the most powerful approaches to abstract atomicity has been
parametrization by auxiliary code. For example, in the case of spin
locks, a way to make $\lockprog$ operate over an extension of $\Spin$
with an unknown STS $X$, is to parametrize $\lockprog$ \emph{a priori}
with an unknown ghost function $\alpha$ over $X$, thus modifying it
into:
\[
\lockprog\ \alpha \eqdef \mathsf{do}\ \!\!\!\begin{array}[t]{l}
  \langle x \leftarrow \mathsf{CAS}(r, \mathsf{false}, \mathsf{true}); 
  \ghostcolor{\mathsf{if}\ x\ \mathsf{then}\ \alpha} \rangle\ \mathsf{while}\ \neg x \end{array}
\]
This differs from our approach, where the modification to $\lockprog$
is done \emph{a posteriori}, and is mediated by the resource type of
the program. Parametrization originated in~\citet{Jacobs-Piessens:POPL11},
and was extended to impredicative
higher-order ghost functions and state in
HOCAP~\cite{Svendsen-al:ESOP13}, iCAP~\cite{Svendsen-Birkedal:ESOP14}
and \Iris~\cite{Jung-al:POPL15,jung:jfp18}).
Parametrization affords abstract specifications that are similar to
specifications that one would ascribe to a data structure in the
sequential setting. It applies to the locking examples, as illustrated
above, but also to many of the examples that we included in the Coq
code. Parametrization relies on higher-order and impredicative ghost
functions to specify a number of concurrency idioms, such as e.g.,
fork/join concurrency, or storing of concurrent programs into the
heap. In contrast, our present
model is predicative~(see~\acite{sec:model}{\apndxModel}{extended});
hence we currently support the more restrictive concurrency by
parallel composition.

Thus our approach doesn't presently extend the range of verifiable
programs. Rather, it proposes novel type-based foundations that
underpin the verification, and that employ morphisms, histories, and
simulations. Histories and simulations are foundational abstractions
in concurrency. Morphisms are similarly so in mathematics, and in our
case, relate to coalgebras and refinement mappings, as
discussed. It's therefore of inherent interest to embed these abstractions
into type theory and obtain a minimalistic proof system for separation logic,
as we have done.

By relying on these foundational abstractions, we achieve some
uniformity and simplicity of reasoning, which we expect to build on in
the future. For example, one challenge to parametrization arises when
the point at which to execute the ghost function can be determined
only after the program has terminated. This is a common pattern when
proving linearizability, as exhibited, say, by the queue
of~\citet{Herlihy-Wing:TOPLAS90}, but has been difficult to address by
parametrization, because it isn't clear at which point in the code to
invoke the parametrizing ghost function.
In our case, histories separate the termination of operations from
their order in the linearization. The order becomes just another ghost
component that can be constructed at run time, with cooperation of
other threads~\cite{del+ser+nan+ban:ecoop17}. A history-based spec can
say that an operation finished executing, but that its exact place in
the linearization is to be fully determined only later, by the action
of other threads.  We thus expect that history-based specs will
support the described pattern of linearizability proofs.  

The \TaDA logic of \citet{ArrozPincho-al:ECOOP14} introduces another
approach to abstract atomicity that doesn't rely on
parametrization. $\TaDA$ defines a new judgment form, $\langle
P\rangle\,e\,\langle Q\rangle$, that captures that $e$ has a
precondition $P$ and postcondition $Q$, but is also abstractly atomic
in the following sense: $e$ and its concurrent environment maintain
the validity of $P$, until at one point $e$ takes an atomic step that
makes $Q$ hold. Afterwards, $Q$ may be invalidated, either by future
steps of $e$, or by the environment. Once judged atomic, programs can
be associated with ghost code of other resources. In this sense,
$\TaDA$'s extension of code is a posteriori, similar to ours.
In contrast to \TaDA, we specify programs using ordinary Hoare triples, but
rely on the PCM of histories to express atomicity: a program is
atomic if it adds a single entry to the self history.
In the Coq files, we have also applied morphisms to algorithms with
helping, such as the flat combiner~\cite{Hendler-al:SPAA10}, where one
thread executes work on behalf of others; helping is an idiom that
\TaDA currently can't express.

\citet{ser+wil+tat:popl18} have designed a logic \Disel for
distributed systems, in which one can combine distributed
protocols---represented as STSs---by means of \emph{hooks}. A hook on
a transition $t$ prevents $t$ from execution, unless the condition $P$
associated with the hook is satisfied. In this sense, hooks implement
an instance of our transition coupling, where one operand is fixed to
the idle transition with a condition $P$, i.e.~$\mathsf{id\_tr}~P =
\lambda s\ s'\ldot P\ s \wedge s' = s$. \Disel doesn't currently
consider hooks where both operands are non-idle, which we used in the
lock examples, or notions of morphism and simulation. On the other
hand, we haven't considered distribution so far.

Finally, while morphisms and simulations provide a solution to
abstract atomicity, they also go beyond it. For example, they apply to
quiescence (Section~\ref{sec:param}), which, unlike abstract
atomicity, doesn't extend functions to resource combinations, but
forgets the histories of a resource. They may also provide a
foundation for answering more basic, categorical, questions about
concurrent structures, such as e.g., ``when are two resources
isomorphic'' (answer: when there are cancelling morphisms between
them). We plan to explore such questions in the future.

\vspace{1mm}

\emph{Automated separation logics for concurrency.\ } A number of
recent automated tools such as Verifast~\cite{Jacobs-al:NFM11},
VerCors~\cite{blo+hui:fm14,ami+hui+blo:metrid18}, and
Viper~\cite{mue+sch+sum:vmcai16}, address the reasoning about
concurrent programs in various extensions of separation logic. In
general, the tools address fragments of Java or C, and completely or
partially automate the discharge of the proof obligations.
Our paper is accompanied with an implementation in Coq as a shallow
embedding. Hence, Coq plays a dual role for us: it's a framework for
mechanizing proofs, but also a concurrent programming language. The
high-level difference from the automated tools is, or course, that
proofs are developed interactively. The scaling of the proving effort
is achieved by the reuse of programs and proofs, enabled by the
compositional nature of the underlying type theory. We haven't
explored automation yet, but the minimalistic nature of our setting
suggests that the underlying abstractions will be useful for both
interactive and automated reasoning.

\vspace{1mm}

\emph{\FCSL.\ }
The current paper adds to \FCSL~\cite{Nanevski-al:ESOP14} the novel
notions of resource morphism, and significantly modifies the notion of
resources. 
In \FCSL, each concurrent resource is a finite map from labels
(natural numbers) to sub-components. For example, using the concepts
from Section~\ref{sec:overview}, one could represent $\SC$ as a finite
map $l_1 \hmapsto \Spin~{\uplus}~l_2 \hmapsto \Cnt$, where $l_1$ and
$l_2$ are labels identifying $\Spin$ and $\Cnt$, respectively. This
approach provides interesting equations on resources; for example, one
can freely rearrange the finite map components by using commutativity
and associativity of disjoint union $\uplus$. However, it also complicates mechanized
proofs, because one frequently, and tediously, needs to show that a
label is in the domain of a map, before extracting the labeled
component. In the current work, states aren't maps, but triples which
are combined by a form of pairing (e.g., the PCM of $\SC$ is a product
of PCMs of $\Spin$ and $\Cnt$). Consequently, if we changed the
definition of $\SC$ in Section~\ref{sec:overview} into $\SC'$ by
commuting $\Spin$ and $\Cnt$ throughout the construction, then $\SC$
and $\SC'$ wouldn't be \emph{equal} resources, but they will be
\emph{isomorphic}, in that we could exhibit cancelling morphisms
between the two. But this requires first having a notion of morphism,
which is one of the technical contributions of this paper. \FCSL
supported quiescence by means of a dedicated and very complex
inference rule, whereas
Section~\ref{sec:param} demonstrates that quiescence is merely an
application of morphism families.

\section{Conclusions}\label{sec:conclusions}
This paper develops novel notions of resource morphisms and associated
simulations, as key mathematical concepts that underpin a separation
logic for fine-grained concurrency. This is a natural development, as
structures in mathematics are always associated with an appropriate
notion of morphism, and simulations are the invariants that the
morphisms preserve.
Morphisms and simulations act on programs, and are integrated into
separation logic via a single inference rule that propagates the
simulation from the precondition to the postcondition of the morphed
program.

Morphisms compose and can support different constructions and
applications. One application is abstract atomicity, whereby a general
spec, such as the one for $\lockprog$ in Section~\ref{sec:overview},
is specialized to a specific ownership discipline, e.g., exclusive
locking in Section~\ref{sec:exclusive}.
Other applications include the managing of scope of ghost state in
quiescent environments, as illustrated in Section~\ref{sec:param}, and
restricting the state space of a resource with additional state and
PCM invariants, as used in Section~\ref{sec:exclusive}.

\section*{Acknowledgments}
We thank Jes\'{u}s Dom\'{i}nguez, Constantin Enea, Franti\v{s}ek
Farka, Joakim \"{O}hman, Exequiel Rivas Gadda, Mihaela Sighireanu, Ana
Sokolova, Gordon Stewart, Anton Trunov and Nikita Zyuzin for their
comments on the various drafts of the paper. We thank the anonymous
reviewers from OOPSLA'19 PC and AEC for their feedback. This research
was partially supported by the Spanish MICINN projects BOSCO
(PGC2018-102210-B-I00) and TRACES (TIN2015-67522-C3-3-R), the European
Research Council projects Mathador (ERC2016-COG-724464) and FOVEDIS
(ERC2015-STG-678177), and the US National Science Foundation
(NSF). Any opinions, findings, and conclusions or recommendations
expressed in the material are those of the authors and do not
necessarily reflect the views of the funding agencies.

\bibliography{bibmacros,references,proceedings}

\ifappendix

\appendix

\newcommand{\fst}{\ensuremath{\mathsf{inl}}}
\newcommand{\snd}{\ensuremath{\mathsf{inr}}}
\newcommand{\lt}{\phi_L}
\newcommand{\rt}{\phi_R}

\section{PCM products, transition couplings and resource combinations}\label{appendix:tensor}

In this section we develop the combination of two resources (the
\emph{combined resource} or \emph{tensor resource}) that has been used
in Sections~\ref{sec:overview} and~\ref{sec:exclusive}. For that, we
need first to give some basic definitions for the product of PCMs and
the coupling of transitions.

\begin{lemma}
Let $A$ and $B$ be PCMs. Then, the Cartesian product $A \times B$ is a
PCM under the point-wise lifting of the join operators, $(a_1, b_1)
\join_{A\times B} (a_2, b_2) = (a_1 \join_A a_2, b_1 \join_B b_2)$,
and units, $\pcmU_{A\times B} = (\pcmU_A, \pcmU_B)$.
\end{lemma}

For example, the PCM for $\SC$ in Figure~\ref{fig:spin} is the product
of the PCMs $\papp{\pcmA}{\Spin}$ and $\papp{\pcmA}{\Cnt}$, that is,
$\papp{\pcmA}{\SC}=\papp{\pcmA}{\Spin}\times\papp{\pcmA}{\Xfer}$.

\begin{definition}\label{def:stprod}
  Let $s_i=(\selfa^i, \jointa^i, \othera^i)$ be an $(M_i, T_i)$-state,
  $i=1,2$, and $s=(\selfa, \jointa, \othera)$ be an
  $(M_1\times M_2, T_1 \times T_2)$-state. The pair state of $s_1$ and
  $s_2$ is
  $\spair{s_1}{s_2} \eqdef ((\selfa^1, \selfa^2), (\jointa^1,
  \jointa^2), (\othera^1, \othera^2))$, and projection states of $s$
  are
  $\sproj{i}{s} \eqdef (\sproj{i}{\selfa}, \sproj{i}{\jointa},
  \sproj{i}{\othera})$, $i=1,2$.
  The usual beta and eta laws for products hold, i.e.:
  $\sproj{i}{\spair{s_1}{s_2}} = s_i$ and
  $s = \spair{\sfst s}{\ssnd s}$.
\end{definition}

\begin{definition}\label{def:products}
Given state spaces $\Sigma_1$ and $\Sigma_2$, the \dt{product state
  space $\Sigma_1 \times \Sigma_2$} is defined by:
%
\[
\begin{array}{rcl}
\hflat{s} & \eqdef & \hflat{\sfst{s}}^{\Sigma_1} \join \hflat{\ssnd{s}}^{\Sigma_2}\\
s \in \Sigma_1 \times \Sigma_2 & \mbox{iff} & \sfst{s} \in \Sigma_1 \wedge \ssnd{s} \in \Sigma_2 \wedge \fapp{\mathsf{defined}}{\hflat{s}}.
\end{array}
\]
\end{definition}

\begin{definition}\label{def:coupling}
Let $t_i$ be a $\Sigma_i$-transition, $i=1,2$. The \dt{coupling of
  $t_1$ and $t_2$} is the $\Sigma_1\times\Sigma_2$-transition defined
as
$(t_1 \couple t_2)\ s\ s' \eqdef t_1\ (\sfst{s})\ (\sfst{s'}) \wedge
t_2\ (\ssnd{s})\ (\ssnd{s'})$
\footnote{In this paper we only consider footprint preserving
  transitions. In Coq code we consider more general notions of
  transitions which need not be footprint preserving. For that case,
  we somewhat modify the definition of coupling to require an
  additional conjunct that $\hflat{s'}$ is a defined heap, similar to
  the definition of product state spaces.}.
\end{definition}

The coupling of transitions is used extensively in our constructions,
e.g.~in $\locktr \couple \opentr$ in Section~\ref{sec:overview}. As we
have already explained there, the formalization in
Definition~\ref{def:coupling} entails that the coupling of two
transitions execute their components simultaneously: $\locktr$ from
$\Spin$ and $\opentr$ from $\Cnt$ in the aforementioned example.


Now we can formalize the notion of combined resource, which the next
definition makes precise.

\begin{definition}\label{def:tensor}
Let $V$, $W$ be resources, $I \subseteq
\TransSet(V)\times\TransSet(W)$ a relation between transitions, such
that for every $(t_v, t_w) \in I$ the transition $t_v \couple t_w$ is
footprint-preserving.  The \dt{combined} (alt.~\dt{tensor}) resource
$V \tensor_I W$ has the state space $\papp{\Sigma}{V\tensor_I W} =
\papp{\Sigma}{V}\times \papp{\Sigma}{W}$ and transitions:
\[
\begin{array}{r@{\,}c@{\,}l}
\transof{V \tensor_I W} & = & \{t_v \couple t_w \mid (t_v, t_w) \in I\}
\end{array}
\]
\end{definition}

For example, resource $\SC$ is the combination of resources $\Spin$
and $\Cnt$, with $I$ defined as:
\[
\begin{array}{r@{\,}c@{\,}l}
I & = & \{\!\!\!\begin{array}[t]{l} (\locktr, \opentr),\ (\unlocktr,
\closetr),\ (\idtr, \idtr)\}
\end{array}
\end{array}
\]
Here, $\mathsf{id}$ is the idle transition defined as
$\lambda s\ s'\ldot s' = s$. We have implicitly assumed that all
resources in this paper possess an idle transition, which is why we
elided this transition from all the figures.

\newcommand{\pval}{\iota}
\newcommand{\psub}{\rho}
\newcommand{\gensep}{\mathbin{R}}
\newcommand{\globalsub}[2]{#1/#2}
\newcommand{\subjectivesub}[2]{\subpcm{#1}{#2}}
\newcommand{\subpcm}[2]{{#1}\,{\!/\!}\,{#2}}

\section{Sub-PCMs, \separateness relations, and PCM morphisms}\label{sec:pcm-morph}

In Section~\ref{sec:exclusive}, we implicitly used a construction to
restrict the state space of the resource $\CSL'$ with an invariant
$\Jone$, to obtain the desired resource $\CSL$. In this section, we
formally develop that construction, which hinges on the concept of
\emph{sub-PCMs}. A sub-PCM is to a PCM, what subset is to a set. It is
customary in abstract algebra to develop the notion of a sub-object of
an algebraic object by first developing the appropriate notion of
morphism on the objects; in our case, this will be PCM morphisms. The
morphisms will model the \emph{injection} of a sub-PCM into the PCM
(similar to how one can inject a subset into a larger set), and a
\emph{retraction} from the larger PCM into the sub-PCM.

\paragraph*{\Separateness relation.}
We start with the auxiliary notion of a \emph{\separateness relation},
which enables us to delineate the elements of the larger PCM that we
want to include into the sub-PCM.
Interestingly, \separateness relations are \emph{binary} relations,
rather than simple unary predicates over the domain of the larger PCM.
This is so, because the sub-PCM construction requires not only
selecting the elements out of the larger PCM, but also restricting its
join operation, i.e., make it ever more partial than it already is. As
the join operation is itself binary, so must be the \separateness
relation. One may thus think of the \separateness relation as being an
abstract algebraic generalization of the idea of \emph{disjointness}
(and indeed, disjointness of histories and heaps are instances of
\separateness relations on the PCMs of histories and heaps,
respectively).

\begin{definition}\label{def:sep}
	Given a PCM $A$, relation $\gensep$ on $A$ is a
        \dt{\separateness relation} if it is:
	\begin{enumerate}
		\item \emph{(unitary)} $\pcmU_A \gensep \pcmU_A$
		\item \emph{(commutative)} $x \gensep y$ iff $y \gensep x$
		\item \emph{(compatible)} if $x \gensep y$ then $x \orth y$
		\item \emph{(unital)} if $x \gensep y$ then $x \gensep \pcmU_A$
		\item \emph{(associative)} if $x \gensep y$ and $(x \join y) \gensep z$ then $x \gensep (y 
		\join 
		z)$ 
		and $y \gensep z$
	\end{enumerate}
\end{definition}

For example, the relation $\orth_\islocked$ defined in
Section~\ref{sec:exclusive} in clause~\eqref{eq:Invcond3} on
page~\pageref{sep_J1}, as
\[
x \orth_\islocked y = \begin{array}[t]{l}
  (\fapp{\islocked}{x} \rightarrow \fapp{\lastkey}{y} < \fapp{\lastkey}{x})\, \wedge \\
  (\fapp{\islocked}{y} \rightarrow \fapp{\lastkey}{x} < \fapp{\lastkey}{y}) \wedge x \orth y
  \end{array}
\]
is a \separateness relation. The critical part of the proof is that of
associativity, which we briefly sketch. Let $x$, $y$ and $z$ be three
histories such that $x \orth_\islocked y$ and $(x \join y)
\orth_\islocked z$, and let the timestamps of their last entries be
$t_x$, $t_y$, and $t_z$, respectively. The interesting case is when
$x$ or $y$ end with a locking entry.
Without loss of generality, let $\fapp{\islocked}{x}$ hold, and thus
~$\fapp{\lastval}{x} = \lockval$.
Then, by $x \orth_\islocked y$, the last entry of $y$ must be
$\unlockval$, and $t_y < t_x$. Similarly, by $(x \join y)
\orth_\islocked z$, the last entry of $z$ must be $\unlockval$ and
$t_z < t_x$. But then, trivially, $y \orth_\islocked z$, because both
$y$ and $z$ end with an $\unlockval$ entry. Similarly, $x
\orth_\islocked (y \join z)$, because $y \join z$ also ends with an
$\unlockval$ entry, and its last timestamp is $\mathsf{max}(t_y, t_z)
\leq t_y, t_z < t_x$.

The clauses~\eqref{eq:Invcond1} and~\eqref{eq:Invcond2} in
Section~\ref{sec:exclusive} also arise as instances of \separateness
relations, in a similar way.
Thus, ultimately, the invariant $\Jone = \eqref{eq:Invcond1} \wedge
\eqref{eq:Invcond2} \wedge \eqref{eq:Invcond3}$ can be defined as a
\separateness relation on the PCM $\aof{\CSL'}$.

\paragraph*{PCM morphisms.} Now we can define morphisms between PCMs
as follows:

\begin{definition}\label{def:pcmmorph}
	A \dt{PCM morphism} $\phi : A \rightarrow B$ with a \separateness relation $\orth_\phi$ is a 
	partial function from $A$ to $B$ such that:
	\begin{enumerate}
		\item $\fapp{\phi}{\pcmU_A} = \pcmU_B$
		
		\item if $x \orth_\phi y$, then $\fapp{\phi}{x}$, $\fapp{\phi}{y}$ exist, and $\fapp{\phi}{x} 
		\orth_B \fapp{\phi}{y}$, and $\fapp{\phi}(x \join y) = \fapp{\phi}{x} \join \fapp{\phi}{y}$
	\end{enumerate}
	The morphism $\phi$ is \emph{total} if $\orth_\phi$ equals $\orth_A$.
\end{definition}
In Section~\ref{sec:exclusive} we have silently used many PCM
morphisms, as the special symbols that we used to name the components
such as $\thepu$, $\thehist$, $\lkv$, $\theheap$, etc., are all PCM
morphisms.  For example, $\thehist$ and $\thepu$ in
Figure~\ref{fig:csl} are projections out of the PCM $\aof{\Spin}$ into
the PCM of histories and permissions, respectively (and projections
out of tuples are morphisms). Then, $\fapp{\selfhist}{s}$ and
$\fapp{\otherhist}{s}$ are abbreviations for
$\papp{\thehist}{\fapp{\selfa}{s}}$ and
$\papp{\thehist}{\fapp{\othera}{s}}$ respectively, and similarly for
$\thepu$.


\paragraph*{sub-PCMs.}
We can now define the notion of \emph{sub-PCMs} as follows:

\begin{definition}\label{def:subpcm}
	PCM $A$ is a \dt{sub-PCM} of a PCM $B$ if there exists a total
        PCM morphism $\pval : A \rightarrow B$ (injection) and a
        morphism $\psub : B \rightarrow A$ (retraction), such that:
	\begin{enumerate}
		\item $\papp{\psub}{\fapp{\pval}{a}} = a$
		
		\item if $b \orth_\psub \pcmU_B$ then $\papp{\pval}{\fapp{\psub}{b}} = b$
		
		\item if $(\fapp{\psub}{x}) \orth_A (\fapp{\psub}{y})$ then $x \orth_\psub y$
	\end{enumerate}
\end{definition}
Property (1) says that $\pval$ is injective, i.e., if we coerce $a \in A$ into $\fapp{\pval}{a}$, we 
can recover $a$ back by $\psub$, since no other element of $A$ maps by $\pval$ into 
$\fapp{\pval}{a}$. The dual property (2) allows the same for a subset of $B$'s elements, that
are related by $\orth_\psub$ to $\pcmU_B$. Hence, intuitively, $A$ is in 1-1 correspondence with 
that subset of $B$. Property (3) extends the correspondence to \separateness relations, i.e., 
$\orth_A$, when considered on images under $\psub$, implies (and hence, by properties of 
morphisms, equals) $\orth_\psub$.

Definition~\ref{def:subpcm} says what it means for $A$ to be a sub-PCM
of $B$. The following lemma shows how to construct a sub-PCM of $B$
given a \separateness relation $\gensep$ on $B$.  We used this
construction in Section~\ref{sec:exclusive} to obtain the PCM for the
resource $\CSL$ out of the PCM $\aof{\CSL'}$.

\begin{lemma}
	Given PCM $A$ and \separateness relation $\gensep$ on $A$, the set $\subpcm{A}{\gensep} = 
	\{x \in A \mid x \mathbin{\gensep} \pcmU_A\}$ forms a PCM under unit $\pcmU_A$, and join 
	operation defined as $x \join_{\subpcm{A}{\gensep}} y = x \join_A y$ if $x\mathbin{\gensep}	y$ 
	and undefined otherwise. The PCM $\subpcm{A}{\gensep}$ is a sub-PCM	of $A$ under the 
	injection $\pval$ and retraction $\psub$ defined as	$\forall x \in \subpcm{A}{\gensep}\ldot 
	\papp{\pval}{x} = x$ and $\forall a \in A\ldot \papp{\psub}{a}=a$ if 
	$a\mathbin{\gensep}\pcmU_A$, and $\papp{\psub}{a}$ undefined otherwise. Moreover, 
	$\gensep = \orth_\psub = \orth_{\subpcm{A}{R}}$.
\end{lemma}
Using the above notation, the PCM of the resource $\CSL$ can be
defined formally as $\subpcm{\aof{\CSL'}}{\Jone}$. 

We now have all the ingredients to formalizing the construction for
restricting a resource that we set out to define.

\begin{definition}\label{def:subresource}
	Let $R$ be an invariant \separateness relation on $\aof{V}$. The \dt{sub-resource} 	
	$\subjectivesub{V}{R}$ is defined with the same type, transitions and erasure as $V$, but with 
	the PCM and the state space defined as
	\begin{enumerate}
		\item $\aof{\subjectivesub{V}{R}} = \subpcm{\aof{V}}{R}$
		\item $s \in \papp{\Sigma}{\subjectivesub{V}{R}}$ iff $s \in \papp{\Sigma}{V} \wedge 
		(\fapp{\selfa}{s}) \mathbin{R} (\fapp{\othera}{s})$
	\end{enumerate}
	There is a generic resource morphism $\iota :
        \morphtp{V}{\subjectivesub{V}{R}}$ that is inclusion on states
        and identity on transitions.
\end{definition}

For example, the resource $\CSL$ from Section~\ref{sec:exclusive} is
the sub-resource of $\CSL'$ taken under the invariant $\Jone$, and the
resource morphism $\iota : \CSL' \rightarrow \CSL$ from
Section~\ref{sec:exclusive} (page~\pageref{iota_morph}), is the
generic morphism defined above.

\paragraph*{Algebraic properties.}
We finish this section by providing some additional evidence that
\separateness relations and PCM morphisms compose, and have pleasant
mathematical properties. For example, the operations of morphism
composition and join come with the \separateness relations as follows.
\[
\begin{array}{r@{\ }c@{\ }lr@{\ }c@{\ }l}
\papp{(f \circ g)}{x} & \eqdef & \papp{f}{\fapp{g}{x}} & x \orth_{f \circ g} y & \eqdef & x \orth_g y \wedge \fapp{g}{x}\mathbin{\orth_f}\fapp{g}{y}\\
\papp{(f \join g)}{x} & \eqdef & \fapp{f}{x} \join \fapp{g}{x} & x \orth_{f \join g} y & \eqdef & x \orth_f y \wedge x \orth_g y \wedge 
\fapp{f}{(x \join y)} \orth \fapp{g}{(x \join y)}
\end{array}
\]
Or, given PCM morphisms $f$ and $g$, we can define \separateness
relation that implements the PCM versions of the algebraic notions of
\emph{kernel} (preimages of unit) and \emph{equalizer} (values on
which two morphisms agree), as follows.
\[
\begin{array}{r@{\ }c@{\ }l}
x \mathbin{(\kernel{f})} y & \eqdef & x \orth_f y \wedge \fapp{f}{x}=\fapp{f}{y}=\pcmU\\
x \mathbin{(\eqlz{f}{g})} y & \eqdef & x \orth_f y \wedge x \orth_g y \wedge 
\fapp{f}{x}=\fapp{g}{x} 
\wedge \fapp{f}{y}=\fapp{g}{y}
\end{array}
\]
Importantly, the above are all \separateness relations, as we have
proved in the Coq files. Similarly, we can restrict a morphism to a
\separateness relation $R$, to define another PCM morphism.
\[
\papp{(\restrict{f}{R})}{x} \eqdef 
    \left\{\begin{array}{ll}
        \fapp{f}{x}, & \mbox{if $x \mathbin{R} \pcmU$}\\
        \mbox{undefined}, & \mbox{otherwise}
    \end{array}\right.\quad \mbox{with}\quad
x \orth_{\restrict{f}{R}} y \eqdef x \orth_f y \wedge x \mathbin{R} y
\]
The import of the above abstract constructions is in the reduction of
proof burden. For example, a morphism equalizer is a \separateness
relation by construction, so the user need not bother proving
compatibility for equalizers. The constructions also combine to
concisely state invariants and assertions. For example, the
\separateness relation that gives rise to $\Jone$, and is thus used to
construct the sub-resource $\CSL$, may be defined as the equalizer
$\eqlz{(\thepu \join \theauth)}{(\islocked' \circ \thehist)}$, where
$\islocked' : \hist \rightarrow O$ is the morphism defined on a
history $h$ as $\fapp{\islocked'}{h} =
\mathsf{if}\ \fapp{\islocked}{h}\ \mathsf{then}\ \own\ \mathsf{else}\ \nown$.

\section{Indexed morphism families}\label{sec:famdef}
In this appendix, we show how the definitions of morphism and
$f$-simulations generalize to indexed families. When $X$ is the unit
type, we recover the morphism-related definitions from
Section~\ref{sec:formal}.

\begin{definition}\label{def:pmorph}
An \dt{indexed family of morphisms} $f : \morphtpX{V}{W}{X}$ (or just
\emph{family}), consists of partial functions $f_\Sigma : X
\rightarrow \sigmaof{W} \rightharpoonup \sigmaof{V}$ (note the
contravariance), and $f_\Delta : X \rightarrow \transof{V}
\rightharpoonup \transof{W}$ on transitions, such that:
\begin{enumerate}
\item[(1)]\label{pmorf:frame} (locality of $f_\Sigma$) there exits a
  function $f_A : \aof{W} \rightarrow \aof{V}$ such that if
  $\xmorpheq{s_v}{(s_w \zig p)}{f_\Sigma}{x}$, then $s_v = s'_v \zig
  \papp{f_A}{p}$, and $\xmorpheq{s'_v \zag \papp{f_A}{p}}{(s_w \zag
    p)}{f_\Sigma}{x}$.
\item[(2)]\label{pmorph:loc} (locality of $f_\Delta$) if
  $\papp{\fapp{f_\Delta}{x}}{s_w \zig p}\,{t} = t'$, then
  $\papp{\fapp{f_\Delta}{x}}{s_w \zag p}\,{t} = t'$.
\item[(3)]\label{pmorf:other} (other-fixity) 
if $\papp{\othera}{s_w} = \papp{\othera}{s'_w}$ and
$\fapp{\fapp{f_\Sigma}{x}}{s_w}$, $\fapp{\fapp{f_\Sigma}{x'}}{s'_w}$ exist, then
$\papp{\othera}{\fapp{\fapp{f_\Sigma}{x}}{s_w}}
= \papp{\othera}{\fapp{\fapp{f_\Sigma}{x'}{s'_w}}}$.
\item[(4)]\label{pmorph:backw} (injectivity on indices) if
  $\fapp{\fapp{f_\Sigma}{x_1}}{s_{w_1}}$ and $\fapp{\fapp{f_\Sigma}{x_2}}{s_{w_2}}$
  exist and $\fapp{\fapp{f_\Sigma}{x_1}}{s_{w_1}} =
  \fapp{\fapp{f_\Sigma}{x_2}}{s_{w_2}}$, then $x_1 = x_2$.
\end{enumerate}
\end{definition}

In most of the properties of Definition~\ref{def:pmorph}, the index
$x$ is simply propagated unchanged. The only property where $x$ is
significant is the new property (4), which requires that
$\fapp{\fapp{f_\Sigma}{x}}{s_w}$ uniquely determines the index $x$. In
the $\Stack$ example in Section~\ref{sec:param}, it is easy to see
that the definition of $f_\Sigma$ satisfies this property, because
equal states have equal histories.

\begin{definition}\label{def:psim}
Given a morphism family $f : \morphtpX{V}{W}{X}$, an
\dt{$f$-simulation} is a predicate $I$ on $X$ and $W$-states such
that:
\begin{enumerate}
\item [(1)]\label{pmorf:simWV} if $\fapp{\fapp{I}{x}}{s_w}$, and
  $\xmorpheq{s_v}{s_w}{f_\Sigma}{x}$, and $t\ s_v\ s'_v$, then there
  exist $x'$, $t' = {f_\Delta}\ {x}\ {s_w}\ {t}$, and $s'_w$ such that
  $\fapp{\fapp{I}{x'}}{s'_w}$ and
  $\xmorpheq{s'_v}{s'_w}{f_\Sigma}{x'}$, and $t'\ s_w\ s'_w$.
\item [(2)]\label{pmorf:simVW} if $\fapp{\fapp{I}{x}}{s_w}$, and $s_v =
  \fapp{\fapp{f_\Sigma}{x}}{s_w}$ exists, and $s_w \osteps[W]{} s'_w$,
  then $\fapp{\fapp{I}{x}}{s'_w}$, and $s'_v =
  \fapp{\fapp{f_\Sigma}{x}}{s'_w}$ exists, and $s_v \osteps[V]{}
  s'_v$.
\end{enumerate}
\end{definition}

Compared to Definition~\ref{def:sim}, property (1) allows that $x$
changes into $x'$ by a transition. In the $\Stack$ example in
Section~\ref{sec:param}, if we lift $e$ by using the index $x =
\mathsf{empty}$ (i.e., write $\mathsf{morph}\ \mathsf{empty}\ f\ e$),
then this index will evolve with $e$ taking the transitions of
$\Stack$ to track how $e$ changes the self history by adding the
entries for pushing 1 and 2. In property (2), the index $x$ simply
propagates.

\section{Denotational semantics}
\label{sec:model}

\newcommand{\ty}{~{:}~}
\newcommand{\Ghi}{\widehat{\Phi}}
\newcommand{\entangle}{\rtimes}
\newcommand{\tsteps}[1]{~{\xrightarrow{#1}}~}
\newcommand{\Path}{\pi}
\newcommand{\Paths}{\zeta}


\newcommand{\ChoiceAct}{\mathsf{ChoiceAct}}
\newcommand{\SeqRet}{\mathsf{SeqRet}}
\newcommand{\SeqStep}{\mathsf{SeqStep}}
\newcommand{\ParRet}{\mathsf{ParRet}}
\newcommand{\ParL}{\mathsf{ParL}}
\newcommand{\ParR}{\mathsf{ParR}}
\newcommand{\MorphStep}{\mathsf{MorphStep}}
\newcommand{\MorphRet}{\mathsf{MorphRet}}
\newcommand{\HideStep}{\mathsf{HideStep}}
\newcommand{\HideRet}{\mathsf{HideRet}}
\newcommand{\InjRet}{\mathsf{InjRet}}
\newcommand{\InjStep}{\mathsf{InjStep}}
\newcommand{\SCST}{our system}

\newcommand{\rely}[1]{\mathcal{R}_{#1}}
\newcommand{\guar}[1]{\mathcal{G}_{#1}}

\newcommand{\unwind}{\mathsf{unwind}}
\newcommand{\pathtp}{\mathsf{path}}
\newcommand{\tsafe}{\mathsf{safe}}

\newcommand{\set}[1]{\left\{{#1}\right\}}

Our semantic model largely relies on the denotational semantic of
\emph{action trees}~\cite{LeyWild-Nanevski:POPL13}. A tree implements
a finite partial approximation of one execution of a program. Thus, a
program of type $\mathsf{ST}\ V\ A$ will be denoted by a set of such
trees. The set may be infinite, as some behaviors (i.e., infinite
loops) can only be represented as an infinite set of converging
approximated executions. Because a program may contain multiple
executions, our denotational semantics can represent non-determinism
in the form of internal choice, though our language currently does not
make use of that feature.

An action tree is a generalization of the Brookes' notion of action
trace~\cite{Brookes:TCS07} in the following sense. Where action trace
semantics represent a program by a set of traces, we represent a
program with a set of trees. A tree differs from a trace in that a
trace is a sequence of actions and their results, whereas a tree
contains an action followed by a \emph{continuation} which itself is a
tree parametrized wrt.~the output of the action.


In this appendix, we first define the denotation of each of our
commands as a set of trees. Then we define the semantic behavior for
trees wrt.~resource states, in a form of operational semantics for
trees, thus formalizing single execution of a program. Then we relate
this low-level operational semantics of trees to high-level
transitions of a resource by an $\mathsf{always}$ predicate
(Section~\ref{sec:modal-pred}) that ensures that a tree is resilient
to any amount of interference, and that all the operational steps by a
tree are \emph{safe}.
The $\mathsf{always}$ predicate will be instrumental in defining the
$\vrf$-predicate transformer from Section~\ref{sec:formal}, and from
there, in defining the type of Hoare triples
$\spec{P}\ A\ \spec{Q}@V$.
Both the $\mathsf{ST}\ V\ A$ type and the Hoare triple type will be
\emph{complete lattices} of sets of trees, giving us a suitable
setting for modeling recursion.
The \emph{soundness} of \FCSL follows from showing that the lemmas
about the $\vrf$ predicate transformer listed in
Section~\ref{sec:formal}, are satisfied by the denotations of the
commands.


We choose the Calculus of Inductive Constructions
(\CiC)~\cite{coq-team,Bertot-Casteran:04} as our meta logic.  This has
several important benefits.  First, we can define a \emph{shallow
  embedding} of our system into \CiC that allows us to program and
prove directly \emph{with the semantic objects}, thus immediately
lifting to a full-blown programming language and verification system
with higher-order functions, abstract types, abstract predicates, and
a module system.  We also gain a powerful dependently-typed
$\lambda$-calculus, which we use to formalize all semantic definitions
and meta theory, including the definition of action trees by
\emph{iterated inductive definitions}~\cite{coq-team},
specification-level functions, and programming-level higher-order
procedures.
Finally, we were able to mechanize the entire semantics and meta theory
in the Coq proof assistant implementation of~\CiC.

\subsection{Action trees and program denotations}
\label{sec:opsem}
\begin{definition}[Action trees]\label{def:trees}
The type $\mathsf{tree}~V~A$ of ($A$-returning) \textbf{action trees}
is defined by the following iterated inductive definition.
\[
\small
\begin{array}{r@{\ }c@{\ }l}
\mathsf{tree}~V~A & \eqdef & \mathsf{Unfinished}\\
& \mid & \mathsf{Ret}\ (v \ty A)\\
& \mid & \mathsf{Act}~(a : \mathsf{action}~V~A)\\
& \mid & \mathsf{Seq}~(T \ty \mathsf{tree}~V~B)~(K \ty B \rightarrow \mathsf{tree}~V~A)\\
& \mid & \mathsf{Par}~(T_1 \ty \mathsf{tree}~V~B_1)\ (T_2 \ty \mathsf{tree}~V~B_2)\ \hbox{}\\
&      & \qquad \!(K \ty B_1 \times B_2 \rightarrow \mathsf{tree}~V~A)\\
& \mid & \mathsf{Morph}~(f : \morphtpX{W}{V}{X})~(x:X)~(T \ty \mathsf{tree}~W~A)
\end{array}
\]  
\end{definition}

Most of the constructors in Definition~\ref{def:trees} are
self-explanatory.
Since trees have finite depth, they can only approximate potentially
infinite computations, thus the $\mathsf{Unfinished}$ tree indicates
an incomplete approximation. %
$\mathsf{Ret}\ v$ is a terminal computation that returns value
$v\,{:}\,A$. %
The constructor $\mathsf{Act}$ takes as a parameter an action $a :
\mathsf{action}~V~A$, as defined in Section~\ref{sec:formal}.
$ \mathsf{Seq}~T~K$ sequentially composes a $B$-returning tree $T$
with a continuation $K$ that takes $T$'s return value and generates
the rest of the approximation.
$\mathsf{Par}\ T_1\ T_2\ K$ is the parallel composition of trees $T_1$
and $T_2$, and a continuation $K$ that takes the pair of their results
when they join.  \CiC's iterated inductive definition permits the
recursive occurrences of $\mathsf{tree}$ to be \emph{nonuniform}
(e.g., $\mathsf{tree}\ B_i$ in $\mathsf{Par}$) and \emph{nested}
(e.g., the \emph{positive} occurrence of $\mathsf{tree}\ A$ in the
continuation). Since the \CiC function space includes case-analysis,
the continuation may branch upon the argument.
%
The $\mathsf{Morph}$ constructor embeds an index $x:X$, morphism $f :
\morphtpX{W}{V}{X}$, and tree $T \ty \mathsf{tree}~W~A$ for the
underlying computation. The constructor will denote $T$ should be
executed so that each of its actions is modified by $f$ with an index
$x$.
We can now define the denotational model of our programs; that is the
type $\mathsf{ST}~V~A$ of sets of trees, containing
$\mathsf{Unfinished}$.
\[
\mathsf{ST}~V~A \eqdef \{e : \mathsf{set}~(\mathsf{tree}~V~A) \mid \mathsf{Unfinished} \in e\}
\]

The denotations of the various constructors combine the trees of the
individual denotations, as shown below.
\[
\small
\begin{array}{r@{\ }c@{\ }l}
\mathsf{ret}\ (r \ty A) & \eqdef & \{\mathsf{Unfinished}, \mathsf{Ret}~r\}
\\
x \leftarrow e_1; e_2 & \eqdef & \{\mathsf{Unfinished}\} \cup \hbox{}\\
& & \qquad \{\mathsf{Seq}~T_1~K \mid T_1 \in e_1 \wedge \forall x \ldot K~x \in e_2\}
\\
e_1 \parallel e_2 & \eqdef & \{\mathsf{Unfinished}\} \cup \hbox{}\\
& & \qquad \{\mathsf{Par}~T_1~T_2~\mathsf{Ret} \mid T_1 \in e_1 \wedge T_2 \in e_2\}
\\
\langle a\rangle & \eqdef & \{\mathsf{Unfinished}, \mathsf{Act}~a\}
\\
\mathsf{morph}~f~x~e & \eqdef & \{\mathsf{Unfinished}\} \cup 
\{\mathsf{Morph}~f~x~T \mid T \in e\}
\end{array}
\]
The denotation of $\mathsf{ret}$ simply contains the trivial
$\mathsf{Ret}$ tree, in addition to $\mathsf{Unfinished}$, and
similarly in the case of $\langle a \rangle$. The trees for sequential
composition of $e_1$ and $e_2$ are obtained by pairing up the trees
from $e_1$ with those from $e_2$ using the $\mathsf{Seq}$ constructor,
and similarly for parallel composition and morphism application.

The denotations of composed programs motivate why we denote programs
by non-empty sets, i.e., why each denotation contains at least
$\mathsf{Unfinished}$. If we had a program $\mathsf{Empty}$ whose
denotation is the empty set, then the denotation of $x \leftarrow
\mathsf{Empty}; e'$, $\mathsf{Empty} \parallel e'$ and
$\mathsf{morph}~f~x~\mathsf{Empty}$ will all also be empty, thus
ignoring that the composed programs exhibit more behaviors. For
example, the parallel composition $\mathsf{Empty} \parallel e'$ should
be able to evaluate the right component $e'$, despite the left
component having no behaviors.
By including $\mathsf{Unfinished}$ in all the denotations, we ensure
that behaviors of the components are preserved in the composition. For
example, the parallel composition $\{\mathsf{Unfinished}\} \parallel
e'$ is denoted by the set:
\[
 \{\mathsf{Unfinished}\} \cup
 \{\mathsf{Par}~\mathsf{Unfinished}~T~\mathsf{Ret} \mid T \in e'\}
\]
which contains an image of each tree from $e'$, thus capturing the
behaviors of $e'$.
%


\subsection{Operational semantics of action trees}

\begin{figure}[t] 
{\small
\begin{mathpar}
\inferrule
 {\mathsf{unwind\_act}~\Delta~\bar{x}~\bar{x}'~a~s~s'~v}
 {\Delta \vdash \bar{x}, s, \mathsf{Act}~a \tsteps{\mathsf{\ChoiceAct}}
   \bar{x}', s'; \mathsf{Ret}~v} 
\\
%
\inferrule
 {~}
 {\Delta \vdash \bar{x}, s, \mathsf{Seq}~(\mathsf{Ret}~v)~K \tsteps{\SeqRet}
   \bar{x}, s, K~v}
\and
%
\inferrule
 {\Delta \vdash \bar{x}, s, T \tsteps{\Path} \bar{x}, s', T'}
 {\Delta \vdash \bar{x}, s, \mathsf{Seq}~T~K \tsteps{\SeqStep~\Path}
   \bar{x}', s', \mathsf{Seq}~T'~K}
\\
%
%
\inferrule
 {\Delta \vdash \bar{x}, s, T_1 \tsteps{\Path} \bar{x}', s', T'_1}
 {\Delta \vdash \bar{x}, s, \mathsf{Par}\ T_1\ T_2\ K \tsteps{\ParL~\Path}
  \bar{x}', s', \mathsf{Par}\ T'_1\ T_2\ K}
\and
\inferrule
 {\Delta \vdash \bar{x}, s, T_2 \tsteps{\Path} \bar{x}', s', T'_2}
 {\Delta \vdash \bar{x}, s, \mathsf{Par}\ T_1\ T_2\ K \tsteps{\ParR~\Path}
  \bar{x}', s', \mathsf{Par}\ T_1\ T'_2\ K}
\\
%
\inferrule
 {~}
 {\Delta \vdash \bar{x}, s, \mathsf{Par}\ (\mathsf{Ret}\ v_1)\ (\mathsf{Ret}\ v_2)\ K
   \tsteps{\ParRet} \bar{x}, s, K\ (v_1, v_2)} 
\\
%
\inferrule
 {~}
 {\Delta \vdash \bar{x}, s, \mathsf{Morph}~f~y~(\mathsf{Ret}~v)
   \tsteps{\MorphRet} \bar{x}, s, \mathsf{Ret}~v} 
\\
\inferrule
 {\Delta, f \vdash (\bar{x}, y), s, T \tsteps{\Path} (\bar{x}', y'), s', T'}
 {\Delta \vdash \bar{x}, s, \mathsf{Morph}~f~y~T \tsteps{\MorphStep~\Path}
   \bar{x}', s', \mathsf{Morph}~f~y'~T'}

\end{mathpar}}
 \caption{Judgment $\Delta \vdash \bar{x}, s, T \tsteps{\Path}
   \bar{x}', s', T'$, for operational semantics on trees, which
   reduces a tree with respect to the path $\Path$.}
 \label{fig:tsteps}
\end{figure}

The judgment for small-step operational semantics of action trees has
the form $\Delta \vdash \bar{x}, s_w, T_v \tsteps{\Path} \bar{x}',
s'_w, T'_v$ (Figure~\ref{fig:tsteps}). We explain the components of
this judgment next.

First, the component $\Delta$ is a morphism context. This is a
sequence, potentially empty, of morphism families
\[
f_0 : \morphtpX{V_1}{W}{X_0},
f_1 : \morphtpX{V_2}{V_1}{X_1}, \ldots, f_n : \morphtpX{V}{V_n}{X_n}
\]
We say that $\Delta$ has resource type $V \rightarrow W$, and index
type $(X_0, \cdots, X_n)$. An empty context $\cdot$ has resource type
$V \rightarrow V$ for any $V$.

Second, the components $\bar{x}$ and $\bar{x}'$ are tuples, of type
$(X_0, \cdots, X_n)$, and we refer to them as indexes. Intuitively,
the morphism context records the morphisms under which a program
operates. For example, if we wrote a program of the form
\[\mathsf{morph}\ f_0\ x_0\ (\cdots (\mathsf{morph}\ f_n\ x_n\ e) \cdots),\]
it will be that the trees that comprise $e$ execute under the morphism
context $f_0, \ldots, f_n$, with an index tuple $(x_0, \ldots, x_n)$.

Third, the components $s_w$ and $s'_w$ are $W$-states, and $T_v, T'_v
: \mathsf{tree}~V~A$, for some type $A$. The meaning of the judgment
is that a tree $T_v$, when executed in a state $s_w$, under the
context of morphisms $\Delta$ produces a new state $s'_w$ and residual
tree $T'_v$, encoding what is left to execute. The resource of the
trees and the states disagree (the states use resource $W$, the trees
use $V$), but the morphism context $\Delta$ relates them as
follows. Whenever the head constructor of the tree is an action, the
action will first be morphed by applying all the morphisms in $\Delta$
in order, to the transitions that constitute the head action,
supplying along the way the projections out of $x$ to the
morphisms. This will produce a new index $x'$ and an action on
$W$-states, which can be applied to $s_w$ to obtain $s'_w$.

Fourth, the component $\Path$ is of $\pathtp$ type, identifying the
position in the tree where we want to make a reduction.
\[
\small
\begin{array}{r@{\ }c@{\ }l@{\ }c@{\ }l@{\ }c@{\ }l@{\ }c@{\ }l}
\pathtp & \eqdef & \ChoiceAct &|& \SeqRet &|& \SeqStep~(\Path : \pathtp) &|&   \\
&&               \ParRet &|& \ParL~(\Path : \pathtp) &|& \ParR~(\Path : \pathtp) &|& \\
&&               \MorphRet &|& \MorphStep~(\Path : \pathtp)
\end{array}
\label{eq:path}
\]
The key are the constructors $\ParL~\Path$ and $\ParR~\Path$. In a
tree which is a $\mathsf{Par}$ tree, these constructors identify that
we want to reduce in the left and right subtree, respectively,
iteratively following the path $\Path$. If the tree is not a
$\mathsf{Par}$ tree, then $\ParL$ and $\ParR$ constructors will not
form a good path; we define further below when a path is good for a
tree. The other path constructors identify positions in other kinds of
trees. For example, $\ChoiceAct$ identifies the head position in the
tree of the form $\mathsf{Act}(a)$, $\SeqRet$ identifies the head
position in the tree of the form $\mathsf{Seq}~(\mathsf{Ret}~v)~K$
(i.e., it identifies a position of a beta-reduction), $\SeqStep~\pi$
identifies a position in the tree $\mathsf{Seq}~T~K$, if $\pi$
identifies a position within $T$, etc. We do not paths for trees of
the form $\mathsf{Unfinished}$ and $\mathsf{Ret}~v$, because these do
not reduce.

\newcommand{\tsquare}[9]{\mathsf{ComSquares}\ #1\ #2\ #3\ #4\ #5\ #6\ #7\ #8\ #9}

In order to define the operational semantics on trees, we next require
a few auxiliary notions.  First, we need the relation
$\tsquare{\Delta}{\bar{x}}{\bar{x}'}{s_v}{s'_v}{t_v}{s_w}{s'_w}{t_w}$
encoding the composition of the commutative diagrams below (each
commutative diagram is an instance of Figure~\ref{fig:spin}(b), for
one of the morphisms $f_0$, $f_1$,\ldots, $f_n$, in the context
$\Delta = f_0; f_1; \cdots; f_n$).
\[
\tikzcdset{arrow style=tikz, diagrams={>=stealth}}
\begin{tikzcd}
s'_v & s'_{n} \arrow[l, "f_{n\Sigma}\ x'_n", swap] & s'_{1} \arrow[l, dotted, dash] & {s'_w} \arrow[l, "f_{0\Sigma}\ x'_0", swap] 
                  \arrow[dl, phantom, "\llcorner", very near start] \\ 
s_v \arrow[u, "t_v"] & s_{n} \arrow[l, "f_{n\Sigma}\ x_n"] \arrow [u, "t_n" description] & s_{1} \arrow[l, dotted, dash] \arrow[u, "t_1" description] & {s_w} \arrow[l, "f_{0\Sigma}\ x_0"] \arrow[u, "t_w = f_{0\Delta}\ x_0\ s_w\ t_1", swap] 
\end{tikzcd} 
\]

The relation captures that $t_v$ is morphed by iterating the state
component of the morphisms in $\Delta$, starting from the state $s_w$,
and passing along the elements of the tuple $\bar{x}$ or type $(X_0,
\cdots, X_n)$, until we reach the state $s_v$. Then transition $t_v$
is executed in $s_v$ to obtain $s'_v$, and the transition components
of the morphisms are ran backwards to obtain $t_w$, and the associated
end state $s'_w$. Formally:
\[
\small
\begin{array}{l}
\tsquare{\cdot}{()}{()}{s_v}{s'_v}{t_v}{s_w}{s'_w}{t_w} \eqdef \hbox{}\\
  \qquad s_v = s_w \wedge s'_v = s'_w \wedge t_v = t_w\\
\tsquare{(f_0 : \morphtpX{V_1}{W}{X_0}), \Delta)}{(x_0, \bar{x})}{(x'_0, \bar{x}')}{}{}{}{}{}{}\\ 
         \hspace{18mm}{s_v}~{s'_v}~{t_v}~{s_w}~{s'_w}~{t_w} \eqdef \hbox{}\\
\qquad\!\!\!\begin{array}[t]{l}
  \exists s_{1}\ s'_{1}\ t_1\ldot \tsquare{\Delta}{\bar{x}}{\bar{x}'}{s_v}{s'_v}{t_v}{s_{1}}{s'_{1}}{t_1} \wedge \hbox{}\\
  \xmorpheq{s_{1}}{s_w}{f_{0\Sigma}}{x_0} \wedge \xmorpheq{s'_{1}}{s'_w}{f_{0\Sigma}}{x'_0}
 \end{array}
\end{array}
\]

We will often abbreviate $\mathsf{ComSquares}$, and write: 
\[
\begin{array}{l}
\mathsf{unwind\_tr}\ \Delta\ \bar{x}\ \bar{x}'\ t_v\ s_w\ s'_w \eqdef \hbox{}\\
\quad \exists s_v\ s'_v\ t_w\ldot \tsquare{\Delta}{\bar{x}}{\bar{x}'}{s_v}{s'_v}{t_v}{s_w}{s'_w}{t_w}\\
\mathsf{unwind\_act}\ \Delta\ \bar{x}\ \bar{x}'\ a\ s_w\ s'_w\ v \eqdef \hbox{}\\
\quad \exists s_v\ s'_v\ t_v\ t_w\ldot \!\!\!\begin{array}[t]{l}
   a\ s_v = (t_v, v) \wedge \hbox{}\\
  \tsquare{\Delta}{\bar{x}}{\bar{x}'}{s_v}{s'_v}{t_v}{s_w}{s'_w}{t_w}
\end{array}
\end{array}
\]
Given a context $\Delta$, input indices $\bar{x}$, input state $s_w$
and input transition $t_v$, $\mathsf{unwind\_tr}$ ``unwinds'' the
transition, in the sense that it computes the output index $\bar{x}'$,
output state $s'_w$, obtained by the morphing iteratively by the
morphisms in $\Delta$. The abbreviation $\mathsf{unwind\_act}$ does
the same, but for input action $a : \mathsf{action}{V}{A}$. In the
process, $\mathsf{unwind\_act}$ also computes the return value $v$.

In the frequent cases where $\Delta$ is the empty context (and
correspondingly, $\bar{x}$ and $\bar{x}'$ are empty tuples $()$), we
will abbreviate the judgment from Figure~\ref{fig:tsteps}, and simply
write $s, T \tsteps{\Path} s', T' $.

The operational semantics on trees in Figure~\ref{fig:tsteps} may not
make a step on a tree for two different reasons. The first, benign,
reason is that the the chosen path $\Path$ does not actually determine
an action or a redex in the tree $T$. For example, we may have $T =
\mathsf{Unfinished}$ and $\Path = \mathsf{ParR}$. But we can choose
the right side of a parallel composition only in a tree whose head
constructor is $\mathsf{Par}$, which is not the case with
$\mathsf{Unfinished}$.  We consider such paths that do not determine
an action or a redex in a tree to be ill-formed.
The second reason arises when $\Path$ is actually well-formed. In that
case, the constructors of the path uniquely determine a number of
rules of the operational semantics that should be applied to step the
tree. However, the premises of the rules may not be satisfies. For
example, in the $\mathsf{ChoiceAct}$ rule, there may not exist a $v$
such that $\unwind~\Delta~\bar{x}~\bar{x}'~a~s~s'~v$.
To differentiate between these two different reasons, we first define
the notion of well-formed, or \emph{good} path, for a given tree.
\begin{definition}[Good paths]\label{def:good}
Let $T \ty \mathsf{tree}~V~A$ and $\Path$ be a path.  Then the
predicate $\mathsf{good}~T~\Path$ recognizes \dt{good} paths for a
tree $T$ as follows:
\[
\small
\begin{array}{r@{\ }l@{}l@{}c@{}l}
  \mathsf{good} & (\mathsf{Act}~a) & \ChoiceAct & ~\eqdef~ & \mathsf{true}   
  \\
  \mathsf{good} & (\mathsf{Seq}~(\mathsf{Ret}~v)~\_) & \SeqRet & ~\eqdef~ & \mathsf{true}  
  \\
  \mathsf{good} & (\mathsf{Seq}~T~\_) & \SeqRet~\Path & ~\eqdef~ & \mathsf{good}~T~\Path 
  \\
  \mathsf{good} & (\mathsf{Par}~(\mathsf{Ret}~\_)~(\mathsf{Ret}~\_)~\_) & \ParRet & ~\eqdef~ & \mathsf{true}  
  \\
  \mathsf{good} & (\mathsf{Par}~T_1~T_2~\_) & \ParL~\Path & ~\eqdef~ & \mathsf{good}~T_1~\Path
  \\
  \mathsf{good} & (\mathsf{Par}~T_1~T_2~\_) & \ParR~\Path & ~\eqdef~ & \mathsf{good}~T_2~\Path
  \\
  \mathsf{good} & (\mathsf{Morph}~f~x~(\mathsf{Ret}~\_)) & \MorphRet & ~\eqdef~ & \mathsf{true}
  \\
  \mathsf{good} & (\mathsf{Morph}~f~x~T) & \MorphStep~\Path & ~\eqdef~ & \mathsf{good}~T~\Path \\
  \mathsf{good} & T & \Path & ~\eqdef~ & \mathsf{false} \text{~~otherwise}
\end{array}
\]
\end{definition}

\begin{definition}[Safe configurations]\label{def:safe}
We say that a state $s$ is \dt{safe} for the tree $T$ and path
$\Path$, written $s \in \tsafe~t~\Path$ if:
\[
\mathsf{good}~T~\Path \rightarrow \exists s'\ T'\ldot s, T \tsteps{\Path} s', T'
\]
\end{definition}

Notice that in the above definition, the trees $\mathsf{Unfinished}$
and $\mathsf{Ret}\ v$ are safe for any path, simply because there are
no good paths for them, as such trees are terminal. On the other hand,
a tree $\mathsf{Act}~a$ does have a good path, namely $\ChoiceAct$,
but may be unsafe, if the action $a$ is not defined on input state
$s$. For example, the $a$ may be an action for reading from some
pointer $x$, but that pointer may not be allocated in the state $s$.

Safety of a tree will be an important property in the definition of
Hoare triples, where we will require that a precondition of a program
implies that the trees comprising the program's denotation are safe
for every path.

The following are several important lemmas about trees and their
operational semantics, which lift most of the properties of
transitions, to trees.

\begin{lemma}[Coverage of stepping by transitions]
Let $\Delta : V \rightarrow W$, and $\Delta \vdash \bar{x}, s, T
\tsteps{\Path} \bar{x}', s', T'$. Then either the step corresponds to
an idle transition, that is, $(\bar{x}, s) = (\bar{x}', s')$, or there
exists a transition $a \in \transof{V}$, such that
$\mathsf{unwind\_tr}~\Delta~\bar{x}~\bar{x}'~a~s~s'~v$.
\end{lemma}

\begin{lemma}[Other-fixity of stepping]
Let $\Delta : V \rightarrow W$ and $\Delta \vdash \bar{x}, s, T
\tsteps{\Path} \bar{x}', s', T'$. Then $a_o(s) = a_o(s')$.
\end{lemma}

\begin{lemma}[Stepping preserves state spaces]
Let $\Delta : V \rightarrow W$ and $\Delta \vdash \bar{x}, s, T
\tsteps{\Path} \bar{x}', s', T'$. If $s \in \papp{\Sigma}{W}$ then $s'
\in \papp{\Sigma}{W}$.
\end{lemma}

\begin{lemma}[Stability of stepping]
Let $\Delta : V \rightarrow W$ and $\Delta \vdash \bar{x}, s, T
\tsteps{\Path} \bar{x}', s', T'$. Then $s^\top \osteps[W]{} s'^\top$.
\end{lemma}

\begin{lemma}[Determinism of stepping]
Let $\Delta : V \rightarrow W$ and $\Delta \vdash \bar{x}, s, T
\tsteps{\Path} \bar{x}', s', T'$, and $\Delta \vdash \bar{x}, s, T
\tsteps{\Path} \bar{x}'', s'', T''$. Then $\bar{x}' = \bar{x}''$, $s'
= s''$ and $T' = T''$.
\end{lemma}

\begin{lemma}[Locality of stepping]
Let $\Delta : V \rightarrow W$ and $\Delta \vdash \bar{x}, (s \zig p),
T \tsteps{\Path} \bar{x}', s', T'$. Then there exists $s''$ such that
$s' = s''\zig p$, and $\Delta \vdash \bar{x}, (s \zag p), T
\tsteps{\Path} \bar{x}', (s'' \zag p), T'$.
\end{lemma}


\begin{lemma}[Safety-monotonicity of stepping]
If $s \zig p \in \tsafe~T~\Path$ then $s \zag p \in \tsafe~T~\Path$.
\end{lemma}

\begin{lemma}[Frameability of stepping]
Let $s \zig p \in \tsafe~T~\Path$, and $s \zag p, T \tsteps{\Path} s',
T'$. Then there exists $s''$ such that $s' = s'' \zag p$ and $s \zig
p, T \tsteps{\Path}, s'' \zig p, T'$.
\end{lemma}

The following lemma is of crucial importance, as it relates stepping
with morphisms. In particular, it says that the steps of a tree are
uniquely determined, no matter the morphism under which it
appears. Intuitively, this holds because each transition that a tree
makes has a unique image under a morphism $f : \morphtpX{V}{W}{X}$.

%

%

\begin{lemma}[Stepping under morphism]
Let $f : \morphtp{V}{W}{X}$ and $\xmorpheq{s_v}{s_w}{f}{x}$. Then the
following hold:
\begin{enumerate}
\item if $s_v, T \tsteps{\Path} s'_v, T'$, then $\exists x'\ s'_w\ldot
  \xmorpheq{s'_v}{s'_w}{f}{x'}$ and $f \vdash (x), s_w, T \tsteps{\Path}
  (x'), s'_w, T'$.
\item if $f \vdash (x), s_w, T \tsteps{\Path} (x'), s'_w, T'$, then
  $\exists x'\ s'_v\ldot \xmorpheq{s'_v}{s'_w}{f}{x'}$ and $s_v, T
  \tsteps{\Path} s'_v, T'$.
\end{enumerate}
\end{lemma}

The first property of this lemma relies on the fact that for a step
over states in $V$, we can also find a step over related states in
$W$, i.e., that $f$ encodes a simulation. The second property relies
on the fact that $f$'s state component is a function in the
contravariant direction. Thus, for each $s_w$ there are unique $x$ and
$s_v$, such that $\xmorpheq{s_v}{s_w}{f}{x}$.

\subsection{Predicate transformers}\label{sec:modal-pred}

In this section we define a number of predicate transformers over
trees that ultimately lead to defining the $\vrf$ predicate
transformer on programs.

\begin{definition}[Modal Predicate Transformers]
  \label{def:modpred} 
  Let $T : \mathsf{tree}~V~A$, and $\Paths$ be a sequence of
  paths. Also, let $X$ be an assertion over $V$-states and $V$-trees,
  and $Q$ be an assertion over $A$-values and $V$-states. We define
  the following predicate transformers: 
\[
\small
\begin{array}{l}
\mathsf{always}^{\Paths}~T~X~s \eqdef \hbox{}\\
\qquad \mathsf{if}~\Paths = \Path :: \Paths'~\mathsf{then}~\hbox{}\\
\qquad\quad \forall s_2\ldot (s \osteps[V]{} s_2) \rightarrow \tsafe~T~\Path~s_2 \wedge X~s_2~T \wedge \hbox{}\\
\qquad\qquad \forall s_3~T'\ldot~~ (s_2, T \tsteps{\Path} s_3, T') \rightarrow
   \mathsf{always}^{\Paths'}~T_2~X~s_3\\
\qquad \mathsf{else}~ \forall s_2\ldot (s \osteps[V]{} s_2) \rightarrow X~s_2~T
\\
\mathsf{always}~T~X~s \eqdef \forall \Paths\ldot \mathsf{always}^{\Paths}~T~X~s
\\\\
\mathsf{after}~T~Q \eqdef 
 \mathsf{always}~T~(\lambda~s'~T'\ldot\forall v\ldot T' = \mathsf{Ret}~v \rightarrow Q~v~s')
\end{array}
\]

\end{definition}

The helper predicate $\mathsf{always}^{\Paths}~T~X~s$ expresses the
fact that starting from the state $s$, the tree $T$ remains safe and
the user-chosen predicate $X$ holds of all intermediate states and
trees obtained by evaluating $T$ in the state $s$ according to the
sequence of paths $\Paths$. The predicate $X$ remains valid under any
any environment steps of the resource $V$.

The predicate $\mathsf{always}~T~X~s$ quantifiers over the path
sequences. Thus, it expresses that $T$ is safe and $X$ holds after any
finite number of steps which can be taken by $T$ in $s$.

The predicate transformer $\mathsf{after}~T~Q$ encodes that $T$ is
safe for any number of steps; however, $Q\ v\ s'$ only holds if $T$
has been completely reduced to $\mathsf{Ret}\ v$ and state $s'$. In
other words $Q$ is a postcondition for $T$, as it is required to hold
only if, and after, $T$ has terminated.

Now we can define the $\vrf$ predicate transformer on programs, by
quantifying over all trees in the denotation of a program.
\[
\vrf~e~Q~s \eqdef \forall T\in e\ldot \mathsf{after}~T~Q~s
\]

This immediately gives us a way to define when a program $e$ has a
precondition $P$ and postcondition $Q$: when all the trees in $T$ have
a precondition $P$ and postcondition $Q$ according to the
$\mathsf{after}$ predicate, or equivalently, when
\[
\forall s \in \papp{\Sigma}{V}\ldot P\ s \rightarrow \vrf~e~Q~s
\]
which is the formulation we used in Section~\ref{sec:formal} to define
the Hoare triples.

We can now state the following soundness theorem, each of whose three
components has been established in the Coq files.

\begin{theorem}[Soundness]~
\begin{itemize}
\item All the properties of $\vrf$ predicate transformer from
  Section~\ref{sec:formal} are valid. 

\item The sets $\mathsf{ST}~V~A$ and $\{P\}\ A\ \{Q\}$ are complete
  lattices under subset ordering with the set
  $\{\mathsf{Unfinished}\}$ as the bottom. Thus one can compute the
  least fixed point of every monotone function by Knaster-Tarski
  theorem.

\item All program constructors are monotone.
\end{itemize}
\end{theorem}

\fi

\end{document}